\documentclass[11pt,a4paper]{article}
\usepackage{jcappub}
\usepackage{amssymb,amsmath,amsfonts}
\usepackage{moreverb}
\usepackage{bm}
\usepackage{amsfonts}
\usepackage{latexsym}
\usepackage{graphicx}
\usepackage{amsmath}
\usepackage{palatino}
\usepackage{mathpazo}
\usepackage{textcomp}
\usepackage{subfigure}
\usepackage{float}
\usepackage{booktabs}
\usepackage{mathtools}
\usepackage{dcolumn}
\usepackage{ragged2e}
\usepackage{amssymb,amsmath,amsfonts}
\usepackage{bm}
\usepackage{amsfonts}
\usepackage{latexsym}
\usepackage{graphicx}
\usepackage{amsmath}
\usepackage{palatino}
\usepackage{mathpazo}
\usepackage{textcomp}
\usepackage{subfigure}
\usepackage{float}
\usepackage{booktabs}
\usepackage{mathtools}
\usepackage{dcolumn}
\usepackage{ragged2e}

\usepackage{amsmath}
\usepackage{xcolor}
\usepackage{orcidlink}
\DeclareMathOperator{\sgn}{sgn}

\usepackage{hyperref}
\hypersetup{
    colorlinks=true,
    linkcolor=red,
    citecolor=blue,
    filecolor=magenta,      
    urlcolor=cyan,
    pdftitle={Scalar field evolution at background and perturbation levels for a broad class of potentials},
    pdfpagemode=FullScreen} 

\begin{document}

\color{black}       %% For one column

\title{Scalar field evolution at background and perturbation levels for a broad class of potentials}

\author[a,b,1]{Genly Leon\orcidlink{0000-0002-1152-6548} \note{Corresponding author.}}

\author[c, d]{Saikat Chakraborty\orcidlink{0000-0002-5472-304X}}

\author[e]{Sayantan Ghosh\orcidlink{0000-0002-3875-0849}}

\author[e]{Raja Solanki\orcidlink{0000-0001-8849-7688}}

\author[e]{P.K. Sahoo\orcidlink{0000-0003-2130-8832}}

\author[f]{and Esteban Gonz\'alez\orcidlink{0000-0001-6769-5722}}

\affiliation[a]{Departamento de Matem\'{a}ticas, Universidad Cat\'{o}lica del 
Norte, Avda.
Angamos 0610, Casilla 1280 Antofagasta, Chile}

\affiliation[b]{Institute of Systems Science, Durban University of Technology, PO 
Box 1334,
Durban 4000, South Africa}

\affiliation[c]{The Institute for Fundamental Study ``The Tah Poe Academia Institute",\\ Naresuan University, Phitsanulok 65000, Thailand}

\affiliation[d]{Center for Space Research, North-West University, Mahikeng 2745, South Africa}

\affiliation[e]{Department of Mathematics, Birla Institute of Technology and
Science-Pilani,\\ Hyderabad Campus, Hyderabad-500078, India}

\affiliation[f]{Direcci\'on de Investigaci\'on y Postgrado, Universidad de Aconcagua, Pedro de Villagra 2265, Vitacura, 7630367 Santiago, Chile}

\emailAdd{genly.leon@ucn.cl} 

\emailAdd{saikat.ch@nu.ac.th}

\emailAdd{sayantanghosh.000@gmail.com}

\emailAdd{rajasolanki8268@gmail.com}

\emailAdd{pksahoo@hyderabad.bits-pilani.ac.in}

\emailAdd{esteban.gonzalez@uac.cl}

%%%%%%%%%%%%%%%%%%%%%%%%%%%%%%%%%%%%  DATE  %%%%%%%%%%%%%%%%%%%%%%%%%%%%%%%%%%%%
\date{\today}
\abstract{In this paper, we investigate a non-interacting scalar field cosmology with an arbitrary potential using the $f$-deviser method that relies on the differentiability properties of the potential. Using this alternative mathematical approach, we present a unified dynamical system analysis at a scalar field's background and perturbation levels with arbitrary potentials. For illustration, we consider a monomial and double exponential potential. These two classes of potentials comprise the asymptotic behaviour of several classes of scalar field potentials, and, therefore, they provide the skeleton for the typical behaviour of arbitrary potentials. Moreover, we analyse the linear cosmological perturbations in the matterless case by considering three scalar perturbations: the evolution of the Bardeen potentials, the comoving curvature perturbation, the so-called Sasaki-Mukhanov variable, or the scalar field perturbation in uniform curvature gauge. Finally, an exhaustive dynamical system analysis for each scalar perturbation is presented, including the evolution of Bardeen potentials in the presence of matter.}


\maketitle

\section{Introduction}\label{sec1}

Scalar fields are prominent in the physical description of the Universe in the inflationary scenario \cite{Guth:1980zm} and can be used to explain the late-time Universe's acceleration. Although $\Lambda$CDM has an excellent concordance with observations, describes the structure formation, and successfully provides a late-time acceleration \cite{Carroll:2000}, $\Lambda$ has yet to succeed in quantifying the quantum vacuum fluctuations \cite{Zeldovich, Weinberg}. That is the primary motivation for introducing Dark Energy as an alternative to $\Lambda$CDM. Some examples are quintessence field \cite{Ratra:1987rm, Rubano:2001su, Parsons:1995kt,  Barrow:2016qkh}, a phantom scalar field (which, however, suffers ghosts instabilities \cite{Urena-Lopez:2005pzi}),  a quintom scalar field model  \cite{Cai:2009zp, Guo:2004fq, Feng:2004ff, Zhang:2005eg, Zhang:2005kj, Lazkoz:2006pa, Lazkoz:2007mx, Setare:2008pz, Setare:2008pc, Leon:2012vt, Leon:2018lnd, Mishra:2018dzq, Marciu:2020vve, Dimakis:2020tzc},   a chiral cosmology \cite{Dimakis:2020tzc, Chervon:2013btx, Paliathanasis:2020wjl}, or multi-scalar field models. The latter describes various epochs of the cosmological history \cite{Elizalde:2004mq, Elizalde:2008yf, Paliathanasis:2018vru}. On the other hand, the Hubble constant value measured with local observations (see SH0ES \cite{Riess:2019cxk}) is in tension with that estimated from early observations (see Planck \cite{Planck:2018}). A possible alternative to solve this tension is considering extensions beyond $\Lambda$CDM \cite{DiValentino:2021izs}. There could be other reasons for the $H_0$ tension, e.g., incomprehension between the SnIa absolute magnitude and the Cepheid-based distance ladder, rather than an exotic late-time physics  \cite{Efstathiou:2021ocp}. Even more,  $H_0$ tension seems to permeate Dark Energy Models (including quintessence), whereby $H_0$ is sent to lower values by any dark energy model with $w_{DE}(z) > -1$, whereas local (model-independent) $H_0$ determinations are biased to more significant values than Planck-$\Lambda$CDM \cite{Banerjee:2020xcn, Lee:2022cyh}. 
Even though the exploration of scalar field models has the attention of several researchers, such that 
Scalar field evolution at the background level was studied in several works, say \cite{Capozziello:2005tf, Nojiri:2005pu, Briscese:2006xu, Nojiri:2006ww, Capozziello:2005mj, Astashenok:2012kb, Astashenok:2012tv, Ito:2011ae, Frampton:2011rh, Bamba:2014daa, Odintsov:2018zai, Odintsov:2018uaw}. 
To analyse the early and late-time dynamics of cosmological problems, the perturbation and averaging methods \cite{verhulst2006method,1981ApJ...250..432B, Fenichel, Fusco1989SlowmotionMD, Berglund, holmes2012introduction, kevorkian2013perturbation} were used in  \cite{Rendall:2006cq, Alho:2015cza, Alho:2019pku, Fajman:2020yjb, Fajman:2021cli, Leon:2021lct, Leon:2021rcx} to single field scalar field cosmologies, and for scalar field cosmologies with two scalar fields which interact only gravitationally with the matter in \cite{Chakraborty:2021vcr}. In reference, \cite{Leon:2020pvt}, scalar field cosmology with a generalised harmonic potential was investigated in Friedmann-Lemaître-Robertson-Walker with flat and negative spatial curvature and for Bianchi I metrics. Besides, an interaction between the scalar field and matter was considered in the conservation equations. In these references, asymptotic methods and the theory of averaging in nonlinear dynamical systems are essential tools to obtain relevant information about the solution space of scalar field with generalised harmonic potential in a vacuum, and adding matter, \cite{Leon:2021lct, Leon:2021rcx, Chakraborty:2021vcr, Leon:2020pvt, Leon:2019iwj, Leon:2020pfy, Leon:2020ovw,   Leon:2021hxc}. The amplitude-angle transformation  was used in \cite{Fajman:2020yjb, Leon:2020pvt,  Leon:2020pfy, Leon:2020ovw, Llibre:2012zz}. In references \cite{Leon:2020pfy, Leon:2020ovw}, scalar field cosmologies with generalised harmonic potentials and exponential couplings to matter in the sense of \cite{Leon:2008de, Giambo:2009byn, Tzanni:2014eja} were examined.   In  \cite{Fajman:2021cli},  a theorem about the large-time behaviour of solutions of Spatially Homogeneous (SH) cosmology with oscillatory behaviour was presented. Moreover, slow-fast methods were used, for example, in analysing theories based on a Generalised Uncertainty Principle (GUP), say in \cite{Paliathanasis:2015cza, Paliathanasis:2021egx}. In \cite{Paliathanasis:2021egx}, a preliminary study of linear perturbations in the matter-dominated phase in the context of GUP was presented. More precisely, the dynamical equations for linear cosmological perturbations were derived, forming a singular differential equations system. In contrast to the usual quintessence, one can explicitly write the perturbed equations' solution in fast and slow manifolds. The extra components enhance the scalar perturbations' growth due to the higher-order derivative terms of the GUP in the fast manifold. However, the scalar perturbations either decay, grow or describe an oscillatory solution in the slow manifold. Consequently, the perturbation equations are also affected by the minimum length  \cite{Paliathanasis:2021egx}. 

Similarly, dynamical system methods are useful for investigating scalar field cosmologies for a wide class of potentials. To use this procedure and to handle the involved
differentiation, it is necessary to determine a specific potential form
$V(\phi)$ of the scalar field $\phi$. This procedure has the disadvantage that for each different potential, one must repeat all the calculations
from the beginning. Therefore, developing an
extended method that could handle the potential differentiation in
a unified way would be beneficial, without the need for any {\it{a priori}} specification. That is the method of $f$-devisers improved in \cite{Escobar:2013js}  and
applied in \cite{delCampo:2013vka} for scalar-field Friedman-Lema\^{i}tre-Robertson-Walker (FLRW) cosmologies in the
presence of a Generalised Chaplygin  Gas. With this method, it can be studied the classes of models discussed in \cite{Ratra:1987rm, Wetterich:1987fm, Yearsley:1996yg, Sahni:1999qe, Sahni:1999gb, Urena-Lopez:2000ewq, Matos:2000ng, Cardenas:2002np, Matos:2009hf, Copeland:2009be, Leyva:2009zz, delCampo:2013vka, Lidsey:2001nj, Pavluchenko:2003ge, Barreiro:1999zs, Gonzalez:2007hw, Gonzalez:2006cj, Peebles:1987ek, Abramo:2003cp, Aguirregabiria:2004xd, Copeland:2004hq, Saridakis:2009pj, Saridakis:2009ej, Leon:2009dt, Chang:2013cba, Skugoreva:2013ooa, Pavlov:2013nra}.

On the other hand, there is an interest in simultaneously investigating cosmological linear perturbations and background equations. One can obtain a unified dynamical system analysis at a scalar field cosmology's background and perturbation levels in cosmological studies. That can be done using the methods by \cite{Bardeen:1980kt, Mukhanov, Brandenberger:1992dw, Brandenberger:1993zc, Brandenberger:1992qj, dunsby_1997, amendola_tsujikawa_2010, Amendola:1999dr, Basilakos:2019dof, Alho:2019jho, Alho:2020cdg} (see references for the notation as well as for the theory to improve the background analysis of a cosmological model). Generally, one can investigate the dynamical system for the model consisting of a system of autonomous nonlinear first-order ordinary differential equations. The state-space $S$ has a product structure $S = B \times P$,  
where $B$ is the background state space, which describes the dynamics of a Robertson-Walker (RW) background, and $P$ are the perturbation state space. This space contains Fourier decomposed
gauge-invariant variables that describe linear cosmological perturbations. In this way, the background dynamics determine the perturbations' dynamics. Several recent studies examine the stability of cosmological perturbations on top or in an extended phase space that incorporates both perturbed scalar quantities and normalised (background) phase space variables \cite{dunsby_1997, amendola_tsujikawa_2010, Amendola:1999dr, Basilakos:2019dof, Alho:2019jho, Alho:2020cdg, Uggla:2011jn, Uggla:2011hs, Uggla:2012gg, Uggla:2013paa, Uggla:2013kya, Uggla:2014hva, Uggla:2018fiy, Uggla:2018cct, Uggla:2019rho, Landim:2019lvl, Uggla:2019zdm, Khyllep:2021wjd, Khyllep:2022spx}. 

In \cite{Basilakos:2019dof}, the authors performed this dynamical system analysis of the background and perturbation equations for a $\Lambda$CDM cosmology and quintessence scenario with exponential potential in a unified way. For the $\Lambda$CDM cosmology, the perturbations do not change the stability of the late-time attractor of the background equations, and the system still results in the dark-energy dominated de Sitter solution, with a transition by a dark-matter era with growth index $\gamma\approx 6/11$. Here $\gamma$ is defined through the relation $d \ln \delta_m/d \ln a \approx \Omega_m^{\gamma}$, where $\delta_m$ is the matter contrast, and $\Omega_m$, the fractional energy density of matter.   In the case of quintessence, incorporating linear perturbations results in a change in the stability and properties of the background evolution. The only conditionally stable points present either an exponentially increasing matter clustering not
favoured by observations or suffering Laplacian instabilities and, thus, are not of physical interest. This result
severely disadvantages quintessence cosmology compared to the $\Lambda$CDM paradigm. In this line, the work \cite{Alho:2019jho} introduced a dynamical system method to describe linear scalar
and tensor perturbations of the $\Lambda$CDM model. That provided pedagogical examples showing the global illustrative powers of dynamical systems in cosmological perturbations. It discussed the validity of the perturbations as approximations to the Einstein field equations. Furthermore, the linear growth rate,  $d \ln \delta_m/d \ln a \approx \Omega_m^{\gamma}$ was corrected to $d \ln \delta_m/d \ln a \approx \Omega_m^{\frac{6}{11}} -\frac{1}{70}(1-\Omega_m)^{\frac{5}{2}}$, and showed that it is much more accurate than the previous ones in the literature. That was the starting point of a series of technical papers. For example, in \cite{Alho:2020cdg}, a new regular dynamical system was derived on a three-dimensional compact state space describing linear scalar perturbations of spatially flat RW geometries for relativistic models with a minimally coupled scalar field with exponential potential. That enables them to construct the global solution space, where known solutions are shown to reside on some invariant sets. They use their dynamical systems approach to obtain new results about the comoving and uniform density curvature perturbations.
Finally, they show how to extend this approach to more general scalar field potentials. That leads to state spaces where the state space of the models with an exponential potential appears as invariant boundary sets, thereby illustrating their role as building blocks in a hierarchy of increasingly complex cosmological models. 
More generalisations appeared in \cite{Landim:2019lvl} and \cite{Sharma:2021ivo, Sharma:2021ayk}, which examined the imprints of interacting dark energy in linear scalar field perturbations. These results extend the analysis of \cite{Basilakos:2019dof}. Moreover, in reference, \cite{Tot:2022dpr} investigated the linear cosmological perturbations for a two-field quintom model interacting through the kinetic terms, following the results of  \cite{Chervon:2013btx} for N-field chiral action.

In \cite{Khyllep:2021wjd}, the authors applied the formalism of \cite{Basilakos:2019dof} to investigate interacting dark energy scenarios at the background and the perturbation levels in a unified way. An extra perturbation variable related to the matter over-density was introduced. The combined analysis found critical points describing the non-accelerating matter-dominated epoch with the proper growth of matter structure. These saddles provide the natural exit from this phase. Furthermore, late-time stable attractors correspond to dark energy-dominated accelerated solutions with constant matter perturbations. It is claimed that interacting cosmology describes the matter and dark energy epochs correctly, both at the background and perturbation levels, which reveals the capabilities of the interaction. 

In \cite{Khyllep:2022spx}, the authors studied cosmological models based on $f(Q)$  gravity, which is based on the non-metricity scalar $Q$ \cite{BeltranJimenez:2017tkd}. The systems were analysed for background and perturbation levels using a dynamical system analysis. Two $f(Q)$  models of the literature are examined: the power law and the exponential ones. Both cases obtained a matter-dominated saddle with the correct growth rate of matter perturbations. This epoch is followed by the transition to a stable dark-energy-dominated accelerated Universe in which matter perturbations remain constant. Furthermore, analysing the behaviour of $f\sigma8$ was deduced that the models fit the
observational data successfully, obtaining a behaviour similar to that of the $\Lambda$CDM scenario. However, the exponential model does not possess  $\Lambda$CDM  as a limit. That is, through the independent approach of dynamical systems, it was verified that $f(Q)$ gravity can be considered an up-and-coming alternative to the $\Lambda$CDM concordance model.

This paper investigates a non-interacting scalar field cosmology with an arbitrary potential using the $f$-deviser method. We present a unified dynamical system analysis at a scalar field's background and perturbation levels with arbitrary potentials using this alternative mathematical approach. Using this procedure, we perform a dynamical system analysis of Background quantities using Hubble-normalised variables. For simplicity, we assume the matterless case for analysing linear cosmological perturbations.
Nonetheless, our analysis with perturbation will be perfectly viable during scalar field-dominated epochs of the Universe, e.g. inflation and late-time acceleration. Following the line of Ref. \cite{Alho:2020cdg}, we investigated the dynamics of linear scalar cosmological perturbations for a generic scalar field model by dynamical systems methods. We considered three types of gauge-invariant scalar perturbation quantities. For the case of a single scalar field, we investigate the Bardeen potentials \cite{Bardeen:1980kt, Mukhanov, Brandenberger:1992dw, Brandenberger:1993zc, Brandenberger:1992qj}, the comoving curvature perturbation  \cite{DeFelice:2010aj}, and the so-called Sasaki-Mukhanov variable or the scalar field perturbation in uniform curvature gauge \cite{Kodama:1984ziu, Mukhanov:1988jd}. An exhaustive dynamical system analysis for each scalar perturbation will be presented.

The paper is organised as follows: In section \ref{sec2}, we present the field equations for a scalar field minimally coupled to gravity, with an arbitrary potential $V(\phi)$ in the presence of matter. 
We discuss there the $f$-devisers method. In section \ref{sect:3}, we perform a dynamical system analysis of background quantities using Hubble-normalised variables and the method of $f$-devisers. For illustration, we consider the monomial potential in subsection \ref{sect:3-1} as a first example. This potential $V(\phi)=\left|\frac{\mu}{n}\right|\phi^{n}$ has been investigated in \cite{Ratra:1987rm, Peebles:1987ek, Abramo:2003cp, Aguirregabiria:2004xd, Copeland:2004hq, Saridakis:2009pj, Saridakis:2009ej, Leon:2009dt, Chang:2013cba, Skugoreva:2013ooa, Pavlov:2013nra}. As a second example, we study the usual exponential potential in section \ref{sect:3-2}. For $f=0$ and $\lambda$ constant, we recover the quintessence scenario with an exponential potential $V=V_0 e^{-\lambda\phi}$ as studied in \cite{Copeland:1997et}. As a final example, in section \ref{sect:3-3}, we investigate the double exponential, say, 
$V(\phi)=V_1 e^{\alpha \phi}+ V_2 e^{\beta \phi}$  \cite{Barreiro:1999zs, Gonzalez:2007hw, Gonzalez:2006cj}. This example contains the particular case of the hyperbolic cosine $V(\phi)=\frac{1}{2} \left(e^{\alpha \phi}+ e^{-\alpha \phi}\right)$ by setting $V_1=
 V_2=1/2$ and $\beta=-\alpha$. When one of the exponents is zero, this corresponds to the exponential potential plus a Cosmological Constant \cite{Yearsley:1996yg, Pavluchenko:2003ge, Cardenas:2002np}. The potentials that are sums of two exponents are interesting in the context of $F(R)$ gravity because the conformal transformation of metric gives
$F(R)$ in analytic form \cite{Vernov:2019ubo}.  These two classes of potentials, monomial (power-law) and exponential (double or single exponential plus a cosmological constant), comprise the asymptotic behaviour of several classes of scalar field potentials. Therefore, they provide the skeleton for the typical behaviour of arbitrary potentials. For simplicity, we assume the matterless case for analysing linear cosmological perturbations in section \ref{sec3}. An exhaustive dynamical system analysis for three types of gauge-invariant scalar perturbation quantities, the Bardeen potentials, the comoving curvature perturbation, and the Sasaki-Mukhanov variable, is presented in section \ref{sect:4}. In section \ref{sect:new}, we investigate cosmological perturbations in the presence of two matter components, e.g. a perfect fluid plus a cosmological constant or perfect fluid plus a scalar field with exponential potential. A widespread practice in literature concentrates on a particular cosmological epoch when only one matter component is dominant. In that sense, even though not generic, our subsequent analysis is still relevant when the Universe is a scalar field dominated, e.g. during the early inflationary epoch or the late-time acceleration. Conclusions are presented in section \ref{sect:6}.

\section{The equations}\label{sec2}
The action we are working with is
\begin{equation}
    \mathcal{S} = \int d^{4}x\sqrt{-g}\left[\frac{R}{2\kappa^2} -\frac{1}{2}\phi_{\mu}\phi^{\mu} -V(\phi)\right] + \mathcal{S}_m,
\end{equation}
where we denote $\phi$ as the scalar field, $R$ is the Ricci scalar, $\mathcal{S}_m$ denotes the matter action, and we use units where $\kappa^2= 8\pi G=1$. Now, for a scalar field $\phi$ with self-interacting potential $V(\phi)$, we have that their energy density and pressure are given by
\begin{align}
   \rho_{\phi}=\frac{\dot \phi^2}{2}+V(\phi), \label{(2)}
\end{align} 
\begin{align}
    p_{\phi}=\frac{\dot \phi^2}{2}-V(\phi),\label{(3)}
\end{align}
respectively. Also, for a pressure-less matter, we can write the Friedman equation as follows
\begin{align}
    3H^2= \rho_m+\rho_{\phi},  \label{(4)}
\end{align}
\begin{align}
   \dot H=-\frac{1}{2}(\rho_m+\dot \phi^2), \label{(5)}
\end{align}
where $H$ is the Hubble parameter defined as $H=\frac{\dot{a}}{a}$, being $a$ the scale factor, and $\rho_m$ is the matter-energy density, whose corresponding conservation equation is given by
\begin{align}
    \dot \rho_m+3H\rho_m=0.  \label{(6)}
\end{align}

On the other hand, given the scalar field Lagrangian, we can get the Klein-Gordon equation as follows
\begin{align}
    \ddot \phi= -3H\dot\phi - \frac{d V}{d\phi}.   \label{(7)}
\end{align}

To extend the standard dynamical analysis method to generic classes of potentials, one uses the  method of $f$-devisers in which it is introduced two new dynamical variables, namely, $\lambda$ and $f$, as 
\begin{eqnarray}
\label{sdef}
\lambda&\equiv&-\frac{V^{\prime}(\phi )}{  V(\phi)},\\
f&\equiv& \frac{V^{\prime \prime}(\phi )}{V(\phi )}-\frac{V^{\prime}(\phi )^2}{V(\phi )^2},
\label{fdef}
\end{eqnarray}
such that
\begin{eqnarray}   
V^{\prime}(\phi ) &=& -  \lambda  V(\phi), \label{eq11}\\
    V^{\prime \prime}(\phi )&=&   \left(f+\lambda ^2\right) V(\phi   ). \label{eq10}
\end{eqnarray}
The only requirement is that  $f$ can be expressed as an explicit function of $\lambda$, that is, $f=f(\lambda)$. Following the above procedure, one can transform a cosmological system into a closed dynamical system for a set of auxiliary normalised variables and the new one $\lambda$. Then, using this procedure, one can investigate a wide range of potentials. In particular, the usual ansatzes of the cosmological literature can be covered by simple forms for $f$, as seen in Tab. \ref{fsform}. Note that the $\lambda$ variable is not required for the single exponential potential since it is a constant, i.e., $f$ is automatically zero. 

\begin{table}[!ht]
\centering
\begin{tabular*}{\columnwidth}{@{\extracolsep{\fill}}ll@{}}
\hline
\multicolumn{1}{@{}l}{Potential  $V(\phi)$}  & $f(\lambda)$\\
\hline
$\left|\frac{\mu}{n}\right|\phi^{n}$ \cite{Ratra:1987rm, Peebles:1987ek, Abramo:2003cp, Aguirregabiria:2004xd, Copeland:2004hq, Saridakis:2009pj, Saridakis:2009ej, Leon:2009dt, Chang:2013cba, Skugoreva:2013ooa, Pavlov:2013nra} & $-\frac{\lambda ^2}{n}$ \\
$V_{0}e^{-\alpha \phi}+V_1$  \cite{Yearsley:1996yg, Pavluchenko:2003ge, Cardenas:2002np} & $-\lambda(\lambda-\alpha)$ \\ 
$V_1 e^{\alpha \phi}+ V_2 e^{\beta \phi}$  \cite{Barreiro:1999zs, Gonzalez:2007hw, Gonzalez:2006cj} & $-(\lambda+\alpha)(\lambda+\beta)$\\ 
$V_{0}/\sinh^{\alpha}(\beta \phi)$  \cite{Ratra:1987rm, Wetterich:1987fm, Copeland:2009be, Leyva:2009zz, Pavluchenko:2003ge, Sahni:1999gb, Urena-Lopez:2000ewq}
& $\frac{\lambda^2}{\alpha}-\alpha\beta^2$\\
$V_{0}\left[\cosh\left( \xi \phi\right)-1\right]$  \cite{Ratra:1987rm, Wetterich:1987fm, Matos:2009hf, Copeland:2009be, Leyva:2009zz, Pavluchenko:2003ge, delCampo:2013vka, Sahni:1999qe, Sahni:1999gb, Lidsey:2001nj, Matos:2000ng}
& $-\frac{1}{2}(\lambda^2-\xi^2)$ \\
\hline
\end{tabular*}
\caption{The function $f(\lambda)$ for the most common quintessence potentials
\protect\cite{Escobar:2013js}.}
\label{fsform}
\end{table}

On the other hand, when the function $f(\lambda)$ is given, we can straightforwardly reconstruct
the corresponding potential form starting with
\begin{eqnarray}
&& \frac{d\lambda}{d\phi}=-  f, \quad  \frac{dV}{d\phi}=-\lambda  V,
\label{dV-dphi}
\end{eqnarray}
which leads to
\begin{eqnarray}
\phi(\lambda)&=&-\int \frac{1}{f} \, d\lambda,\label{quadphi}\\
V(\lambda)&=&V_0 e^{\int \frac{\lambda}{f} \, d\lambda}\label{quadV}.
\end{eqnarray} 
Note that the relations
(\ref{quadphi}) and (\ref{quadV}) are always valid, giving the
potential in an implicit form. However, for the usual cosmological cases of
Tab. \ref{fsform}  we can additionally eliminate  $\lambda$  between
\eqref{quadphi} and \eqref{quadV}, and write the potential explicitly as
$V=V(\phi)$.
Finally, note that the $f$-devisers method  
also allows   reconstructing a scalar field potential from a model with
stable equilibrium points. In particular, choosing a function $f$ with
the requested properties (existence of minimum, intervals of monotony,
differentiability) to have late-time stable attractors,
one uses \eqref{quadphi} and \eqref{quadV} to explicitly obtain  
$V(\phi)$. That is similar to the superpotential construction
method  \cite{Arefeva:2009tkq}, which allows for the construction of stable
kink-type solutions in scalar-field cosmological models, starting from the
dynamics, and specifically for the Lyapunov stability. One field model with a stable kink solution was considered
earlier in \cite{Arefeva:2004odl}. 

Nevertheless, this method is not universal. That means it cannot be applied to any arbitrary potential. The procedure can be fully implemented only when  $f$ is an explicit function of
$\lambda$. For instance, in some specific forms in the inflationary context, such as 
$V(\phi)\propto \phi ^p \ln ^q(\phi )$ \cite{Barrow:1995xb} and  $V(\phi)\propto \phi^n e^{-q\phi^m}$
\cite{Parsons:1995ew}, the expression $f$ cannot be expressed as a 
single-valued function of $\lambda$. In general, for a wide
range of potential the introduction of the
variables $f$ and $\lambda$ add an extra direction in the phase space, whose
neighbouring points correspond to ``neighbouring'' potentials. 

\section{Dynamical system in terms of Background quantities}
\label{sect:3}

It is well-known that for the investigation of cosmological models, one can introduce auxiliary
variables which transform the cosmological equations into an autonomous
dynamical system
\cite{Burd:1988ss, tavakol_1997, Copeland:1997et, Coley:2003mj, Leon2012CosmologicalDS, Gong:2006sp, Setare:2008sf, Chen:2008ft, Gupta:2009kk, Farajollahi:2011ym, Urena-Lopez:2011gxx, Escobar:2011cz, Escobar:2012cq, Xu:2012jf, Leon:2013qh}. Hence, we obtain a system of the form  $\textbf{X}'=\textbf{f(X)}$, where $\textbf{X}$ is the
column vector of the auxiliary variables and $\textbf{f(X)}$ is a vector field for autonomous equations. Prime denotes the differentiation with respect to a logarithmic time scale. The stability analysis comprises several steps. First, the critical points $\bf{X_c}$ are extracted under the requirement  of $\bf{X}'=0$. Then, one consider linear perturbations around $\bf{X_c}$ as $\bf{X}=\bf{X_c}+\bf{U}$, with $\textbf{U}$ the column vector of the  auxiliary variable's perturbations. Therefore, up
to first order we obtain $\textbf{U}'={\bf{\Xi}}\cdot \textbf{U}$, where  the matrix ${\bf {\Xi}}$ contains coefficients of the perturbed equations. Finally, the type and stability of each hyperbolic critical point are determined by the eigenvalues of ${\bf {\Xi}}$. The point is stable (unstable) if the reals parts of the eigenvalues are negative (positive) or saddle
if the eigenvalues have real parts with different signs.

To proceed forward, we can take equation \eqref{(4)} and divide them by $3H^2$, and also putting the value of $\rho_{\phi}$ from equation \eqref{(2)}, we get 
\begin{align}
    1=\frac{\rho_m}{3H^2}+\frac{ \dot \phi^2}{6H^2}+\frac{  V}{3H^2}.  \label{(8)}
\end{align}
Now we denote the following
\begin{align}
    x^2=\frac{ \dot \phi^2}{6H^2},\quad y^2=\frac{  V}{3H^2},\quad \Omega_m=\frac{ \rho_m}{3H^2}.\label{(9)}
\end{align}
So the equation \eqref{(8)} becomes
\begin{align}
    1=\Omega_m+x^2+y^2 \quad \textrm{or}\quad 1-x^2-y^2=\Omega_m.  \label{(10)}
\end{align}
As we see from equation \eqref{(10)}, $x^2+y^2\leq 1$ and $x^2+y^2\geq 0$, i.e., the system is bounded for a non-negative fluid density $ \rho_m\geq 0$. Then, the evolution of this system is completely described by trajectories within the unit disc, where the lower half-disc, $y<0$, corresponds to contracting universes. As the system is symmetric under the reflection $(x,y)\mapsto (x, -y)$ and time reversal $t\mapsto -t $, we only consider the
upper half-disc, $y\geq 0 $ in the following discussion.

Now we write a dynamical equation for each of the variables. 
Using the dynamical variable $N=\ln(a)$ with $dN=Hdt$, we write our dynamical system for $(x, y,\lambda)$ as a system of first-order equations. 
\begin{align}
x^{\prime}& =-\frac{3}{2}x\left(y^2-x^2+1\right)+\sqrt{\frac{3}{2}}\lambda y^2, \label{(18)}
\\
y^{\prime}& =-\sqrt{\frac{3}{2}}\lambda x y-\frac{3}{2}y\left(y^2-x^2-1\right), \label{(19)}
\\
\lambda^{\prime}& =-\sqrt{6} x f, \label{(20)}
\end{align}
where to close the system, we assume that  
$f$ can be written as an implicit function of $\lambda$. That is, $f(\lambda)$ can be explicitly obtained by inverting \eqref{sdef} and 
\eqref{fdef}. This procedure only gives a closed dynamical system when we can explicitly obtain $f=f(\lambda)$. In Tab. \ref{fsform}, we present cases where this approach can be completely implemented.

An important cosmological parameter is the deceleration parameter which can be written in terms of the dynamical variables as
\begin{equation}\label{dec_param}
    q \equiv -1-\frac{\dot{H}}{H^2} = \frac{1}{2}\left(1 + 3x^2 - 3y^2\right).
\end{equation}
From the above equation, we can see that, at the equilibrium points, the deceleration parameter is constant. Then, we can obtain an expression for the scale factor $a(t)$ that is valid asymptotically according to whether the constant   $q=-1$ or $q\neq -1$. Indeed, for the $q$ constant, integrating the expression
\begin{equation}
    \frac{a\ddot{a}}{\dot{a}^2}=-q,
\end{equation}
with the initial values $a(t_U)=1$, $\dot{a}(t_U)=H_0$, where $t_U$ is the age of the Universe and $H(t_U)=H_0$ is the current value of the Hubble parameter, we can obtain $a(t)$. Then, by definition, we obtain $H(t)= \dot{a}(t)/a(t)$. Summarising,
\begin{align}
  a(t)& = \left\{ \begin{array}{cc}
      \left(1+H_0 \left(q+1\right) \left(t-t_U\right)\right)^{\frac{1}{q+1}}, &  q\neq -1 \\
      e^{H_0 \left(t-t_U\right)}, & q=-1
  \end{array} \right.,\\
  H(t)& = \left\{ \begin{array}{cc}\frac{H_0}{H_0 \left(q+1\right) \left(t-t_U\right)+1}, & q\neq -1 \\
      H_0, & q=-1
  \end{array} \right..
\end{align}
Finally, because $x$ is a constant at the equilibrium points, we have 
\begin{equation}
\phi(t)= \phi_0 + \sqrt{6} x_c \int_{t_U}^t H(s) ds =  \phi_0 + \left\{\begin{array}{cc}                \ln \left(\left(H_0 \left(q+1\right)
   \left(t-t_U\right)+1)\right)^{\frac{\sqrt{6}x_c}{(1+q)}}\right), & x_c \neq 0\\         
    0, &  x_c= 0      \end{array}\right.. 
\end{equation}

The asymptotic behaviours of the scale factor, the Hubble scalar and the scalar field as a function of $t$ depend on the value of $x_c$ at the equilibrium points. 
\begin{itemize}
\item Case $x_c= \lambda^*/\sqrt{6}$ for any $\lambda^*$ satisfying $f(\lambda^*)=0, -\sqrt{6}<\lambda^*<\sqrt{6}$:      
\begin{align}
    & a(t)=\left\{\begin{array}{cc}                \left(\frac{H_0}{2}{\lambda^*}^2 \left(t-t_U\right)+1\right)^{\frac{2}{{\lambda_*}^2}}, & \lambda^* \neq 0\\                e^{H_0\left(t-t_U\right)}, &  \lambda^*= 0      \end{array}\right., \label{A1}
    \\
    & H(t)=\left\{\begin{array}{cc}  \frac{H_0}{\frac{H_0}{2}{\lambda^*}^2 \left(t-t_U\right)+1},  & \lambda^* \neq 0\\               H_0, &  \lambda^*= 0      \end{array}\right., \label{A1b}\\
    & \phi(t)=   \phi_0 + \left\{\begin{array}{cc}                \ln \left(\left(\frac{1}{2} H_0 {\lambda^*}^2 \left(t-t_U\right)+1\right)^{\frac{2}{|\lambda^*|}}\right), & \lambda^* \neq 0\\        0, &  \lambda^*= 0      \end{array}\right.. \label{A1c}
\end{align}
\item Case $x_c= \pm 1$: 
\begin{align}
    & a(t)=\left(3 H_0  \left(t-t_U\right)+1\right)^{\frac{1}{3}}, \label{A2}\\
    & H(t)=\frac{H_0}{3 H_0 \left(t-t_U\right)+1}, \label{A2b}\\
    & \phi(t)=   \phi_0 \pm \frac{\sqrt{6}}{3} \ln \left(3 H_0 \left(t-t_U\right)+1\right). \label{A2c}
\end{align} 
\item Case $-1< x_c<1$:  \begin{align}
    & a(t)=\left\{\begin{array}{cc}     \left(3 H_0 {x_c}^2 \left(t-t_U\right)+1\right)^{\frac{1}{3{x_c}^2}},  & x_c \neq 0\\                e^{H_0\left(t-t_U\right)}, &  x_c= 0      \end{array}\right., \label{CASE-A}\\ 
    & H(t)=\left\{\begin{array}{cc}  \frac{H_0}{3 H_0 {x_c}^2 \left(t-t_U\right)+1},  & x_c \neq 0\\               H_0, &  x_c= 0      \end{array}\right.,\label{CASE-Ab}\\
    & \phi(t)= \phi_0 + \sqrt{6} x_c \int_{t_U}^t H(s) ds =  \phi_0 + \left\{\begin{array}{cc}                \ln \left(\left(3 H_0 {x_c}^2 \left(t-t_U\right)+1\right)^{\frac{\sqrt{6}}{3|x_c|}}\right), & x_c \neq 0\\         
    0, &  x_c= 0      \end{array}\right.. \label{CASE-Ac}
\end{align}
\item Case $x_c= \pm \sqrt{3}/3$: 
\begin{align}
    & a(t)=H_0 \left(t-t_U\right)+1, \label{A3}\\
    & H(t)=\frac{H_0}{  H_0   \left(t-t_U\right)+1},    \label{A3b} \\
    & \phi(t)=  \phi_0 \pm  \sqrt{2} \ln \left( H_0   \left(t-t_U\right)+1\right). \label{A3c}
\end{align}
\end{itemize}

\subsection{Physical interpretation and stability of the equilibrium points}

The equilibrium points of the system \eqref{(18)}, \eqref{(19)}, and \eqref{(20)},  in the finite region for an arbitrary function $f(\lambda)$ are presented in Tab. \ref{Background_a}. For arbitrary potentials, we have the following. 

\begin{enumerate}
    \item The set of equilibrium points $O$ corresponds to matter-dominated solutions, which, as expected, are saddles, i.e. intermediate cosmological epochs. The deceleration parameter is $q= \frac{1}{2}$. Then, we have the asymptotic solutions 
   $a(t)=\left(\frac{3}{2} H_0 \left(t-t_U\right)+1\right)^{2/3}$, 
 $ H(t)  =\frac{H_0}{\frac{3}{2} H_0 \left(t-t_U\right)+1}$,   $\rho_m(t)= \rho_{m0} \left(\frac{3}{2} H_0    \left(t-t_U\right)+1\right)^{-2}$, and  $\phi(t)= 0$.

\item $K_\pm(\lambda^*)$  exist for $f({\lambda^*})=0$, and they represent kinetic-dominated solutions. They are associated with the Universe's early stages and correspond to stiff solutions. 
    
\item $K_-(\lambda^*)$  is a source for  $\lambda^*>-\sqrt{6}, \; f'({\lambda^*})>0$. It is a saddle for $\lambda^*<-\sqrt{6}$ or $ f'({\lambda^*})<0$. Non-hyperbolic for  $\lambda^*=-\sqrt{6}$ or $ f'({\lambda^*})=0$. 
 
\item $K_+(\lambda^*)$  is a source for  $\lambda^*<\sqrt{6}, \; f'({\lambda^*})<0$. It is a saddle for $\lambda^*>\sqrt{6}$ or $ f'({\lambda^*})>0$. Non-hyperbolic for  $\lambda^*= \sqrt{6}$ or $ f'({\lambda^*})=0$. 
    
For these solutions, the value of the deceleration parameter is $q=2$. Then,  we have the same asymptotic behaviour   $a(t)=(3 H_0  \left(t-t_U\right)+1)^{\frac{1}{3}}$,  $H(t)=\frac{H_0}{3 H_0 \left(t-t_U\right)+1}$, $\phi(t)=   \phi_0 \pm \frac{\sqrt{6}}{3} \ln \left(3 H_0 \left(t-t_U\right)+1\right)$,  and $\rho_m(t)=0$. 
 
\item $MS_{-}(\lambda^*)$ exists for  $f({\lambda^*})=0, \lambda^*<-\sqrt{3}$. It represents a matter-scalar field scaling solution where neither the scalar field nor the matter field dominates.   \newline It  is a sink for ${\lambda^*}\leq -2 \sqrt{\frac{6}{7}}, f'({\lambda^*})<0$ (stable spiral) or $ -2 \sqrt{\frac{6}{7}}<{\lambda^*}<-\sqrt{3}, f'({\lambda^*})<0$ (stable node). It is non-hyperbolic for ${\lambda^*}=-2 \sqrt{\frac{6}{7}}$ or $f'({\lambda^*})=0$. It is a saddle otherwise. 
   
\item  $MS_{+}(\lambda^*)$ exists for  $f({\lambda^*})=0, \lambda^*>\sqrt{3}$. It represents a matter-scalar field scaling solution where neither the scalar field nor the matter field dominates. \newline It  is a sink for ${\lambda^*}\geq 2 \sqrt{\frac{6}{7}}, f'({\lambda^*})>0$ (stable spiral) or $\sqrt{3}<{\lambda^*}<2 \sqrt{\frac{6}{7}}, f'({\lambda^*})>0$ (stable node). It is non-hyperbolic for ${\lambda^*}=2 \sqrt{\frac{6}{7}}$ or $f'({\lambda^*})=0$. It is a saddle otherwise. 
    
For these solutions, the deceleration parameter is $q= \frac{1}{2}$. Then, we have the asymptotic solutions 
    $\rho_m(t)= 0$, $a(t)=\left(\frac{3}{2} H_0 \left(t-t_U\right)+1\right)^{2/3}$, 
 $ H(t)  =\frac{H_0}{\frac{3}{2} H_0 \left(t-t_U\right)+1}$. Since $x_c=\frac{\sqrt{\frac{3}{2}}}{{\lambda^*}}$, we have $\phi(t)=   \phi_0 + \ln \left(\left(\frac{3}{2} H_0
   \left(t-t_U\right)+1\right)^{2/\lambda^* }\right)$. For $\lambda=\lambda^*$ and $f(\lambda^*)=0$, the potential asymptotically behaves as $V(\phi)= 3 y_c^2 H(t)^2= \frac{3 H_0^2}{2 {\lambda^{*2}} \left(\frac{3}{2} H_0
   \left(t-t_U\right)+1\right)^2}\sim  e^{-\lambda^*(\phi-\phi_0)}$. 

\item $Sf(\lambda^*)$ exists $-\sqrt{6}<\lambda^*<\sqrt{6}$. It represents an scalar-field dominated solution. It is a sink for $-\sqrt{3}<{\lambda^*}<0,  f'({\lambda^*})<0$ or $0<{\lambda^*}<\sqrt{3}, f'({\lambda^*})>0$. It is non-hyperbolic for  ${\lambda^*}\in \left\{-\sqrt{3}, 0, \sqrt{3}\right\}$ or   $f'({\lambda^*})=0$. It is a saddle otherwise. 
    For this solution, the deceleration parameter is 
   $q= \frac{1}{2} \left({\lambda^{*2}}-2\right)$. Then, $a(t)=\left(\frac{1}{2} H_0\lambda^{*2}
   \left(t-t_U\right)+1\right)^{\frac{2}{\lambda^{*2}}}$, $H(t)= \frac{2 H_0}{H_0 \lambda^{*2} \left(t-t_U\right)+2}$,  $\phi(t)=  \phi_0 + \ln \left(\left(\frac{1}{2} H_0 {\lambda^{*2}}
   \left(t-t_U\right)+1\right)^{2/{\lambda^*} }\right)$, and $\rho_m(t)=0$.  For $\lambda=\lambda^*$ and $f(\lambda^*)=0$, the potential asymptotically behaves as $V(\phi)= 3 y_c^2 H(t)^2= {2 H_0^2 \left(6- \lambda^{*2}\right)}/{\left(H_0 \lambda^{*2} \left(t-t_U\right)+2\right)^2} \sim  e^{-\lambda^*(\phi-\phi_0)}$. 

\item $dS$ is a potential dominated solution representing de Sitter solutions. It is stable for $f(0)> 0$ or a saddle for $f(0)<0$.  For this solution, the deceleration parameter is 
   $q=-1$. Then, $a(t)= e^{H_0 \left(t-t_U\right)}$, $H(t)= H_0$,  $\phi(t)=\phi_0$, $V(\phi)= 3 H_0^2$, and $\rho_m(t)=0$. 
   
\end{enumerate}

\begin{table}[]
    \centering
            \resizebox{\textwidth}{!}{%
    \begin{tabular}{|c|c|c|c|c|c|c|c|}
    \hline 
Label &    $x$ & $y$  & $\lambda$ & Existence & $k_1$ & $k_2$ & $k_3$ \\\hline
$O$ & $ 0$ & $0$ & $\lambda_c$ & $\lambda_c\in \mathbb{R}$  & $-\frac{3}{2}$ & $\frac{3}{2}$ & $0$ \\\hline
$K_-(\lambda^*)$ & $-1$ & $0$ & ${\lambda^*}$ & $f({\lambda^*})=0$ & $3$ & $\sqrt{\frac{3}{2}} {\lambda^*}+3$ & $\sqrt{6} f'({\lambda^*})$ \\\hline
$K_+(\lambda^*)$ & $ 1$ &$ 0$ & ${\lambda^*}$ & $f({\lambda^*})=0$ &$ 3$ & $3-\sqrt{\frac{3}{2}} {\lambda^*}$ &$ -\sqrt{6} f'({\lambda^*})$ \\\hline
$MS_{-}(\lambda^*)$ & $ \frac{\sqrt{\frac{3}{2}}}{{\lambda^*}}$ &$ -\frac{\sqrt{\frac{3}{2}}}{{\lambda^*}}$ &$ {\lambda^*}$ & $f({\lambda^*})=0, \lambda^*<-\sqrt{3}$ &$ -\frac{3 \left(\sqrt{24-7 {\lambda^*}^2}+{\lambda^*}\right)}{4 {\lambda^*}}$ &$ \frac{3}{4} \left(\frac{\sqrt{24-7 {\lambda^*}^2}}{{\lambda^*}}-1\right) $&$ -\frac{3 f'({\lambda^*})}{{\lambda^*}}$ \\\hline
$MS_{+}(\lambda^*)$  & $ \frac{\sqrt{\frac{3}{2}}}{{\lambda^*}}$ & $\frac{\sqrt{\frac{3}{2}}}{{\lambda^*}}$ & ${\lambda^*}$ & $f({\lambda^*})=0, \lambda^*>\sqrt{3}$ &$ -\frac{3 \left(\sqrt{24-7 {\lambda^*}^2}+{\lambda^*}\right)}{4 {\lambda^*}} $& $\frac{3}{4} \left(\frac{\sqrt{24-7 {\lambda^*}^2}}{{\lambda^*}}-1\right)$ & $-\frac{3 f'({\lambda^*})}{{\lambda^*}}$ \\\hline
$Sf(\lambda^*)$ &  $\frac{{\lambda^*}}{\sqrt{6}}$ &$ \sqrt{1-\frac{{\lambda^*}^2}{6}}$ & ${\lambda^*}$& $f({\lambda^*})=0,-\sqrt{6}<\lambda^*<\sqrt{6}$ & $\frac{1}{2} \left({\lambda^*}^2-6\right)$ & ${\lambda^*}^2-3 $&
  $ -{\lambda^*} f'({\lambda^*}) $\\\hline
$dS$ &  $0$ &$ 1$ & $0$ & always & $-3$ &$ \frac{1}{2} \left(-3-\sqrt{9-12 f(0)}\right)$ & $\frac{1}{2} \left(-3+ \sqrt{9-12 f(0)}\right)$\\\hline
    \end{tabular}}
    \caption{Equilibrium points of the system \eqref{(18)}, \eqref{(19)}, and \eqref{(20)}, and their eigenvalues $k_1$, $k_2$ and $k_3$, in the finite region for an arbitrary function $f(\lambda)$. $\lambda^*$ represent zeros of the function $f(\lambda)$. Note that $O$ is actually a line of fixed points whereas all the others are isolated fixed points.}
    \label{Background_a}
\end{table}

To illustrate the potentiality of the $f$-devisers method, we consider some examples. 

\subsection{First Example: monomial potential}
\label{sect:3-1}

Consider the potential $V(\phi)=\left|\frac{\mu}{n}\right|\phi^{n}$ \cite{Ratra:1987rm, Peebles:1987ek, Abramo:2003cp, Aguirregabiria:2004xd, Copeland:2004hq, Saridakis:2009pj, Saridakis:2009ej, Leon:2009dt, Chang:2013cba, Skugoreva:2013ooa, Pavlov:2013nra}, which produces the function $f(\lambda)=-\frac{\lambda ^2}{n}$. For this potential, the evolution equations are \eqref{(18)}, \eqref{(19)}, together with 
\begin{align}
\lambda^{\prime}& = \frac{\sqrt{6}}{n} x \lambda ^2. \label{(20b)}
\end{align}
For the function $f(\lambda)$ we have $f^{\prime}(\lambda)= -\frac{2\lambda}{n}$ and $f(\lambda)=0 \Longleftrightarrow\lambda=0$. Then, $\lambda^*=0$ and $f'(\lambda^*)=0$. The equilibrium points of this example are the following, summarised in Tab. \ref{Background_a-powerlaw}.

\begin{table}[]
    \centering
    \begin{tabular}{|c|c|c|c|c|c|c|c|c|}
    \hline 
Label &    $x$ & $y$  & $\lambda$ & Existence & $k_1$ & $k_2$ & $k_3$ & Stability \\\hline
$O$ & $ 0$ & $0$ & $\lambda_c$ & $\lambda_c\in \mathbb{R}$  & $-\frac{3}{2}$ & $\frac{3}{2}$ & $0$ & saddle \\\hline
$K_-(0)$ & $-1$ & $0$ & ${0}$ & always & $3$ & $3$ & $0$ & unstable \\\hline
$K_+(0)$ & $ 1$ &$ 0$ & ${0}$ & always  &$ 3$ & $3$ & $0$ & unstable \\\hline
$dS$ &  $0$ &$ 1$ & $0$ & always & $-3$ &$ -3$ & $0$ & saddle ($n>0$); sink ($n<0$)  \\\hline
    \end{tabular}
    \caption{Equilibrium points of the system \eqref{(18)}, \eqref{(19)}, and \eqref{(20)},  in the finite region for $f(\lambda)=-\frac{\lambda ^2}{n}$.}
    \label{Background_a-powerlaw}
\end{table}

Analysing the case of the de Sitter solution $dS$ in detail, note that the eigenvalues are $-3, -3, 0$, i.e., non-hyperbolic. Using the Centre Manifold theorem, we obtain that the graph locally gives the centre manifold of the origin
\begin{align}
& \Big\{(x, y, \lambda)\in [-1,1] \times [0,1] \times \mathbb{R}: x=\frac{\lambda}{\sqrt{6}}+h_1(\lambda), y=1+ h_2(\lambda),  \nonumber \\
& h_1(0)=0, h_2(0)=0,  h_1'(0)=0, h_2'(0)=0, |\lambda|<\delta\Big\},
\end{align}
for a small enough $\delta$, where the functions $h_1$ and $h_2$ satisfy the differential equations

\begin{align}
   & -24 \lambda ^2 \left(\left(\sqrt{6} h_1(\lambda )+\lambda \right) h_1'(\lambda )+h_1(\lambda
   )\right) +6 n \Big(\sqrt{6} \lambda  \left(3 h_1(\lambda )^2+h_2(\lambda ) (h_2(\lambda
   )+2)\right) \nonumber \\
   & +6 h_1(\lambda ) \left(h_1(\lambda )^2-h_2(\lambda ) (h_2(\lambda
   )+2)-2\right)+3 \lambda ^2 h_1(\lambda )\Big)+\sqrt{6} \lambda ^3 (n-4)=0,
\\
   & -\frac{\lambda ^2 \left(\sqrt{6}
   h_1(\lambda )+\lambda \right) h_2'(\lambda )}{n}-\frac{1}{4} (h_2(\lambda )+1) \left(-6
   h_1(\lambda )^2+6 h_2(\lambda ) (h_2(\lambda )+2)+\lambda ^2\right)=0.
\end{align}
Then, using the Taylor series, we have the solutions
\begin{align}
 x(\lambda) & =  \frac{\lambda }{\sqrt{6}}-\frac{\lambda ^3}{3 \left(\sqrt{6} n\right)}-\frac{(n-8) \lambda ^5}{18
   \left(\sqrt{6} n^2\right)}-\frac{((n-18) (n-6)) \lambda ^7}{108 \left(\sqrt{6} n^3\right)}\nonumber \\
   &+\frac{(1984-n ((n-48)
   n+592)) \lambda ^9}{648 \sqrt{6} n^4} +\frac{(-n (n ((n-80) n+1880)-16320)-45280) \lambda ^{11}}{3888 \sqrt{6}
   n^5}\nonumber \\
   & +\frac{(1223424-n (n (n ((n-120) n+4560)-72896)+504624)) \lambda ^{13}}{23328 \sqrt{6} n^6}+O\left(\lambda
   ^{14}\right), \label{expansion-x-t}
\\
y(\lambda) & =1-\frac{\lambda ^2}{12}-\frac{(n-16) \lambda ^4}{288 n}-\frac{((n-48) n+288) \lambda ^6}{3456
   n^2} \nonumber \\
   &+\frac{(n (-5 (n-96) n-8064)+31744) \lambda ^8}{165888 n^3} +\frac{(n (376832-7 n ((n-160) n+5184))-1159168)
   \lambda ^{10}}{1990656 n^4} \nonumber \\
   &+\left(\frac{n (5147648-21 n ((n-240) n+12672))-40069120}{47775744 n^4}+\frac{59}{27
   n^5}\right) \lambda ^{12}+O\left(\lambda ^{14}\right). 
\end{align}

The 1D dynamical system dictates the dynamics at the centre manifold is
\begin{align}
    \frac{d \lambda}{d N} & =- U'(\lambda),
\end{align}
which corresponds to a gradient-like equation with potential 
\begin{align}
 U(\lambda)& =-\frac{\lambda ^4}{4 n}+\frac{\lambda ^6}{18 n^2}+\frac{\lambda ^8 (n-8)}{144 n^3}+\frac{\lambda ^{10} (n-18)
   (n-6)}{1080 n^4}   -\frac{\lambda ^{12} (1984-n ((n-48) n+592))}{7776 n^5} \nonumber \\
   & +\frac{\lambda ^{14} (n (n ((n-80)
   n+1880)-16320)+45280)}{54432 n^6}\nonumber \\
   & -\frac{\lambda ^{16} (1223424-n (n (n ((n-120) n+4560)-72896)+504624))}{373248
   n^7}+O\left(\lambda ^{17}\right). \label{Gradient-Potential}
\end{align}
Therefore, since $U^{(4)}(0)=-6/n\neq 0$, the origin is a degenerate maximum of the potential for $n>0$, and the centre manifold of the origin and the origin are unstable (saddle). At the same time, it is stable if $n<0$. In this example, the points $MS_{-}(\lambda^*)$ and 
$MS_{+}(\lambda^*)$  do not exist and $Sf(\lambda^*)$ reduces to $dS$. 
However, there are equilibrium points at the invariant sets 
$\lambda= \pm\infty$, where the dynamics are given, under a time re-scaling which does not affect the orbits of the phase space, by  
\begin{align}
\frac{d x}{d \tau}& = \pm \sqrt{\frac{3}{2}} y^2,  \quad
\frac{d y}{d \tau} =\mp \sqrt{\frac{3}{2}}  x y. \label{PowerlawInf-b}
\end{align}
The orbits as $\lambda= \pm\infty$ are semicircles $x^2+y^2= x_0^2+y_0^2$.

One can define 
\begin{equation}
u=\frac{2 \tan ^{-1}(\lambda )}{\pi }, -1<u<1, \label{eqU}
\end{equation}
obtaining a compactification of phase space and the vector field, which defines a global phase space that  comprises the dynamics at finite $\lambda$, and, i.e., 

\begin{align}
\frac{d x}{d N}& = \left\{\begin{array}{cc}
      \sqrt{\frac{3}{2}} y^2, & u=1\\
    \sqrt{\frac{3}{2}} y^2 \tan \left(\frac{\pi  u}{2}\right)+\frac{3}{2} x \left(x^2-y^2-1\right), &  -1<u<1 \\
     -\sqrt{\frac{3}{2}} y^2, & u=-1
     \end{array} \right.,  \label{Powerlaw-2a}
\\
\frac{d y}{d N}& =  \left\{\begin{array}{cc}
     - \sqrt{\frac{3}{2}}  x y, & u=1\\
    -\frac{1}{2} y \left(\sqrt{6} x \tan \left(\frac{\pi  u}{2}\right)-3 x^2+3 y^2-3\right),&  -1<u<1 \\
      \sqrt{\frac{3}{2}}  x y, & u=-1
\end{array} \right., \label{Powerlaw-2b}
\\
\frac{d u}{d N}& = -\frac{\sqrt{6} x (\cos (\pi  u)-1)}{\pi  n}. \label{Powerlaw-2c}
\end{align}

\begin{figure}[]
    \centering
    \includegraphics[scale=0.65]{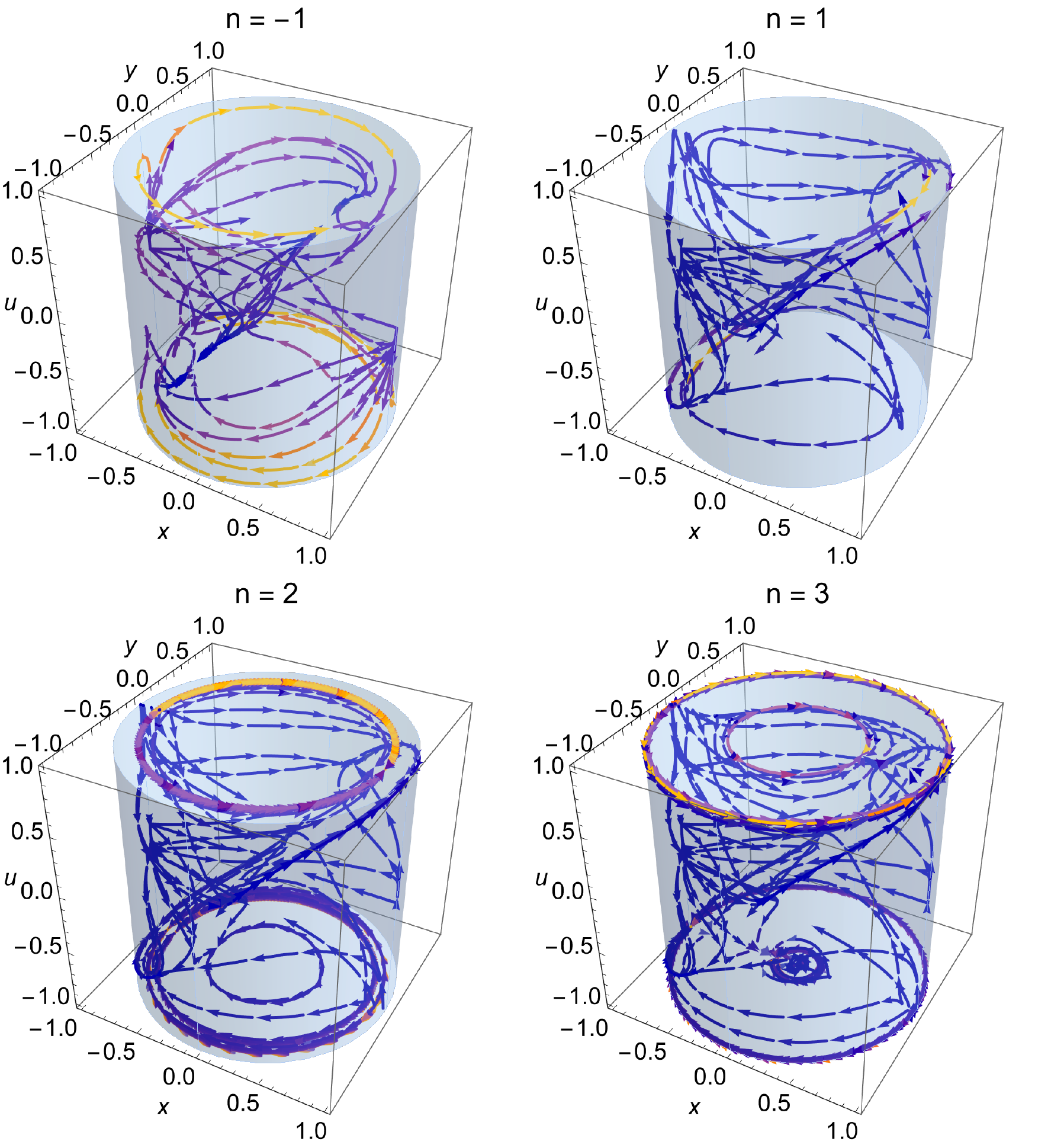}
    \caption{Compact 3D phase space of the system \eqref{Powerlaw-2a}, \eqref{Powerlaw-2b}, and \eqref{Powerlaw-2c}, for $n=-1$, $1$, $2$, and $3$.}
    \label{fig:PL2}
\end{figure}

In Fig. \ref{fig:PL2} is represented the flow of the system \eqref{Powerlaw-2a}, \eqref{Powerlaw-2b}, and \eqref{Powerlaw-2c}, for  $n=-1$, $n=1$, $2$, and $3$. The dynamics as   $\lambda \rightarrow \pm \infty$ is represented on the top and bottom disks ($u=\pm 1, x^2+y^2 \leq 1$), where the orbits are concentric semicircles.   The planes $u=\pm1$ correspond to the limiting cases when the scalar field potential has an infinite negative/positive slope (see the definition of $\lambda$ in Eq.\eqref{sdef}). One half of this plane acts as an attracting invariant submanifold while the other half acts as a repelling invariant submanifold, with $x=0$ separating the two regions (see Eq.\eqref{Powerlaw-2c}).

\subsection{Second Example: exponential potential}
\label{sect:3-2}
For $f=0$ and $\lambda$ constant, we recover the quintessence scenario with an exponential potential $V=V_0 e^{-\lambda\phi}$ as studied in \cite{Copeland:1997et},  where for generality,  we have considered a perfect fluid with linear equation of state $p_m=  w_m \rho_m$, and barotropic index $\gamma_m\equiv w_m+1$. 

The dynamical system equations are  
\begin{align}
   \frac{d x}{d N} & =   -3x + \lambda \sqrt{\frac{3}{2}} y^2
 + \frac{3}{2} x \left[ 2x^2 + \gamma_m \left( 1 - x^2 - y^2 \right)
\right], \label{eomx}\\
\frac{d y}{d N} & =  - \lambda \sqrt{\frac{3}{2}} xy
 + \frac{3}{2} y \left[ 2x^2 + \gamma_m \left( 1 - x^2 - y^2 \right)
\right], \label{eomy}
\end{align}
defined in the phase space 
\begin{equation}
\left\{(x,y)\in \mathbb{R}^2: x^2+y^2\leq 1, y\geq 0\right\}.
\end{equation}

\begin{table}[]
\begin{center}
\resizebox{\textwidth}{!}{
\begin{tabular}{|c|c|c|c|c|c|c|}
\hline
C.P. & $x$ & $y$ & Existence & Stability & $\Omega_d$ & $w_d$ \\ \hline
A & $0$ & $0$ & Always & Saddle for $0 < \gamma_m < 2$ & $0$ & Undefined \\ \hline
B & $1$ & $0$ & Always & Unstable node for $\lambda < \sqrt{6}$ & $1$ & $1$ \\ 
&  &  &  & Saddle for $\lambda > \sqrt{6}$ &  &  \\ \hline
C & $-1$ & $0$ & Always & Unstable node for $\lambda > -\sqrt{6}$ & $1$ & $1 $\\ 
&  &  &  & Saddle for $\lambda < -\sqrt{6}$ &  &  \\ \hline
D & $\lambda/\sqrt{6}$ & $[1-\lambda^2/6]^{1/2}$ & $\lambda^2 < 6$ & Stable
node for $\lambda^2 < 3\gamma_m$ & 1 & $\frac{\lambda^2}{3}-1$ \\ 
&  &  &  & Saddle for $3\gamma_m < \lambda^2 < 6$ &  &  \\ \hline
E & $(3/2)^{1/2} \, \gamma_m/\lambda$ & $[3(2-\gamma_m)\gamma_m/2%
\lambda^2]^{1/2}$ & $\lambda^2 > 3\gamma_m$ & Stable node for $3\gamma_m <
\lambda^2 < 24 \gamma_m^2/(9\gamma_m -2)$ & $3\gamma_m/\lambda^2$ & $w_m$ \\ 
&  &  &  & Stable spiral for $\lambda^2 > 24 \gamma_m^2/(9\gamma_m -2)$ &  & 
\\ \hline
\end{tabular}}%
\end{center}
\caption[Quintback]{The critical points, their stability conditions (the
corresponding eigenvalues are given in \protect\cite{Copeland:1997et}), and
the values of $\Omega_\phi$ and $w_\phi$, for the quintessence scenario, with $%
\protect\gamma_m\equiv w_m+1$ and $w_d=\gamma_\phi -1$. }
\label{Quintback}
\end{table}
In table \ref{Quintback} are presented the critical points, their stability conditions (the
corresponding eigenvalues are given in \protect\cite{Copeland:1997et}), and
the values of $\Omega_\phi$ and $w_\phi$, for the quintessence scenario, with a perfect fluid with linear equation of state $p_m=  (\gamma_m-1) \rho_m$ and $\gamma_\phi \equiv {\rho_\phi+p_\phi \over \rho_\phi}  = {\dot\phi^2 \over V + \dot\phi^2/2}
 = {2x^2 \over x^2 + y^2}$.

\begin{figure}[]
    \centering
    \includegraphics[scale=0.8]{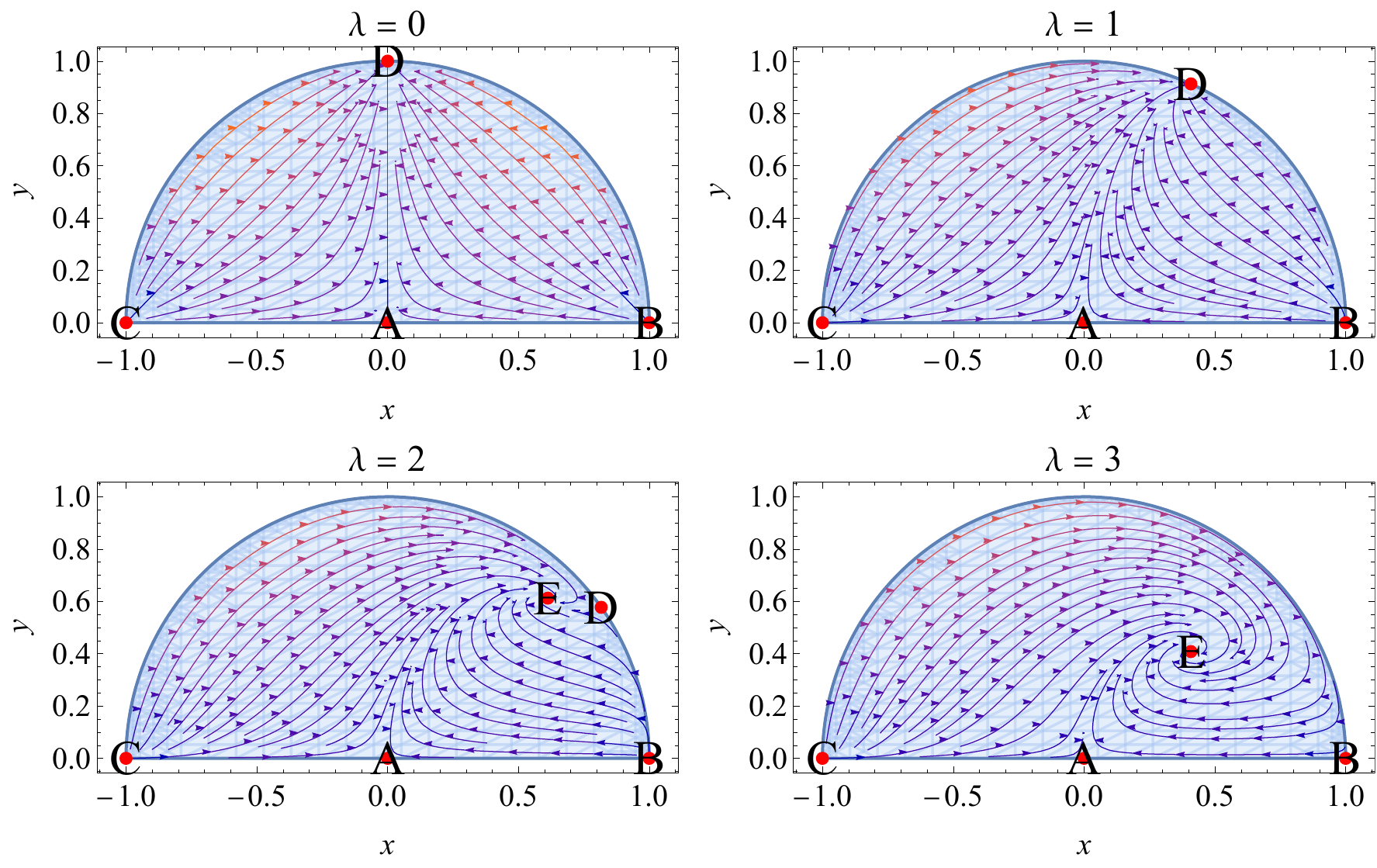}
    \caption{Compact 2D phase space of the system \eqref{eomx} and \eqref{eomy}, for different values of $\lambda$ and $w_m=0$. }
    \label{fig:exp}
\end{figure}
In Fig. \ref{fig:exp}  is presented a compact 2D phase space of the system \eqref{eomx} and \eqref{eomy}, for different values of $\lambda$ and $w_m=0$.

The equilibrium points at finite values of $x$ and $y$ in the phase-plane
correspond to solutions where the scalar field has a barotropic
equation of state and the scale factor of the universe evolves as
$a\propto t^p$ where $p={2/3\gamma_\phi}$.

Two of the fixed points ($B$ and $C$) correspond to solutions where the Friedman constraint
Eq.~\eqref{(10)} is dominated by the kinetic energy of the
scalar field with a stiff equation of state, $\gamma_\phi=2$. As
expected these solutions are unstable and are only expected to be
relevant at early times.

Moreover, we find that the barotropic
fluid dominated solution (A) where $\Omega_\phi=0$ is a saddle for all values of $\gamma_m>0$ (saddle). For any $\gamma_m>0$, and however steep
the potential (i.e.~whatever the value of $\lambda$), the energy density of
the scalar field  never vanishes with respect to the other matter
in the universe. Generically, the system admits only two possible late-time attractor solutions. One
of these is the well-known scalar field-dominated solution
($\Omega_\phi=1$) which exists for sufficiently flat potentials,
$\lambda^2<6$. The scalar field has an effective barotropic index
$\gamma_\phi=\lambda^2/3$ giving rise to a power-law inflationary
expansion~\cite{Lucchin:1984yf,Kitada:1992uh} ($\ddot{a}>0$) for
$\lambda^2<2$. Previous phase-plane
analyses to \cite{Copeland:1997et}, as \cite{Halliwell:1986ja, Burd:1988ss,Coley:1997nk} have shown that a wide class of
homogeneous vacuum models approach the spatially-flat FRW model for
$\lambda^2<2$.This scalar field-dominated solution
is a late-time attractor in the presence of a barotropic fluid when  $\lambda^2<3\gamma_m$.

However, for $\lambda^2>3\gamma_m$, we find a different late-time attractor
where neither the scalar field nor the barotropic fluid entirely
dominates the evolution. Instead, there is a scaling solution where the
energy density of the scalar field remains proportional to that of the
barotropic fluid with $\Omega_\phi=3\gamma_m/\lambda^2$. This solution was
first found by Wetterich~\cite{Wetterich:1987fm} and shown to be the global
attractor solution for $\lambda^2>3\gamma_m$ in Ref.~\cite{Wands:1993zm}.

\begin{enumerate}
\item $\lambda^2<3\gamma_m$. 
Both kinetic-dominated solutions are unstable nodes.
The fluid-dominated solution is a saddle point.
The scalar field-dominated solution is the late-time attractor and is 
inflationary in parameter region~$\lambda^2<\min\{2, 3\gamma_m\}$ and non-inflationary in region~$2<\lambda^2<3\gamma_m$.
\item $3\gamma_m<\lambda^2<6$. 
Both kinetic-dominated solutions are unstable nodes.
The fluid-dominated solution is a saddle point.
The scalar field-dominated solution is a saddle point.
The scaling solution is a stable node/spiral.
\item $6<\lambda^2$. 
The kinetic-dominated solution with $\lambda x<0$ is an unstable
node.
A saddle point is a kinetic-dominated solution with $\lambda x>0$.
The fluid-dominated solution is a saddle point.
The scaling solution is a stable spiral.
\end{enumerate}

The bifurcation value $\gamma=0$ was studied in \cite{Copeland:1997et}, where the largest eigenvalue for linear perturbations
vanishes. Thus,  higher-order perturbations about the
critical point are involved to determine its stability. The result is that $x=y=0$ is a stable attractor, but that trajectory only approaches this as the
logarithm of the scale factor, $N$. The late-time evolution is 
given by $y^2 = {\sqrt{6}\over \lambda} x \approx {1 \over \lambda^2 N} $.

\subsection{Third Example: double exponential potential}
\label{sect:3-3}

Consider the potential
$V(\phi)=V_1 e^{\alpha \phi}+ V_2 e^{\beta \phi}$  \cite{Barreiro:1999zs, Gonzalez:2007hw, Gonzalez:2006cj} which provides the function  $f(\lambda)=-(\lambda+\alpha)(\lambda+\beta)$. This example contains the particular case of the hyperbolic cosine $V(\phi)=\frac{1}{2} \left(e^{\alpha \phi}+ e^{-\alpha \phi}\right)$ by setting $V_1=
 V_2=1/2$ and $\beta=-\alpha$. 
 
 For this potential we have $f^{\prime}(\lambda)=-\alpha -\beta -2 \lambda$ and $f(\lambda)=0 \Longleftrightarrow\lambda\in\left\{-\alpha, -\beta\right\}$, with $ f'(-\alpha)=\alpha-\beta$ and $ f'(-\beta)=-(\alpha-\beta)$. Moreover, we have $f(0)=-\alpha \beta$ and $f'(0)=-\alpha- \beta$. Without losing generality, we can assume $\alpha<\beta$. The equilibrium points of this example are the following, summarised in  Tab. \ref{Background_a-double-exp}.

 \begin{table}[]
    \centering
        \resizebox{\textwidth}{!}{%
    \begin{tabular}{|c|c|c|c|c|c|c|c|c|c|c|}
    \hline 
Label &    $x$ & $y$  & $\lambda$ & Existence & $k_1$ & $k_2$ & $k_3$ & Stability \\\hline
$O$ & $ 0$ & $0$ & $\lambda_c$ & $\lambda_c\in \mathbb{R}$  & $-\frac{3}{2}$ & $\frac{3}{2}$ & $0$ & saddle \\\hline
$K_-(-\alpha)$ & $-1$ & $0$ & $-\alpha$ & always & $3$ & $3 -\sqrt{\frac{3}{2}}\alpha$ & $\sqrt{6} (\alpha -\beta )$ & saddle \\\hline
$K_-(-\beta)$ & $-1$ & $0$ & $-\beta$ & always & $3$ & $3 -\sqrt{\frac{3}{2}}\beta$ & $-\sqrt{6} (\alpha -\beta )$ & source for \\
 &&&&&&&& $\alpha<\beta<\sqrt{6}$\\
 &&&&&&&& saddle for \\
 &&&&&&&& $\alpha<\beta, \beta> \sqrt{6}$ \\\hline
$K_+(-\alpha)$ & $ 1$ &$ 0$ & $-\alpha$ & always &$ 3$ & $3 + \sqrt{\frac{3}{2}}\alpha$ &$ -\sqrt{6} (\alpha -\beta )$  & source for \\
 &&&&&&&& $-\sqrt{6}<\alpha<\beta$\\
 &&&&&&&& saddle for  \\
 &&&&&&&& $\alpha<\beta, \alpha < -\sqrt{6}$ \\\hline
$K_+(-\beta)$ & $ 1$ &$ 0$ & $-\beta$ & always &$ 3$ & $3 + \sqrt{\frac{3}{2}} \beta$ &$ \sqrt{6} (\alpha -\beta )$ & saddle \\\hline
$MS_{-}(-\alpha)$ & $ -\frac{\sqrt{\frac{3}{2}}}{{\alpha}}$ &$ \frac{\sqrt{\frac{3}{2}}}{{\alpha}}$ &$-\alpha$ & $ \alpha>\sqrt{3}$ &$ -\frac{3 \left(\alpha-\sqrt{24-7 {\alpha}^2}\right)}{4 {\alpha}}$ &$ -\frac{3 \left(\alpha+\sqrt{24-7 {\alpha}^2}\right)}{4 {\alpha}}$&$ 3 \left(1-\frac{\beta }{\alpha }\right)$ & sink for \\
 &&&&&&&& $ \beta >\alpha  >\sqrt{3}$\\\hline
$MS_{-}(-\beta)$ & $ -\frac{\sqrt{\frac{3}{2}}}{{\beta}}$ &$ \frac{\sqrt{\frac{3}{2}}}{\beta}$ &$-\beta$ & $\beta>\sqrt{3}$ &$ -\frac{3 \left(\beta-\sqrt{24-7 {\beta}^2}\right)}{4 {\beta}}$ &$ -\frac{3 \left(\beta+\sqrt{24-7 {\beta}^2}\right)}{4 {\beta}}$&$ 3 \left(1-\frac{\alpha }{\beta }\right)$ & saddle \\\hline
$MS_{+}(-\alpha)$  & $- \frac{\sqrt{\frac{3}{2}}}{{\alpha}}$ & $-\frac{\sqrt{\frac{3}{2}}}{{\alpha}}$ & $-\alpha$ & $\alpha<-\sqrt{3}$ &$ -\frac{3 \left(\alpha-\sqrt{24-7 {\alpha}^2}\right)}{4 {\alpha}}$ &$ -\frac{3 \left(\alpha+\sqrt{24-7 {\alpha}^2}\right)}{4 {\alpha}}$ & $3 \left(1-\frac{\beta }{\alpha }\right)$ & saddle \\\hline
$MS_{+}(-\beta)$  & $-\frac{\sqrt{\frac{3}{2}}}{{\beta}}$ & $-\frac{\sqrt{\frac{3}{2}}}{{\beta}}$ & $-\beta$ & $\beta <-\sqrt{3}$ &$ -\frac{3 \left(\beta-\sqrt{24-7 {\beta}^2}\right)}{4 {\beta}}$ &$ -\frac{3 \left(\beta+\sqrt{24-7 {\beta}^2}\right)}{4 {\beta}}$& $3 \left(1-\frac{\alpha }{\beta }\right)$ & sink for \\
 &&&&&&&& $\alpha<\beta < -\sqrt{3}$ \\\hline
$Sf(-\alpha)$ &  $-\frac{{\alpha}}{\sqrt{6}}$ &$ \sqrt{1-\frac{{\alpha}^2}{6}}$ & $-\alpha$& $-\sqrt{6}<\alpha<\sqrt{6}$ & $\frac{1}{2} \left({\alpha}^2-6\right)$ & ${\alpha}^2-3 $&
  $\alpha  (\alpha -\beta )$ & sink for \\
 &&&&&&&&  $0<\alpha <\sqrt{3}, \beta >\alpha$\\
 &&&&&&&&  saddle otherwise\\\hline
  $Sf(-\beta)$ &  $-\frac{{\beta}}{\sqrt{6}}$ &$ \sqrt{1-\frac{{\beta}^2}{6}}$ & $-\beta$& $-\sqrt{6}<\beta<\sqrt{6}$ & $\frac{1}{2} \left({\beta}^2-6\right)$ & ${\beta}^2-3 $&
  $-\beta (\alpha -\beta ) $ & sink for \\
 &&&&&&&& $-\sqrt{3}<\beta <0, \alpha <\beta$\\
 &&&&&&&&  saddle otherwise\\\hline
$dS$ &  $0$ &$ 1$ & $0$ & always & $-3$ &$ \frac{1}{2} \left(-3-\sqrt{9+12 \alpha \beta}\right)$ & $\frac{1}{2} \left(-3+ \sqrt{9+12 \alpha \beta}\right)$ & stable for $\alpha \beta< 0$\\\hline
    \end{tabular}}
    \caption{Equilibrium points of the system \eqref{(18)}, \eqref{(19)}, and \eqref{(20)},  in the finite region for $f(\lambda)=-(\lambda+\alpha)(\lambda+\beta), \; \alpha \neq \beta$. Without loss generality, we can assume $\alpha<\beta$. }
    \label{Background_a-double-exp}
\end{table}

As in section \ref{sect:3-1}, using the same compact variable as defined in Eq.\eqref{eqU}, we obtain a compactification of the phase space and the vector field, which defines a global phase space that  comprises the dynamics at finite $\lambda$, and the dynamics at infinity under a time re-scaling, which does not affect the orbits of the phase space, i.e., 
 \begin{align}
\frac{d x}{d N}& = \left\{\begin{array}{cc}
    \sqrt{\frac{3}{2}} y^2, & u=1\\
    \sqrt{\frac{3}{2}} y^2 \tan \left(\frac{\pi  u}{2}\right)+\frac{3}{2} x \left(x^2-y^2-1\right), &  -1<u<1 \\
     -\sqrt{\frac{3}{2}} y^2, & u=-1
\end{array} \right.,  \label{double-exp-2a}
\\
\frac{d y}{d N}& =  \left\{\begin{array}{cc}
      - \sqrt{\frac{3}{2}}  x y, & u=1\\
    -\frac{1}{2} y \left(\sqrt{6} x \tan \left(\frac{\pi  u}{2}\right)-3 x^2+3 y^2-3\right),&  -1<u<1 \\
      \sqrt{\frac{3}{2}}  x y, & u=-1
\end{array} \right., \label{double-exp-2b}
\\
\frac{d u}{d N}& =  \frac{2 \sqrt{6} x \cos ^2\left(\frac{\pi  u}{2}\right) \left(\alpha +\tan \left(\frac{\pi  u}{2}\right)\right) \left(\beta +\tan \left(\frac{\pi  u}{2}\right)\right)}{\pi}. \label{double-exp-2c}
\end{align}
\begin{figure}[]
    \centering
    \includegraphics[scale=0.65]{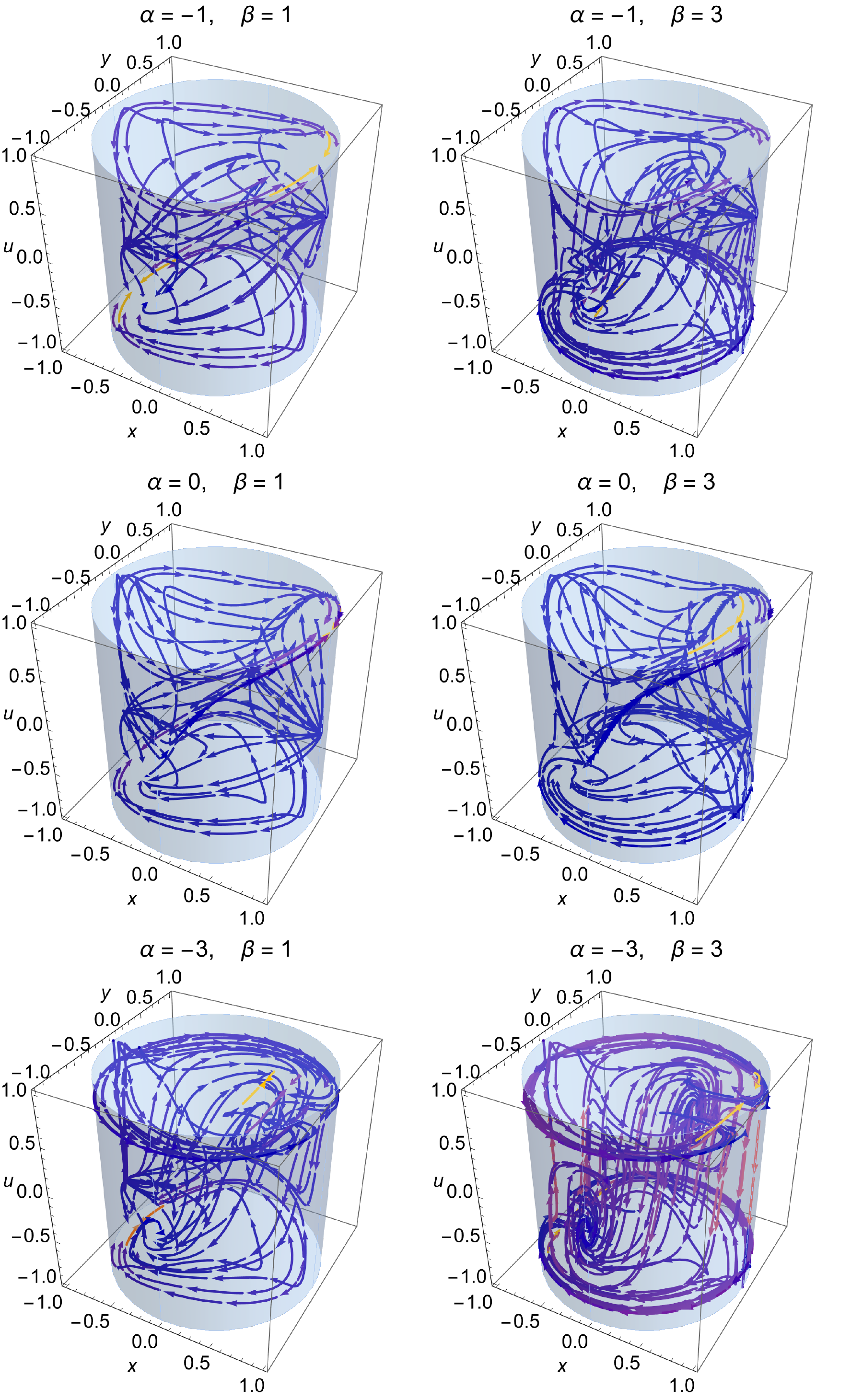}
    \caption{Compact 3D phase space of the system \eqref{double-exp-2a}, \eqref{double-exp-2b}, and \eqref{double-exp-2c}, for different values of $\alpha$ and $\beta$.   }
    \label{fig:double-exp}
\end{figure}

In Fig. \ref{fig:double-exp} is represented the flow of the system \eqref{double-exp-2a}, \eqref{double-exp-2b}, and \eqref{double-exp-2c}, for different values of $\alpha$ and $\beta$.  The planes $u=\pm1$ correspond to the limiting cases when the scalar field potential has an infinite negative/positive slope (see the definition of $\lambda$ in Eq. \eqref{sdef}).

When $\beta=-\alpha$, the potential reduces to a hyperbolic cosine, and when one parameter is zero, it reduces to an exponential potential plus a Cosmological Constant. Therefore, we cover three of the most common quintessence potentials displayed in the Tab. 
\ref{fsform}. 
\section{Evolution of Cosmological perturbations}\label{sec3}

In this section, following the line of Ref. \cite{Alho:2020cdg}, we investigate the dynamics of linear scalar cosmological perturbations for a generic scalar field model by the methods of dynamical systems. We use the perturbation of a scalar field $\phi_0$ in the background. The most generic scalar perturbed FLRW metric can be written as \cite{Bardeen:1980kt}
\begin{equation}
    ds^2 = - \left(1+\alpha\right)dt^2 - 2a(t)\left(\beta,_{i}-S_{i}\right)dtdx^i + a^{2}(t)\left[\left(1+2\psi\right)\delta_{ij} + 2\partial_{i}\partial_{j}\gamma + 2\partial_{(i}F_{j)} + h_{ij}\right]dx^{i}dx^{j},
\end{equation}
where the inhomogeneous perturbation quantities $\alpha,\,\beta,\,\psi,\,\gamma,\,F_i,\,h_{ij}$ are functions of both $t$ and $\bar{x}$. The quantity $\psi(t,\bar{x})$ is directly related to the 3-curvature of the spatial hyper-surface
\begin{equation}
    ^{(3)}R = - \frac{4}{a^2}\nabla^{2}\psi.
\end{equation}
For a scalar field, one also needs to take into account the perturbation of the scalar field $\delta\phi(t,\bar{x})$ and, for a perfect fluid, the perturbed energy-momentum tensor is
\begin{equation}
    T^0_0 = -\left(\rho(t) + \delta\rho(t,\bar{x})\right), \quad
    T^0_i = -\left(\rho(t) + P(t)\right)\partial_{i}v(t,\bar{x}), \quad
    T^i_j = \left(P(t) + \delta P(t,\bar{x})\right)\delta^i_j,
\end{equation}
being $v(t,\bar{x})$ the velocity potential. In what follows, we will restrict ourselves to the case when there is no matter, but only a scalar field is present. The reason is simplicity. 

Suppose one wants to investigate cosmological perturbations in the presence of two matter components, e.g. a perfect fluid and a scalar field. In that case, one needs to consider entropy perturbations as well. A widespread practice in literature concentrates on a particular cosmological epoch when only one matter component is dominant. In that sense, even though not generic, our subsequent analysis is still relevant when the Universe is a scalar field dominated, e.g. during the early inflationary epoch or the late-time acceleration. 
Of course, there is the gauge issue; the perturbation quantities defined above are not gauged invariant \cite{Mukhanov, Brandenberger:1992dw, Brandenberger:1993zc, Brandenberger:1992qj, DeFelice:2010aj}. In this sense, various gauge-invariant perturbation quantities have been introduced in the literature. In this article, we will consider the following three gauge-invariant perturbation quantities.
\begin{itemize}

\item \textbf{Bardeen potential $\Phi$}:  
James Bardeen introduced Bardeen potentials \cite{Bardeen:1980kt}, who gave the first-ever gauge-invariant formulation for cosmological perturbations. These quantities are gauge-invariant perturbation quantities constructed solely out of metric perturbations. There are two such quantities
\begin{equation}
    \Phi \equiv \alpha - \frac{d}{dt}\left[a\left(\beta+a\gamma\right)\right], \qquad \Psi \equiv - \psi + aH\left(\beta+a\dot{\gamma}\right).
\end{equation}
It can be shown that for the case of a single scalar field, both the Bardeen potentials are equal and follows the equation \cite{Bardeen:1980kt, Mukhanov, Brandenberger:1992dw, Brandenberger:1993zc, Brandenberger:1992qj}
\begin{align}\label{1h}
    \Phi^{\prime\prime} + 2\left(\mathcal{H}-\frac{\phi_0^{\prime\prime}}{\phi_0^{\prime}}\right)\Phi^{\prime} + 2\left(\mathcal{H}^{\prime}-\mathcal{H}\frac{\phi_0^{\prime\prime}}{\phi_0^{\prime}}\right)\phi - {\mathcal{H}^{-2}}\nabla^2  \Phi = 0,
\end{align}
where $\mathcal{H} = a H$, and $'$ denotes derivative with respect to the conformal time $\eta$ given by
 \begin{equation}\label{1b}
      d\eta =\frac{dt}{a(t)}.
 \end{equation}
Time derivatives with respect to $t$ and $\eta$ are related as 
 \begin{equation}\label{1d}
     \frac{d}{d\eta} =a \frac{d}{dt}, \qquad \frac{d^2}{d\eta^2}=a^2H\frac{d}{dt}+a^2\frac{d^2}{dt^2}.
 \end{equation}
 The variable $\psi$ gives the 3-curvature perturbation of the otherwise spatially flat constant time slice: $^{(3)}R=-\frac{4}{a^2}\nabla^{2}\psi$.
 
\item \textbf{Comoving curvature perturbation $\mathcal{R}$}:
For single scalar field models, comoving curvature perturbation is defined as
\begin{equation}
    \mathcal{R} \equiv \psi - \frac{H}{\dot{\phi}}\delta\phi
\end{equation}
The name comes from the fact that this variable coincides with the 3-curvature perturbation of the spatial slice in the \emph{comoving} gauge, which, for single scalar field models, is given by $\delta\phi=0$. At the linear level, comoving curvature perturbation evolves according to the following equation \cite{DeFelice:2010aj}:
\begin{align}\label{2h}
    \ddot{\mathcal{R}} + \frac{\left(a^{3}\frac{\dot{\phi}^2}{H^2}\right)^.}{\left(a^{3}\frac{\dot{\phi}^2}{H^2}\right)}\dot{\mathcal{R}} - \frac{1}{a^2}\nabla^2\mathcal{R}=0
\end{align}

\item \textbf{Sasaki-Mukhanov variable $\varphi_c$}:
Another gauge-invariant perturbation variable that we will consider is the so-called Sasaki-Mukhanov variable \cite{Kodama:1984ziu, Mukhanov:1988jd}, or the scalar field perturbation in uniform curvature gauge, defined as
\begin{equation}
    \varphi_c \equiv \delta\phi - \frac{\dot{\phi}}{H}\psi.
\end{equation}
At the linear level, this variable follows the perturbation equation
\begin{align}\label{3h}
& \frac{d^2\varphi_c}{dN^2}+\frac{d\varphi_c}{dN}\left(\frac{V}{H^2}\right) + \left( \frac{V_{,\phi \phi} + 2\frac{\dot{\phi}}{H} V_{,\phi}+\left(\frac{\dot{\phi}}{H}\right)^2 V}{H^2}\right)\varphi_c - {\mathcal{H}^{-2}}\nabla^2 \varphi_c = 0.
 \end{align}
\end{itemize}
The perturbation equations \eqref{1h}, \eqref{2h}, and \eqref{3h} are valid strictly only in the absence of matter.

To obtain a dynamical system that describes the evolution of perturbations, we first introduce Cartesian spatial coordinates and make the Fourier transform of the perturbation variables. This results in 
\begin{equation}
   \mathcal{H}^{-2} \nabla^2 \longrightarrow - k^2 \mathcal{H}^{-2}.
\end{equation}
We now consider the evolution of the three perturbation quantities in separate sections.

\subsection{Evolution of perturbed quantities}
\label{Section-6.3} 

We have  confirmed that, generically, for $q\neq -1$, the scale factor a has a power law
dependence on conformal/cosmic time, and thereby a constant deceleration parameter. 

In analysing the solutions, we need the following properties of conformal time that follow
from the assumption that $q$ is a non-zero constant (and different from $-1$). 

For fixed $x_c\neq 0$, $a(t)= (3 H_0 {x_c}^2 \left(t-t_U\right)+1)^{\frac{1}{3{x_c}^2}}$ and $q=-1+3 x_*^2\neq 0$ and $q\neq -1$. Moreover, from \eqref{1d} it follows
     \begin{align}
    \eta & = \int \frac{d\eta}{dt} dt = \int a^{-1} d t= \int \left(3 H_0 {x_c}^2 \left(t-t_U\right)+1\right)^{-\frac{1}{3{x_c}^2}} dt \nonumber \\
    & = \frac{\left(3 {H_0} {x_c}^2
   \left(t-t_U\right)+1\right)^{1-\frac{1}{3 {x_c}^2}}}{{H_0}
   \left(3 {x_c}^2-1\right)}  = \frac{\left({H_0} (q+1)
   \left(t-t_U\right)+1\right)^{\frac{q}{q+1}}}{{H_0} q}.
     \end{align}
     On the other hand, 
 \begin{align}
   a= \left({H_0} (q+1) \left(t-t_U\right)+1\right)^{\frac{1}{q+1}}, \quad
   H= \frac{{H_0}}{{H_0} (q+1) \left(t-t_U\right)+1}.
 \end{align}
Then, 
\begin{align}
  \mathcal{H}= a H=  {H_0} \left({H_0} (q+1) \left(t-t_U\right)+1\right)^{-\frac{q}{q+1}}= H_0 a^{-q}.
\end{align}
Finally,
\begin{equation}
  \mathcal{H} \eta=  \frac{1}{q}, \quad  \eta =  \frac{a^{q}}{{H_0} q} = \eta_0 e^{q N}, \quad  \mathcal{H}=\mathcal{H}_0 e^{-q N}. 
\end{equation}
where $\eta_0=\frac{1}{{H_0} q}$, $\mathcal{H}_0= a_0 H_0=H_0$, recall that we have taken $a_0=1$ such that $N=\ln a$. 

In this case, it is convenient to introduce a new variable
\begin{equation}
    \nu  = a^p \times \text{Perturbed quantity},
\end{equation}
(where $p$ is chosen to remove the first order derivative of  $\nu$) 
and to use conformal time $\eta$ instead of e-fold time $N$.  In making the transition from N to $\eta$, we use the relations 
\begin{align}
 \frac{d}{dN} & = \mathcal{H}^{-1}  \frac{d}{d\eta}, \quad 
   \frac{d^2}{dN^2} =   \mathcal{H}^{-2} \frac{d^2}{d\eta^2} +  q  \mathcal{H}^{-1}  \frac{d}{d\eta}.
\end{align}

\subsubsection{Bardeen potential}\label{Section-6.3.1}

Using the dynamical variables, we can write the perturbation equation \eqref{1h} as
\begin{align}\label{ptbn_bardeen3}
& \frac{d^2\Phi_k}{dN^2} + \left[7-3x^{2}+\sqrt{6}\lambda\left(\frac{1-x^2}{x}\right)\right]\frac{d\Phi_k}{dN} + \left[6\left(1-x^2\right)+\sqrt{\frac{3}{2}}\lambda\left(\frac{1-x^2}{x}\right) + \frac{k^2}{a^2H^2}\right]\Phi_k=0.
\end{align} 

When $q$ is a constant, $q\notin\{0,-1\}$, $dq/dN=0$, $dx/dN=0$, hence, 
$\left(6x - \sqrt{6} \lambda\right)\left(1-x^2\right)=0$. Therefore, either $x=\lambda/\sqrt{6}$ or $x=\pm 1$, such $x\neq 0$. Then, 
$6x \left(1-x^2\right)= \sqrt{6} \lambda\left(1-x^2\right)$ implies $\lambda\left(\frac{1-x^2}{x}\right)=\sqrt{6} \left(1-x^2\right)$. Then, equation \eqref{ptbn_bardeen3} becomes
\begin{align}
& \frac{d^2\Phi_k}{dN^2} + \left(13-9 x^2\right)\frac{d\Phi_k}{dN} + \left[9\left(1-x^2\right) + \frac{k^2}{a^2H^2}\right]\Phi_k=0.
\end{align}
But $q=-1+3 x^2$ implies
\begin{align}\label{ptbn_bardeen33}
& \frac{d^2\Phi_k}{dN^2} + \left[10-3 q\right]\frac{d\Phi_k}{dN} + \left[6 -3 q + \frac{k^2}{a^2H^2}\right]\Phi_k=0.
\end{align}
Then, passing to the variable $\eta$, and using the relation $  \mathcal{H} \eta=  1/{q}$,  equation \eqref{ptbn_bardeen33} becomes 
\begin{align}
&    \frac{d^2\Phi_k}{d\eta^2} +  \left[10-2 q\right](q \eta)^{-1}  \frac{d\Phi_k}{d\eta} + \left[(6 -3 q)(q \eta)^{-2} + k^2\right]\Phi_k=0.
\end{align}

Defining \begin{equation}
    v_k = a^p \Phi_{k},
\end{equation}
we have
\begin{equation}
\frac{d^2\Phi_k}{d\eta^2}=\frac{p v_k a^{-p}   (p+q+1)}{\eta ^2 q^2}-\frac{2 p  a^{-p}}{\eta  q} \frac{d  v_k}{d\eta} +a^{-p} \frac{d^2 v_k}{d\eta^2},
\end{equation}
and
\begin{equation}
  \frac{d  \Phi_k}{d\eta}=a^{-p}  \frac{d  v_k}{d\eta}-\frac{p v_k   a^{-p}}{\eta  q}
\end{equation}
Then,
\begin{align}
& \frac{d^2 v_k}{d\eta^2} -\frac{2 (p+q-5)}{\eta  q} \frac{d  v_k}{d\eta} + v_k \left(k^2+\frac{p (p+3 q-9)-3 q+6}{\eta ^2 q^2}\right)=0. 
\end{align}
Because $q$ is constant at the fixed points, we can eliminate the first-order derivative of $v_k$ by defining the constant $p$ such that 
\begin{equation}
  (p+q-5)=0. 
\end{equation}
Then we obtain the Bessel equation for the function $v_k$, 
\begin{align}
 \frac{d^2 v_k}{d\eta^2} +    v_k \left(k^2-\frac{(q-2) (2 q-7)}{\eta ^2 q^2}\right)=0,
\end{align}
This equation can be written as 
\begin{equation}
  \frac{d^2 v_k}{d\eta^2} +    v_k \left(k^2-\left({\nu ^2-\frac{1}{4}}\right){\eta ^{-2}}\right)=0.
\end{equation}
by defining
\begin{align}
    \nu^2= \frac{14-11 q}{q^2}+\frac{9}{4}.
\end{align}
The resulting equation admits the solution 
\begin{align}
  v_k(\eta )= C_+ \sqrt{\eta } J_{\nu }(k \eta )+C_- \sqrt{\eta } Y_{\nu }(k \eta). 
\end{align}
 $C_+$ and $C_-$ are complex constants depending on $k$. 
 
In the special case $x_c = \pm \sqrt{3}/3$, $q=0$, the previous asymptotic analysis fails. 
To analyse this case, we use the relation $ \mathcal{H}= a H=  {H_0}$ (constant), and that $\lambda$ is constant at the equilibrium point. Then, equation
\eqref{ptbn_bardeen3} becomes 
\begin{align}\label{ptbn_bardeen3-special}
& \frac{d^2 \Phi_k}{d\eta^2} + 2\left[3 \pm \sqrt{2}\lambda\right] \mathcal{H}  \frac{d\Phi_k}{d\eta} + \left[ \left(4\pm \sqrt{2}\lambda\right)\mathcal{H}^{2}+  k^2\right]\Phi_k=0.
\end{align}
Defining 
\begin{equation}\label{2ptbn_bardeen3-special} 
 \Phi_k(\eta)=  v_k(\eta )   e^{-H_0\eta   \left(3 \pm \sqrt{2} \lambda\right)},
\end{equation}
equation \eqref{ptbn_bardeen3-special} becomes 
\begin{align}\label{3ptbn_bardeen3-special} 
 \frac{d^2 v_k}{d\eta^2}  + v_k  \left(k^2-H_0^2 \left(2 \lambda ^2+5 \sqrt{2} \lambda  \epsilon +5\right)\right)=0,
\end{align}
with $\epsilon=\pm 1$. The solution
is
\begin{equation} \label{4ptbn_bardeen3-special}
 v_k(\eta)=   C_+ e^{\eta  \sqrt{H_0^2 \left(2 \lambda ^2+5 \sqrt{2} \lambda  \epsilon +5\right)-k^2}}+C_- e^{-\eta 
   \sqrt{H_0^2 \left(2 \lambda ^2+5 \sqrt{2} \lambda  \epsilon +5\right)-k^2}},
\end{equation}
where  $C_+$ and $C_-$ are complex constants depending on $k$.

\subsubsection{Comoving curvature perturbation}
\label{Section-6.3.2}

Using the dynamical variables, we can write the perturbation equation \eqref{2h} as
 \begin{align}\label{ptbn_comov}
 & \frac{d^2\mathcal{R}_k}{dN^2} + \sqrt{6}\lambda\left(\frac{1-x^2}{x}\right)\frac{d\mathcal{R}_k}{dN} + \left(\frac{k^2}{a^2H^2}\right)\mathcal{R}_k = 0.
\end{align}

Using the relation $\lambda\left(\frac{1-x^2}{x}\right)=\sqrt{6} \left(1-x^2\right)$, valid for constant $q$, equation \eqref{ptbn_comov} becomes 

\begin{align}
 &  \frac{d^2 \mathcal{R}_k}{d\eta^2} +  \left(4-q\right) \frac{1}{q \eta}  \frac{d  \mathcal{R}_k}{d\eta} +   k^2 \mathcal{R}_k = 0.
\end{align}
Defining \begin{equation}
    v_k = a^p \mathcal{R}_k,
\end{equation}
we have
\begin{align}
\frac{d^2 v_k}{d\eta^2} -\frac{(2 p+q-4)}{\eta  q} \frac{dv_k}{d\eta} +  v_k \left(k^2+\frac{p (p+2 q-3)}{\eta ^2 q^2}\right)=0.
\end{align}
Defining 
\begin{equation}
    p=  2-\frac{q}{2},
\end{equation}
we have 
\begin{equation}
 \frac{d^2 v_k}{d\eta^2} +  v_k \left(k^2-\frac{(q-4) (3 q-2)}{4 \eta ^2 q^2}\right)=0.
\end{equation}
Defining 
\begin{equation}
    \nu^2=-\frac{7}{2 | q| }+\frac{2}{q^2}+1,
\end{equation}
the equation can be written as 
\begin{equation}
  \frac{d^2 v_k}{d\eta^2} +    v_k \left(k^2-\left({\nu ^2-\frac{1}{4}}\right){\eta ^{-2}}\right)=0,
\end{equation}
that admits the solution 
\begin{align}
  v_k(\eta )= C_+ \sqrt{\eta } J_{\nu }(k \eta )+C_- \sqrt{\eta } Y_{\nu }(k \eta). 
\end{align}
 $C_+$ and $C_-$ are complex constants depending on $k$.

In the special case $x_c = \pm \sqrt{3}/3$, $q=0$, the previous asymptotic analysis fails. 
To analyse this case, we use the relation $ \mathcal{H}= a H=  {H_0}$ (constant), and that $\lambda$ is constant at the equilibrium point. In this case, the equation
\eqref{ptbn_comov} becomes 
\begin{align}\label{ptbn_comov3-special}
& \frac{d^2 \mathcal{R}_k}{d\eta^2}\pm  2\sqrt{2}\lambda \mathcal{H}  \frac{d\mathcal{R}_k}{d\eta} +   k^2\mathcal{R}_k=0.
\end{align}
Defining 
\begin{equation}\label{2ptbn_comov3-special} 
 \mathcal{R}_k(\eta)=  v_k(\eta )   e^{\mp H_0  \eta \sqrt{2} \lambda},
\end{equation}
equation \eqref{ptbn_comov3-special} becomes 
\begin{align}\label{3ptbn_comov33-special} 
 \frac{d^2 v_k}{d\eta^2}  + v_k  \left(k^2-2H_0^2 \lambda ^2\right)=0.
\end{align}
The solution is
\begin{equation} \label{4ptbn_comov3-special}
 v_k(\eta)=   C_+ e^{\eta  \sqrt{2 H_0^2 \lambda ^2-k^2}}+C_- e^{-\eta  \sqrt{2 H_0^2 \lambda ^2-k^2}},
\end{equation}
where  $C_+$ and $C_-$ are complex constants depending on $k$.

\subsubsection{Sasaki-Mukhanov variable}
\label{Section-6.3.3}

Using the dynamical variables, we can write the perturbation equation \eqref{3h} as
\begin{align}
    \frac{d^2\varphi_{ck}}{dN^2} + 3 \left( 1-x^2\right)\frac{d\varphi_{ck}}{dN} + \left[ 18 \left(1-x^2\right)\left( \frac{f}{6}+\left(x-\frac{\lambda }{\sqrt{6}}\right)^2\right) +\frac{k^2}{a^2H^2}\right]\varphi_{ck} = 0. \label{eq:Uggla} 
\end{align}

As before, when $q$ is constant,  $dx/dN=0$, hence, 
$\left(6x - \sqrt{6} \lambda\right)\left(1-x^2\right)=0$. Therefore, either $x=\lambda/\sqrt{6}$ or $x=\pm 1$ (note that $x\neq 0$). Using $x\neq 0$ in equation \eqref{(20)}, it follows at the fixed point that $\lambda$ is constant and $f(\lambda)=0$. At equilibrium, $\left(1-x^2\right) \left(\frac{f}{6}+\left(x-\frac{\lambda }{\sqrt{6}}\right)^2\right)=0$. Then, passing to the time variable $\eta$, equation \eqref{eq:Uggla}  becomes 
\begin{align}
 \frac{d^2 \varphi_{ck}}{d\eta^2} +  2 (\eta q)^{-1} \frac{d\varphi_{ck} }{d\eta} +  k^2\varphi_{ck} = 0.
\end{align}

Defining \begin{equation}
    v_k = a \varphi_{ck},
\end{equation}
we have
\begin{align}
 \frac{d^2 v_k}{d\eta^2}+    v_k  \left(k^2+\eta ^{-2} |q|^{-1}\right)=0,
 \end{align}
with the solution 
\begin{align}
  v_k(\eta )= C_+ \sqrt{\eta } J_{\nu}(k \eta )+C_- \sqrt{\eta } Y_{\nu}(k \eta ),
\end{align}
where 
 \begin{equation}
   \nu=\frac{1}{2} \sqrt{1-4 |q|^{-1}}.
 \end{equation}
 $C_+$ and $C_-$ are complex constants depending on $k$.

In the special case $x_c = \pm \sqrt{3}/3, \lambda=\pm \sqrt{2}$, with $f(\pm \sqrt{2})=0$ and $q=0$, the previous asymptotic analysis fails. 
To analyse this case, we use the relation $ \mathcal{H}= a H=  {H_0}$ (constant). In this case, the equation
\eqref{eq:Uggla} becomes 
\begin{align}\label{eq:Uggla-special}
&  \frac{d^2\varphi_{ck}}{d\eta^2} + 2H_0\frac{d\varphi_{ck}}{d\eta} + k^2\varphi_{ck} = 0.
\end{align}
Defining 
\begin{equation}\label{2eq:Uggla-special} 
\varphi_{ck}(\eta)=  v_k(\eta )   e^{-H_0 \eta}, 
\end{equation}
equation \eqref{eq:Uggla-special} becomes 
\begin{align}\label{3eq:Ugglaspecial} 
 \frac{d^2 v_k}{d\eta^2}  + v_k  \left(k^2-H_0^2\right)=0,
\end{align}
with $\epsilon=\pm 1$. The solution
is
\begin{equation} \label{4eq:Uggla-special}
 v_k(\eta)=   C_+ e^{\eta  \sqrt{H_0^2-k^2}}+C_- e^{-\eta  \sqrt{H_0^2 -k^2}},
\end{equation}
where  $C_+$ and $C_-$ are complex constants depending on $k$.

\section{Dynamical system analysis at background and perturbation levels for the matterless case}
\label{sect:4}

Defining
\begin{align}
    Z=k^2(aH)^{-2}, \label{EQ:(63)}
\end{align}
then $Z^{\prime}$ satisfies the following equation via the deceleration parameter $q$
\begin{align}
    Z^{\prime}=2qZ,
\end{align}
this last one is related to the Hubble parameter through the expression $\dot{H}=-(1+q)H^2$ or, as $\frac{d}{dN}=H^{-1}\frac{d}{dt}$,
\begin{align}
    H^{\prime}=-(1+q)H.
\end{align}
Finally, the above expression can be written as
\begin{align}
    1+q = \frac{1}{2}\frac{\dot{\phi}^2}{H^2} = 3x^2.
\end{align}

For $q\neq -1$ and $x\neq 0$, we have at the equilibrium points 
\begin{align}
    1+q_*  = 3{{x}^*}^2.
\end{align}
Then, 
\begin{align}
    & \frac{d \ln H}{d \ln a}=-3{{x}^*}^2 \implies H=H_0 a^{-3{{x}^*}^2}  \nonumber \\
    & \implies a(t)=\left(3 H_0 {{x}^*}^2  \left(t-t_U\right)+1\right)^{\frac{1}{3{{x}^*}^2}}, \;  H(t)=\frac{H_0}{3 H_0 {{x}^*}^2 \left(t-t_U\right)+1},
\end{align}
where $t_U$ is the age of the Universe, and we have assumed $H(t_U)=H_0$, and  $a(t_U)=1$. 

When ${{x}^*}=0$, $q_*=-1$ we have a de Sitter expansion with 
\begin{align}
a(t)= e^{H_0 \left(t-t_U\right)}, \;  H(t)= H_0. 
\end{align}
Deepening into the interpretation of the variable $Z= \frac{k^2}{\mathcal{H}^2}$, with $\mathcal{H}= a H$. Perturbations with $k^2 \mathcal{H}^{-2}\ll 1$ are called long wavelength or super-horizon. Those with $k^2 \mathcal{H}^{-2}\gg 1$ are considered short wavelengths or sub-horizon. Long wavelength perturbations are usually studied by choosing the idealised limiting value $k=0$, corresponding to $Z = 0$.
On the other hand, short wavelength perturbations correspond to $Z\rightarrow \infty$. We also note that in
choosing $Z= \frac{k^2}{\mathcal{H}^2}$ as a dynamical variable, we have that the wave number
$k$ is absorbed in the definition when formulating the dynamical system. However, if we choose the reference time $t=t_U$ (i.e.,
when $N:= \ln a = 0$) to be the time for setting initial data in the state space, then different choices of $Z_0= \frac{k^2}{H_0^2}$ for a given $H_0$ (we assume $a(t_U)=1$) yield solutions with different wave
number $k$ \cite{Alho:2020cdg}.

The evolution of the background quantities leads to the (not bounded) dynamical system 
\begin{align}
& x^{\prime} = -\left(3x - \sqrt{\frac{3}{2}} \lambda\right)\left(1-x^2\right), \quad  \lambda^{\prime} =-\sqrt{6} x f, \quad  Z^{\prime} = 2(3x^2 -1)Z,
\end{align}
To do the compactification,   we denote
\begin{align}
    \bar{Z}=\frac{Z}{1+Z}=\frac{k^2}{k^2+(aH)^2},\quad 
    Z=\frac{\bar{Z}}{1-\bar{Z}}.
\end{align}
We note that
\begin{align}
    \bar{Z}^{\prime} = 2(3x^2 -1)\bar{Z}\left(1-\bar{Z}\right).
\end{align}

However, as we can see from that, when $\bar{Z}\to 1$, we have a singularity, so we change the ``$e$ folding time'' from $N$ to $\bar{N}$ through
\begin{align}
    \frac{d\bar{N}}{dN}=\frac{1}{1-\bar{Z}}=1+Z.
\end{align}

Recall that we refer to the invariant set $\bar{Z}= 0$ as the long
wavelength boundary (or the super-horizon boundary), and $\bar{Z} = 1$ as the short wavelength boundary (or sub-horizon boundary). 
After compactification of $Z$ to $\bar{Z}$,  we get the final set of equations  as
\begin{subequations}
\label{(eq:99)}
\begin{align}
& \frac{dx}{d\bar{N}} =-\left(3x - \sqrt{\frac{3}{2}} \lambda\right)\left(1-x^2\right)\left(1-\bar{Z}\right), \\
& \frac{d\lambda}{d\bar{N}} = - \sqrt{6}xf\left(1-\bar{Z}\right), \\
& \frac{d\bar{Z}}{d\bar{N}} = 2(3x^2 - 1)\bar{Z}\left(1-\bar{Z}\right)^2. 
\end{align}
\end{subequations}

The function $\lambda$ can be negative, zero, positive, or unbounded. However, in some exceptional cases,  say, when $f(\lambda)$ is an even function, $f(\lambda)= f(-\lambda)$, there is no loss of generality in assuming that $\lambda$ is non-negative. In this case, the field equations are invariant under the transformation $(x, \phi) \rightarrow -(x, \phi)$ and $\lambda \rightarrow -\lambda$. That is the case of the exponential potential where $f\equiv 0$. 

For an arbitrary potential, and given $\lambda^*$, with $f(\lambda^*)=0$, the equilibrium point  $Sf(\lambda^*)$, with ${{x}^*}= \lambda^*/\sqrt{6}$ exists for $-\sqrt{6}<\lambda^*<\sqrt{6}$ and is a sink for $-\sqrt{3}<{\lambda^*}<0,  f'({\lambda^*})<0$ or $0<{\lambda^*}<\sqrt{3}, f'({\lambda^*})>0$. Also, 
$dS$ is  stable for $f(0)> 0$. That is, at the future attractor, we have  ${{x}^*}= \lambda^*/\sqrt{6}$ or zero at the 
background state space. At these equilibrium points the deceleration parameter $q= -1+ {{\lambda}^*}^2/2$ or $q=-1$ and is thus constant. The range
$-\sqrt{6} \leq \lambda^* \leq \sqrt{6}$ corresponds to the range $-1 \leq q \leq 2$ for $q$. This range of $q$ also describes a space-time with a perfect fluid with a linear equation of state $p = w \rho$, with $w:= (2 q-1)/3$ in the range $-1\leq w \leq 1$. Thus ${\lambda}^*=0$ corresponds to a cosmological
constant while the bifurcation value $\lambda^*= \pm \sqrt{6}$ corresponds to a stiff fluid with the speed of sound equal to that of light. On the other hand, ${\lambda^*}^2>6$ yields an equation of state with superluminal speed. Therefore, at the physically interesting late-time attractors, $\lambda$ is bounded. However, we can handle the cases $\lambda \rightarrow \pm \infty$ using the new variable \eqref{eqU}.

\subsection{Stability analysis of the fixed points on the background space $B$}
\label{sect:A.1}

The dynamics at the background space  $B = \left\{\left(x, \lambda, \bar{Z}\right)\in [-1,1]\times \mathbb{R} \times [0,1]\right\}$ is  given by 
\begin{align}
\label{background(eq:99)}
& \frac{dx}{d {N}} =-\left(3x - \sqrt{\frac{3}{2}} \lambda\right)\left(1-x^2\right), \quad  \frac{d\lambda}{d {N}} = - \sqrt{6}xf, \quad  \frac{d\bar{Z}}{d {N}} = 2(3x^2 - 1)\bar{Z}\left(1-\bar{Z}\right), 
\end{align}
where it is convenient to use the e-folding variable as the time variable.

\begin{table}[]
    \centering
               \resizebox{\textwidth}{!}{
    \begin{tabular}{|c|c|c|c|c|c|c|c|c|}
    \hline 
Label &    $x$ &$\lambda$ & $\bar{Z}$ & Existence & $k_1$ & $k_2$ & $k_3$ & Stability \\\hline
$P_1({\lambda^*})$ & $\frac{{{\lambda^*}}}{\sqrt{6}}$ & ${{\lambda^*}}$ & $0$ & $-\sqrt{6} \leq \lambda^*\leq \sqrt{6}$ &  $\frac{1}{2} \left({{\lambda^*}}^2-6\right)$ & ${{\lambda^*}}^2-2$ &  $-{{\lambda^*}} f'({{\lambda^*}})$&  Saddle  for \\
 &&&&&&&& $f'(\lambda^*)<0, -\sqrt{2}<\lambda^*<0$, \\
 &&&&&&&& or $f'(\lambda^*)>0, 0<\lambda^*<\sqrt{2}$, \\
 &&&&&&&& or $2 <\lambda^{*^2}<6$, \\
 &&&&&&&& or ${\lambda^{*}} f'({\lambda^{*}})<0$. \\
 &&&&&&&& N. H.  otherwise. \\
\hline
$P_2({\lambda^*})$ & $-1$ & ${{\lambda^*}}$ & $0$ & always & $4$ & $\sqrt{6} {{\lambda^*}}+6$ & $\sqrt{6} f'({{\lambda^*}})$ & Source for \\
&&&&&&&& $ {{\lambda^*}}>- \sqrt{6}, f'({{\lambda^*}})>0$. \\
&&&&&&&& Saddle for  $ {{\lambda^*}}<- \sqrt{6}$ \\
&&&&&&&& or  $f'({{\lambda^*}})<0$.\\
&&&&&&&& N. H.  otherwise. \\\hline  
$P_3({\lambda^*})$ & $1$ & $ {{\lambda^*}}$ & $ 0$ & always & $4$ & $6-\sqrt{6} {{\lambda^*}}$ & $-\sqrt{6} f'({{\lambda^*}})$ & Source for \\
&&&&&&&& $ {{\lambda^*}}< \sqrt{6},  f'({{\lambda^*}})<0$, \\
&&&&&&&& saddle for  $ {{\lambda^*}}> \sqrt{6}$, \\
&&&&&&&& or $f'({{\lambda^*}})>0$. \\
&&&&&&&& N. H.  otherwise.\\\hline 
$P_4({\lambda^*})$ & $\frac{{{\lambda^*}}}{\sqrt{6}}$ & ${{\lambda^*}}$ & $1$ & $-\sqrt{6} \leq\lambda^*\leq \sqrt{6}$ &  $\frac{1}{2} \left({{\lambda^*}}^2-6\right)$ & $ 2-{{\lambda^*}}^2$ & $ -{{\lambda^*}} f'({{\lambda^*}})$& sink for \\
&&&&&&&& $2 <\lambda^{*^2}<6, {\lambda^{*}} f'({\lambda^{*}})>0$. \\
&&&&&&&& saddle for $0\leq \lambda^{*^2}<2$, \\
&&&&&&&& or ${\lambda^{*}} f'({\lambda^{*}})<0$. \\
&&&&&&&& N. H.  otherwise. \\\hline
$P_5({\lambda^*})$ & $-1$ & ${{\lambda^*}}$ & $ 1$ & always & $-4$ & $\sqrt{6} {\lambda^{*}}+6$ & $\sqrt{6} f'({\lambda^{*}})$ &  sink for \\
&&&&&&&& ${\lambda^{*}}<-\sqrt{6}, f'({\lambda^{*}})<0$. \\
&&&&&&&& saddle for ${\lambda^{*}}<-\sqrt{6} $, \\
&&&&&&&& or $f'({\lambda^{*}})>0$.  \\
&&&&&&&& N. H.  otherwise. \\\hline 
$P_6({\lambda^*})$ & $1$ & $ {{\lambda^*}}$ & $ 1$ & always & $-4$ & $6-\sqrt{6} {\lambda^{*}}$ & $-\sqrt{6} f'({\lambda^{*}})$ & sink for \\
&&&&&&&& ${\lambda^{*}}>\sqrt{6}, f'({\lambda^{*}})>0$.\\
&&&&&&&& saddle for ${\lambda^{*}}<\sqrt{6} $.\\
&&&&&&&& or $f'({\lambda^{*}})<0$ \\
&&&&&&&& N. H.  otherwise \\\hline 
$P_7$ & $-\frac{1}{\sqrt{3}}$ & $ -\sqrt{2}$ & $ \bar{Z}_c$ & $\begin{array}{c}
   f(-\sqrt{2})=0, \\
   0\leq \bar{Z}_c \leq 1
\end{array}$ & $-2 $ & $0$ & $\sqrt{2} f'\left(-\sqrt{2}\right)$ & saddle if $f'\left(-\sqrt{2}\right)>0$ \\
&&&&&&&& sink for $f'\left(-\sqrt{2}\right)<0$\\\hline
$P_8$ & $\frac{1}{\sqrt{3}}$ & $ \sqrt{2}$ & $ \bar{Z}_c$ & $\begin{array}{c}f(\sqrt{2})=0,\\ 0\leq \bar{Z}_c \leq 1
\end{array}$ & $-2 $ &  $0$ & $-\sqrt{2} f'\left(\sqrt{2}\right)$.  &  saddle if $f'\left(\sqrt{2}\right)<0$. \\
&&&&&&&& sink for $f'\left(\sqrt{2}\right)>0$.\\\hline
$P_9$ & $0$ & $0$ & $0$ & always & $-2$ & $-\frac{1}{2} \left(3+\sqrt{9-12 f(0)}\right)$ & $ -\frac{1}{2} \left(3-\sqrt{9-12 f(0)}\right)$ & sink for $f(0)> 0$.\\
&&&&&&&& saddle for $f(0)<0$.\\\hline 
$P_{10}$ & $0$ & $0$ & $1$ & always & $2$  & $-\frac{1}{2} \left(3+\sqrt{9-12 f(0)}\right)$ & $ -\frac{1}{2} \left(3-\sqrt{9-12 f(0)}\right)$ & saddle. \\\hline
\end{tabular}}
    \caption{Equilibrium points of system \eqref{background(eq:99)} in the finite region for an arbitrary function $f(\lambda)$. N. H. stands for Non-hyperbolic.}
    \label{Background_b}
\end{table}

The equilibrium points in the background space are the following. 
\begin{enumerate}
 \item  $P_1({\lambda^*}): \left(x, \lambda, \bar{Z}\right)=\left(\frac{{{\lambda^*}}}{\sqrt{6}}, {{\lambda^*}}, 0\right)$ that  exists for $-\sqrt{6} \leq \lambda^*\leq \sqrt{6}$. The eigenvalues are $\frac{1}{2} \left({{\lambda^*}}^2-6\right), {{\lambda^*}}^2-2, -{{\lambda^*}} f'({{\lambda^*}})$. It is a saddle  for $f'(\lambda^*)<0, -\sqrt{2}<\lambda^*<0$, or $f'(\lambda^*)>0, 0<\lambda^*<\sqrt{2}$, or $2 <\lambda^{*^2}<6$ or ${\lambda^{*}} f'({\lambda^{*}})<0$. It is non-hyperbolic otherwise.

 \item $P_2({\lambda^*}): \left(x, \lambda, \bar{Z}\right)=\left(-1, {{\lambda^*}}, 0\right)$ that always exists. The eigenvalues are \newline $4, \sqrt{6} {{\lambda^*}}+6, \sqrt{6} f'({{\lambda^*}})$. It is a source for $ {{\lambda^*}}>- \sqrt{6}, f'({{\lambda^*}})>0$. It is a saddle for  $ {{\lambda^*}}<- \sqrt{6}$ or  $f'({{\lambda^*}})<0$.  It is non-hyperbolic otherwise.

 \item $P_3({\lambda^*}): \left(x, \lambda, \bar{Z}\right)=\left(1, {{\lambda^*}}, 0\right)$ that always exists. The eigenvalues are \newline $4, 6-\sqrt{6} {{\lambda^*}}, -\sqrt{6} f'({{\lambda^*}})$. It is a source for $ {{\lambda^*}}< \sqrt{6},  f'({{\lambda^*}})<0$. It is a saddle for  $ {{\lambda^*}}> \sqrt{6}$ or  
   $f'({{\lambda^*}})>0$.  It is non-hyperbolic otherwise.

\item $P_4({\lambda^*}): \left(x, \lambda, \bar{Z}\right)=\left(\frac{{{\lambda^*}}}{\sqrt{6}}, {{\lambda^*}}, 1\right)$ that exists for $-\sqrt{6} \leq\lambda^*\leq \sqrt{6}$. The eigenvalues are $\frac{1}{2} \left({{\lambda^*}}^2-6\right), 2-{{\lambda^*}}^2, -{{\lambda^*}} f'({{\lambda^*}})$. It is a sink for $2 <\lambda^{*^2}<6$ and ${\lambda^{*}} f'({\lambda^{*}})>0$. It is a saddle for $0\leq \lambda^{*^2}<2$ or ${\lambda^{*}} f'({\lambda^{*}})<0$. It is non-hyperbolic otherwise. 

 \item  $P_5({\lambda^*}): \left(x, \lambda, \bar{Z}\right)=\left(-1, {{\lambda^*}}, 1\right)$ that always exists. The eigenvalues are \newline $-4, \sqrt{6} {\lambda^{*}}+6, \sqrt{6} f'({\lambda^{*}})$.  It is a sink for ${\lambda^{*}}<-\sqrt{6} $ and $f'({\lambda^{*}})<0$. It is a saddle for ${\lambda^{*}}<-\sqrt{6} $ or $f'({\lambda^{*}})>0$.  It is non-hyperbolic otherwise.

 \item  $P_6({\lambda^*}): \left(x, \lambda, \bar{Z}\right)=\left(1, {{\lambda^*}}, 1\right)$ that always exists. The eigenvalues are \newline $-4, 6-\sqrt{6} {\lambda^{*}}, -\sqrt{6} f'({\lambda^{*}})$. It is a sink for ${\lambda^{*}}>\sqrt{6} $ and $f'({\lambda^{*}})>0$. It is a saddle for ${\lambda^{*}}<\sqrt{6} $ or $f'({\lambda^{*}})<0$.  It is non-hyperbolic otherwise. 

\item  The line
$P_7: \left(x, \lambda, \bar{Z}\right)=\left(-\frac{1}{\sqrt{3}}, -\sqrt{2}, \bar{Z}_c\right)$ exists for $f(-\sqrt{2})=0$ and $0\leq \bar{Z}_c \leq 1$. 
The eigenvalues are $-2, 0, \sqrt{2} f'\left(-\sqrt{2}\right)$. The eigenvector associated with the zero eigenvalues is tangent to the line. Then, it is normally hyperbolic. This implies it is a  saddle if $f'\left(-\sqrt{2}\right)>0$ or a sink for $f'\left(-\sqrt{2}\right)<0$.

\item  The line 
$P_8: \left(x, \lambda, \bar{Z}\right)=\left(\frac{1}{\sqrt{3}}, \sqrt{2}, \bar{Z}_c\right)$ exists for $f(\sqrt{2})=0$ and $0\leq \bar{Z}_c \leq 1$. The eigenvalues are $-2, 0, -\sqrt{2} f'\left(\sqrt{2}\right)$. The eigenvector associated with the zero eigenvalues is tangent to the line. Then, it is normally hyperbolic. This implies it is a  saddle if $f'\left(\sqrt{2}\right)<0$ or a sink for $f'\left(\sqrt{2}\right)>0$.

\item $P_9: \left(x, \lambda, \bar{Z}\right)=\left(0,0,0\right)$.  The eigenvalues are \newline $-2, -\frac{1}{2} \left(3+\sqrt{9-12 f(0)}\right), -\frac{1}{2} \left(3-\sqrt{9-12 f(0)}\right)$. It is a sink for $f(0)>0$ or a saddle for $f(0)<0$.

\item $P_{10}: \left(x, \lambda, \bar{Z}\right)=\left(0,0,1\right)$.  The eigenvalues are \newline $2, -\frac{1}{2} \left(3+\sqrt{9-12 f(0)}\right), -\frac{1}{2} \left(3-\sqrt{9-12 f(0)}\right)$. It is a saddle.

\end{enumerate}

Now we present some numerical solutions. As we commented, $\lambda$ is generically bounded at late-time attractors. However,  we handle the cases $\lambda \rightarrow \pm \infty$ using the new variable  \eqref{eqU}.

\subsubsection{First Example: monomial potential}

Substituting the function $f(\lambda)=-\frac{\lambda ^2}{n}$ in \eqref{background(eq:99)} we obtain
\begin{subequations}
\label{background(eq:99)-monomial}
\begin{align}
\frac{dx}{d {N}} &=-\left(3x - \sqrt{\frac{3}{2}} \lambda\right)\left(1-x^2\right), \\ \frac{d\lambda}{d N} &= \frac{\sqrt{6}}{n} x \lambda ^2, \\  \frac{d\bar{Z}}{d {N}} &= 2(3x^2 - 1)\bar{Z}\left(1-\bar{Z}\right), 
\end{align}
\end{subequations}
defined on the background space  $B = \left\{\left(x, \lambda, \bar{Z}\right)\in [-1,1]\times \mathbb{R} \times [0,1]\right\}$.

\begin{table}[]
    \centering
        
    \begin{tabular}{|c|c|c|c|c|c|c|c|c|}\hline
 Label & $x$ & $\lambda$ & $\bar{Z}$ & Existence & $k_1$ & $k_2$ & $k_3$ & Stability \\    \hline 
 $P_1(0)$ & $0$ & $0$ & $0$ & always & $-3$ & $-2$ & $0$ & saddle ($n>0$); sink ($n<0$) \\\hline
 $P_2(0)$ & $-1$ & $ 0$ & $ 0$ & always & $4$ & $ 6$ & $0$ & unstable  \\\hline
 $P_3(0)$ & $1$ & $ 0$ & $ 0$  & always  & $4$ & $6$ & $ 0$ & unstable \\\hline
 $P_4(0)$ & $0$ & $0$ & $1$ & always  & $-6$ & $ 2$ & $0$ &  saddle \\\hline
 $P_5(0)$ & $-1$ & $ 0$ & $ 1$& always & $-4$ & $ 6$ & $ 0$ &  saddle \\\hline 
 $P_6(0)$ & $1$ & $0$ & $ 1$ & always & $-4$ & $ 6$ & $ 0$ & saddle \\\hline
\end{tabular}
       \caption{Equilibrium points of system \eqref{background(eq:99)-monomial}   in the finite region for $f(\lambda)=-\frac{\lambda ^2}{n}$.}
    \label{Background_a-powerlawb}
    \end{table}

Tab. \ref{Background_a-powerlawb} presents the equilibrium points of system \eqref{background(eq:99)-monomial} in the finite region.

Interestingly, the de Sitter point $P_1(0): \left(x, \lambda, \bar{Z}\right)=\left(0, 0, 0\right)$ always exists. The eigenvalues are $-3, -2, 0$. It is non-hyperbolic. Using the Centre Manifold theorem, we obtain that the graph locally gives the centre manifold of the origin
\begin{align}
& \Big\{\left(x, \lambda, \bar{Z}\right)\in [-1,1]\times \mathbb{R} \times [0,1]: x=\frac{\lambda}{\sqrt{6}}+h_1(\lambda), \bar{Z}=h_2(\lambda), \nonumber \\
& h_1(0)=0, h_2(0)=0,  h_1'(0)=0, h_2'(0)=0, |\lambda|<\delta\Big\}
\end{align}
for a small enough $\delta$. The functions $h_1$ and $h_2$ satisfy the differential equations
\begin{align}
   & 6 \lambda ^2 \left(\left(\sqrt{6} h_1(\lambda )+\lambda \right) h_1'(\lambda )+h_1(\lambda )\right)  -3
   n h_1(\lambda ) \left(2 \sqrt{6} \lambda  h_1(\lambda )+6 h_1(\lambda )^2+\lambda
   ^2-6\right)+\sqrt{6} \lambda ^3=0,
\\
   & -\frac{\lambda ^2 \left(\sqrt{6} h_1(\lambda )+\lambda \right) h_2'(\lambda )}{n}-\left(2 \sqrt{6} \lambda 
   h_1(\lambda )+6 h_1(\lambda )^2+\lambda ^2-2\right) h_2(\lambda )^2 \nonumber \\
   & +\left(2 \sqrt{6} \lambda 
   h_1(\lambda )+6 h_1(\lambda )^2+\lambda ^2-2\right) h_2(\lambda )=0.
\end{align}
Using the Taylor series, we have the solution for $ x(\lambda)$ given by \eqref{expansion-x-t} and 
\begin{align}
Z(\lambda) & =   O\left(\lambda ^{14}\right). 
\end{align}

The 1D dynamical system dictates the dynamics at the centre manifold 
\begin{align}
    \frac{d \lambda}{d N} & =- U'(\lambda).
\end{align}
That is a gradient-like equation with potential 
$U(\lambda)$ defined  through \eqref{Gradient-Potential}. 
Since $U^{(4)}(0)=-6/n\neq 0$, the origin is a degenerate maximum of the potential for $n>0$. Therefore, the centre manifold of the origin and the origin are unstable (saddle), and if $n<0$, it is stable. 

\begin{figure}
    \centering
    \includegraphics[scale=0.6]{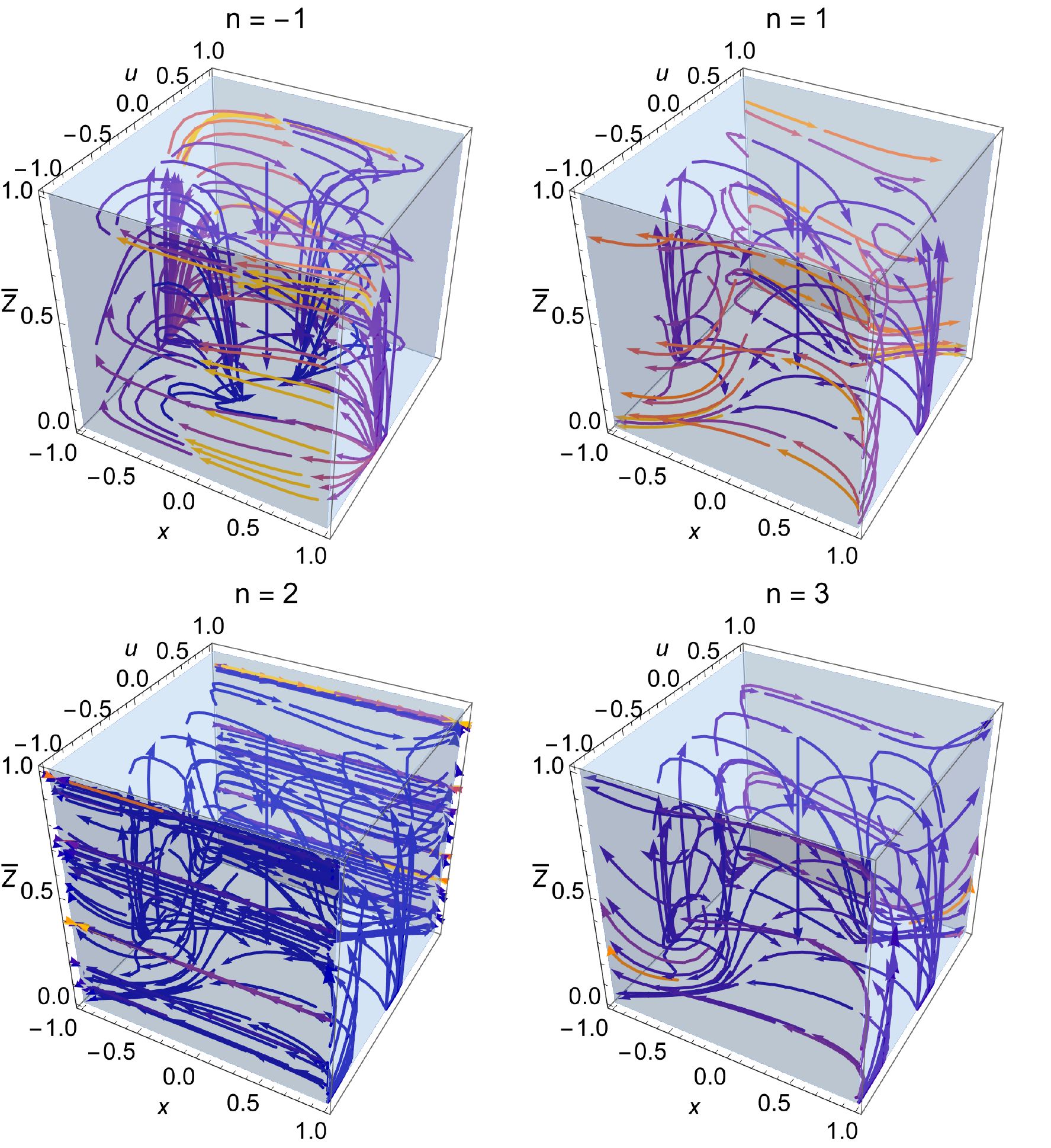}
    \caption{Flow of  the system \eqref{background(eq:99)-monomial} in the representations   $\left(x, u\right)$ for $n=-1,1,2,3$.}
    \label{fig:Powerlaw-background}
\end{figure}
In Fig.  \ref{fig:Powerlaw-background} is represented the flow of the system \eqref{background(eq:99)-monomial} in the phase space $\left(x, u, \bar{Z}\right)$ for $n=-1,1,2,3$. 

\begin{figure}
    \centering
    \includegraphics[scale=0.6]{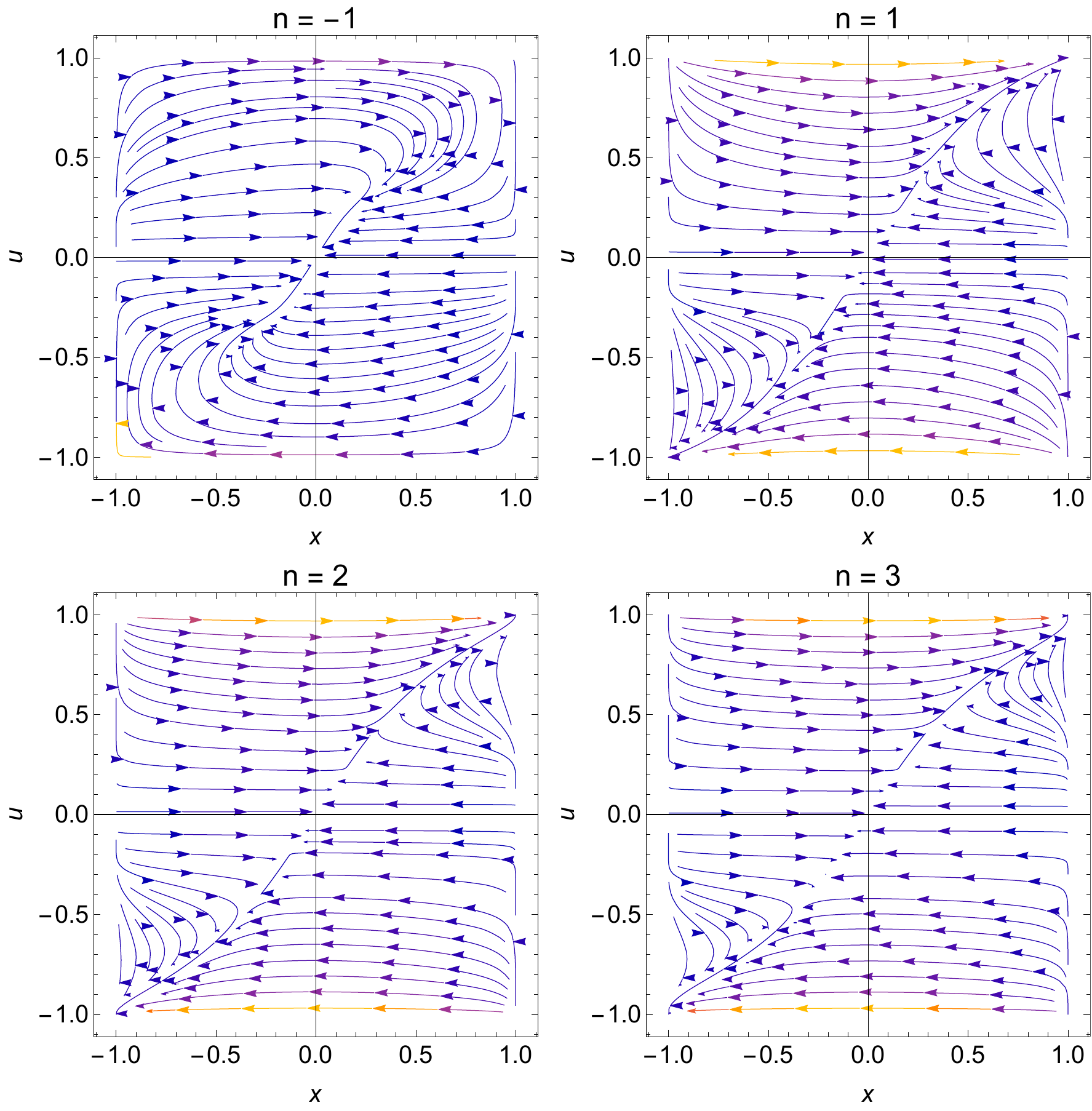}
    \caption{Flow of the system \eqref{background(eq:99)-monomial} in the plane $\left(x, u\right)$ for $n=-1,1,2,3$.
}
    \label{fig:powerlaw-background2D}
\end{figure}
In Fig.  \ref{fig:powerlaw-background2D} is represented the flow of the system \eqref{background(eq:99)-monomial} restricted to the plane $\left(x, u\right)$ for $n=-1,1,2,3$.

As shown in these Figs. \ref{fig:Powerlaw-background} and
\ref{fig:powerlaw-background2D}, for $n>0$ the late time attractor corresponds to $\lambda\rightarrow \pm\infty$ (along the centre manifold of the origin).  Which corresponds to $(x, \lambda,\bar{Z})=\left(\pm 1, \pm \infty, 1\right)$. For $n<0$, the attractor is the origin. For $n>0$, the origin is a saddle point. 
These numerical results illustrate the analytical results in Tab. \ref{Background_a-powerlawb}, therefore, the stability analysis is numerically confirmed.   
\subsubsection{Second Example: double exponential}

Substituting the function  $f(\lambda)=-\left(\lambda+\alpha\right)\left(\lambda+\beta\right)$ in \eqref{background(eq:99)} we obtain
\begin{subequations}
\label{background(eq:99)-double-exp}
\begin{align}
\frac{dx}{d {N}} & =-\left(3x - \sqrt{\frac{3}{2}} \lambda\right)\left(1-x^2\right), \\  \frac{d\lambda}{d {N}} & =  \sqrt{6}x \left(\lambda+\alpha\right)\left(\lambda+\beta\right), \\  \frac{d\bar{Z}}{d {N}} & = 2\left(3x^2 - 1\right)\bar{Z}\left(1-\bar{Z}\right), 
\end{align}
\end{subequations}
defined on the background space  $B = \left\{\left(x, \lambda, \bar{Z}\right)\in [-1,1]\times \mathbb{R} \times [0,1]\right\}$.

Recall that $f^{\prime}(\lambda)=-\alpha -\beta -2 \lambda$, and $f(\lambda)=0 \Longleftrightarrow\lambda\in\left\{-\alpha, -\beta\right\}$ and $ f'(-\alpha)=\alpha-\beta$,  $ f'(-\beta)=-\left(\alpha-\beta\right)$. Moreover, we have $f(0)=-\alpha \beta$, and $f'(0)=-\alpha- \beta$. Without losing generality, we can assume $\alpha<\beta$. 

\begin{table}[]
    \centering
               \resizebox{\textwidth}{!}{
    \begin{tabular}{|c|c|c|c|c|c|c|c|c|}\hline
 Label & $x$ & $\lambda$ & $\bar{Z}$ & Existence & $k_1$ & $k_2$ & $k_3$ & Stability \\\hline
$P_1(-\alpha)$ & $-\frac{\alpha }{\sqrt{6}}$ & $-\alpha $  &$ 0$ & $-\sqrt{6}<\alpha < \sqrt{6}$ & $\frac{1}{2} \left(\alpha ^2-6\right) $& $\alpha ^2-2$ &$ \alpha  (\alpha  -\beta )$ &   N. H.  for  \\
 &&&&&&&& $\alpha \in\left\{-\sqrt{2}, 0, \sqrt{2}\right\}$\\
 &&&&&&&&  sink for \\
 &&&&&&&&  $ 0<\alpha <\sqrt{2}, 
      \beta >\alpha$\\
 &&&&&&&&  saddle otherwise\\\hline
$P_1(-\beta)$ &  $-\frac{\beta }{\sqrt{6}}$ & $-\beta $ & $0$ & $-\sqrt{6}<\beta < \sqrt{6}$  & $\frac{1}{2} \left(\beta ^2-6\right)$& $\beta ^2-2$ & $\beta  (\beta -\alpha )$ &    N. H.  for \\
 &&&&&&&& $\beta \in\left\{-\sqrt{2}, 0, \sqrt{2}\right\}$\\
  &&&&&&&&  sink for \\
 &&&&&&&& $-\sqrt{2}<\beta <0, \alpha <\beta$\\
 &&&&&&&&  saddle otherwise\\\hline
$P_2(-\alpha)$ &  $-1 $& $-\alpha$  & $0 $ & always & $4$ &$ 6-\sqrt{6} \alpha$  & $\sqrt{6} (\alpha -\beta ) $ & N. H.  for $\alpha=\sqrt{6}$\\
  &&&&&&&&  saddle otherwise\\\hline
$P_2(-\beta)$ &$ -1$ & $-\beta$  & $0 $ & always & $4$  & $6-\sqrt{6} \beta $ & $\sqrt{6} (\beta -\alpha )$  &  
N. H.  for $\beta=\sqrt{6}$\\
 &&&&&&&& source for $\alpha <\beta <\sqrt{6}$\\
  &&&&&&&&  saddle otherwise\\\hline
$P_3(-\alpha)$ & $1$ & $-\alpha $ &$ 0$ & always & $4 $&$ \sqrt{6} \alpha +6$ & $\sqrt{6} (\beta -\alpha ) $ & N. H.  for $\alpha=-\sqrt{6}$\\
 &&&&&&&& source for $-\sqrt{6}<\alpha <\beta$\\
  &&&&&&&&  saddle otherwise\\\hline
$P_3(-\beta)$ & $ 1$ & $-\beta$  &$ 0$ & always &$ 4$ & $\sqrt{6} \beta +6 $ &$ \sqrt{6} (\alpha -\beta ) $ & N. H.  for $\beta=-\sqrt{6}$\\
  &&&&&&&&  saddle otherwise\\\hline
$P_4(-\alpha)$ &$ -\frac{\alpha }{\sqrt{6}}$ & $-\alpha$  & $1 $  & $-\sqrt{6}<\alpha < \sqrt{6}$  & $\frac{1}{2} \left(\alpha ^2-6\right)$ &$ 2-\alpha ^2$ & $\alpha  (\alpha -\beta )$  &  N. H.  for \\
 &&&&&&&& $\alpha \in\left\{-\sqrt{2}, 0, \sqrt{2}\right\}$\\
 &&&&&&&& sink for \\
 &&&&&&&& $\sqrt{2}<\alpha <\sqrt{6}, \beta >\alpha$\\
 &&&&&&&&  saddle otherwise\\\hline
$P_4(-\beta)$ &  $-\frac{\beta }{\sqrt{6}}$ &$ -\beta $ & $1$ & $-\sqrt{6}<\beta < \sqrt{6}$   & $\frac{1}{2} \left(\beta ^2-6\right)$ & $2-\beta ^2$ & $\beta  (\beta -\alpha )$  &  N. H.  for \\
 &&&&&&&& $\beta \in\left\{-\sqrt{2}, 0, \sqrt{2}\right\}$\\
 &&&&&&&& sink for \\
 &&&&&&&& $-\sqrt{6}<\beta <-\sqrt{2}, \alpha <\beta$ \\
 &&&&&&&&  saddle otherwise\\\hline
$P_5(-\alpha)$ & $ -1$ &$ -\alpha$  & $1$ & always & $-4 $& $6-\sqrt{6} \alpha$  & $\sqrt{6} (\alpha -\beta ) $ &  N. H.  for $\alpha= \sqrt{6}$\\
 &&&&&&&& sink for $\sqrt{6}<\alpha <\beta$\\
  &&&&&&&&  saddle otherwise\\\hline
$P_5(-\beta)$ &  $-1 $&$ -\beta$  &$ 1$ & always & $ -4 $  & $6-\sqrt{6} \beta  $ & $\sqrt{6} (\beta -\alpha ) $  & N. H.  for  $\beta=\sqrt{6}$\\
  &&&&&&&&  saddle otherwise\\\hline
$P_6(-\alpha)$ &$ 1 $& $-\alpha$  &$ 1 $& always & $-4 $&$ \sqrt{6} \alpha +6 $&$ \sqrt{6} (\beta -\alpha ) $ & N. H.  for $\alpha=-\sqrt{6}$\\
  &&&&&&&&  saddle otherwise\\\hline
$P_6(-\beta)$ & $1 $&$ -\beta $ &$ 1$ & always & $-4$  &$ \sqrt{6} \beta +6$ & $\sqrt{6} (\alpha -\beta )$ &  N. H.  for $\beta=-\sqrt{6}$\\
  &&&&&&&&  sink for $\alpha <\beta <-\sqrt{6}$ \\
  &&&&&&&& saddle otherwise\\\hline
 $P_7$ & $-\frac{1}{\sqrt{3}}$ & $ -\sqrt{2}$ & $ \bar{Z}_c$ & $\begin{array}{c}
  \beta>\alpha=\sqrt{2}, \\
   0\leq \bar{Z}_c \leq 1
\end{array}$ & $-2 $ & $0$ & $\sqrt{2} \left(\sqrt{2}-\beta \right)$ & sink \\ 
&  & &   & $\begin{array}{c} \alpha<\beta=\sqrt{2}, \\
   0\leq \bar{Z}_c \leq 1
\end{array}$ & $-2 $ & $0$ & $\sqrt{2} \left(\sqrt{2}-\alpha \right)$ & saddle\\\hline
$P_8$ & $\frac{1}{\sqrt{3}}$ & $ \sqrt{2}$ & $ \bar{Z}_c$ & $\begin{array}{c}
  \beta>\alpha=-\sqrt{2}, \\
   0\leq \bar{Z}_c \leq 1
\end{array}$ & $-2 $ &  $0$ & $\sqrt{2} \left(\sqrt{2}+\beta\right)$  &  saddle \\
&  & &   & $\begin{array}{c} \alpha<\beta=-\sqrt{2}, \\
   0\leq \bar{Z}_c \leq 1
\end{array}$ &&&  $\sqrt{2} \left(\sqrt{2}+\alpha\right)$ & sink\\\hline
$P_{9}$ & $0$ & $0$ & $0$ & always & $-2$ & $\frac{1}{2} \left(-\sqrt{12 \alpha  \beta +9}-3\right)$ & $\frac{1}{2} \left(\sqrt{12 \alpha  \beta
   +9}-3\right)$ & stable for $\alpha \beta< 0$ \\\hline
$P_{10}$ & $0$ & $0$ & $1$ & always  &$ 2$ & $\frac{1}{2} \left(-\sqrt{12 \alpha  \beta +9}-3\right)$ &$ \frac{1}{2} \left(\sqrt{12 \alpha  \beta
   +9}-3\right)$ & saddle \\\hline
\end{tabular}}
    \caption{Equilibrium points of system \eqref{background(eq:99)-double-exp}  in the finite region for $f(\lambda)=-(\lambda+\alpha)(\lambda+\beta), \; \alpha \neq \beta$. Without losing generality, we can assume $\alpha<\beta$. N.H. stand for Non-hyperbolic }
    \label{Background_Zb}
\end{table}
Tab. \ref{Background_Zb} presents the equilibrium points of system \eqref{background(eq:99)-double-exp}  in the finite region.

\begin{figure}
    \centering
    \includegraphics[scale=0.6]{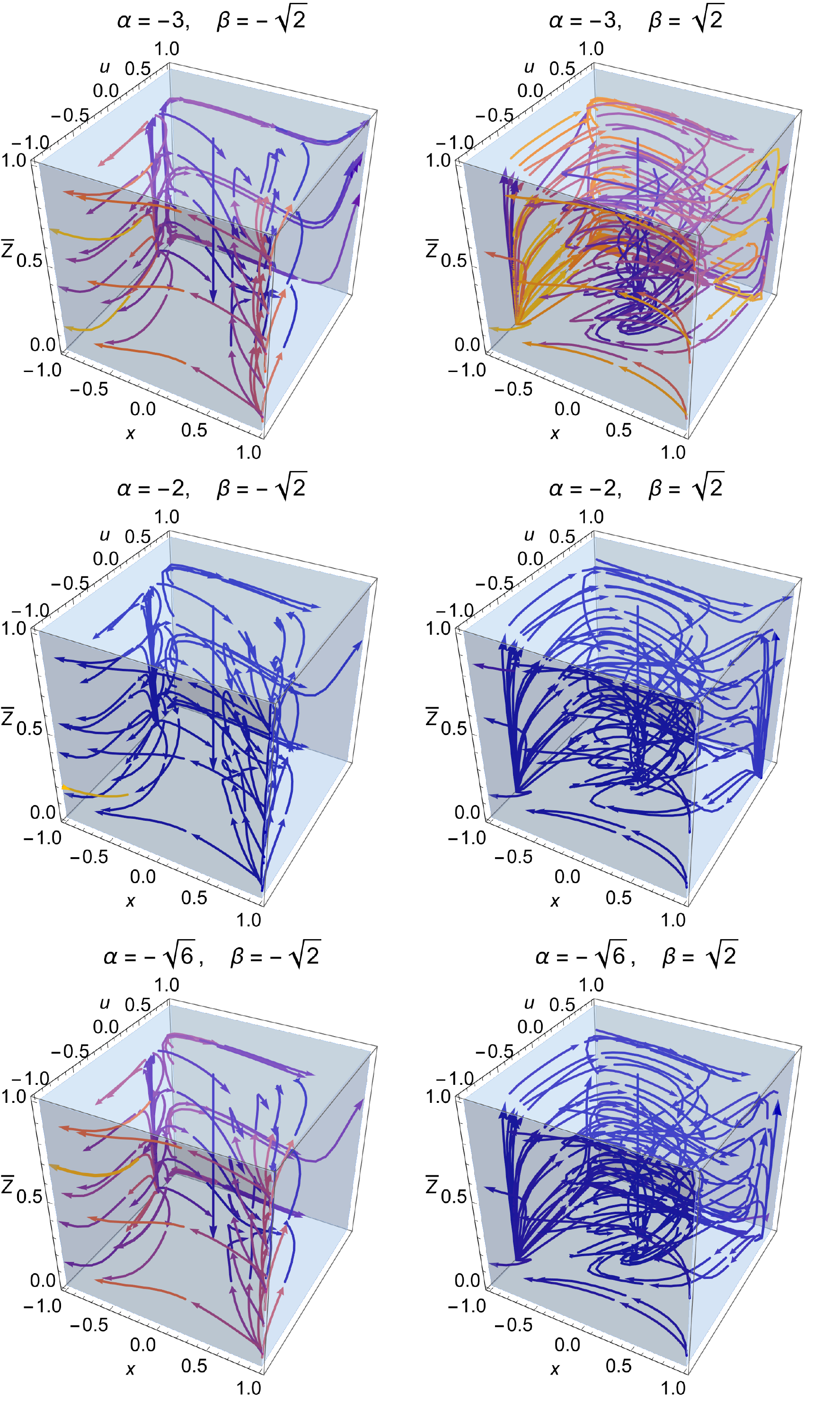}
    \caption{Flow of the system \eqref{background(eq:99)-double-exp} for different values of $\alpha$ and $\beta$.}
    \label{fig:double-exp-background}
\end{figure}

\begin{figure}
    \centering
    \includegraphics[scale=0.6]{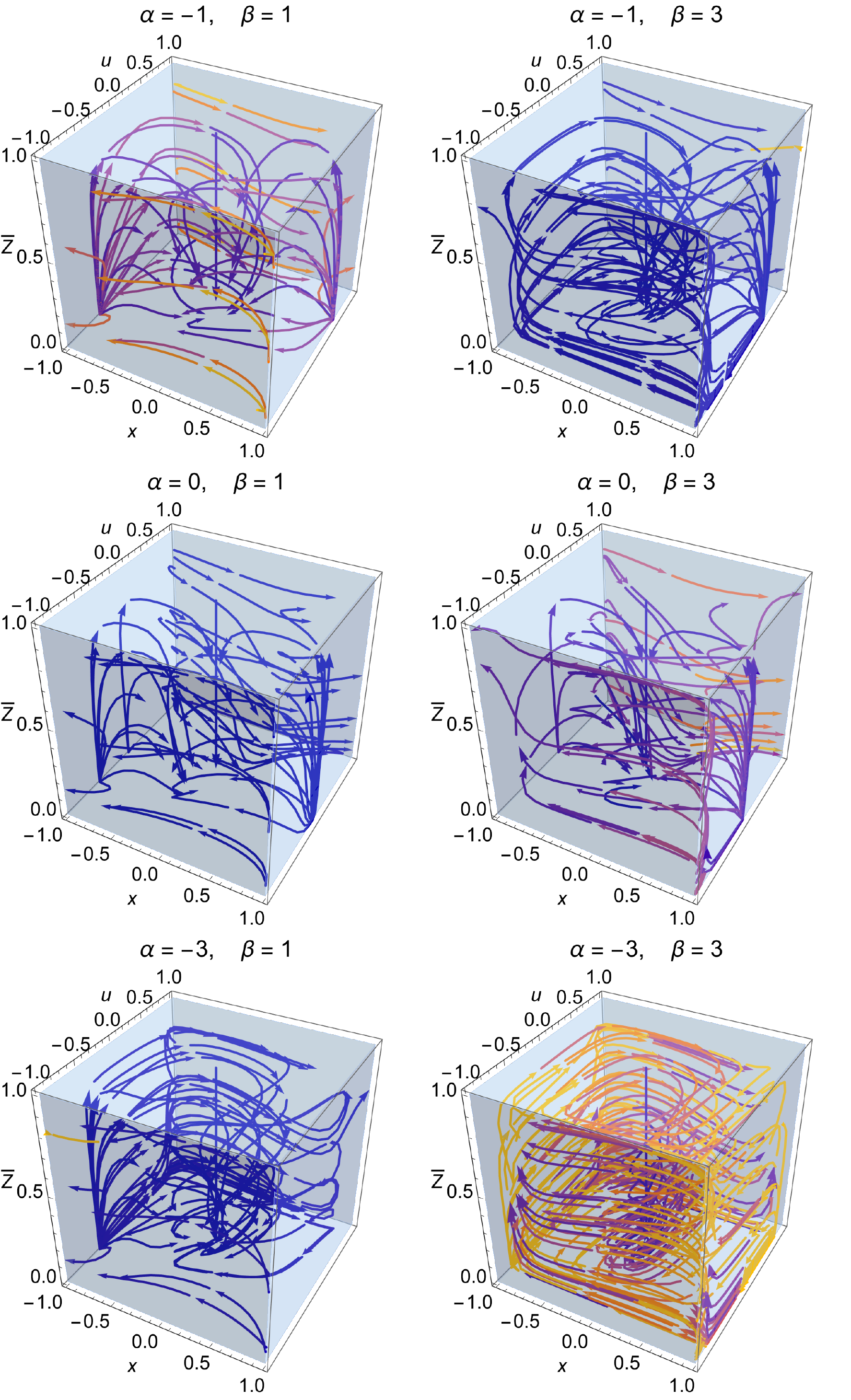}
    \caption{Flow of the system \eqref{background(eq:99)-double-exp} for different values of $\alpha$ and $\beta$.}
    \label{fig:double-exp-background2}
\end{figure}
Figs.  \ref{fig:double-exp-background} and  \ref{fig:double-exp-background2}  represent the flow of the system \eqref{background(eq:99)-double-exp} for different values of $\alpha$ and $\beta$ in the phase space $(x, u, \bar{Z})$. 

\begin{figure}
    \centering
    \includegraphics[scale=0.6]{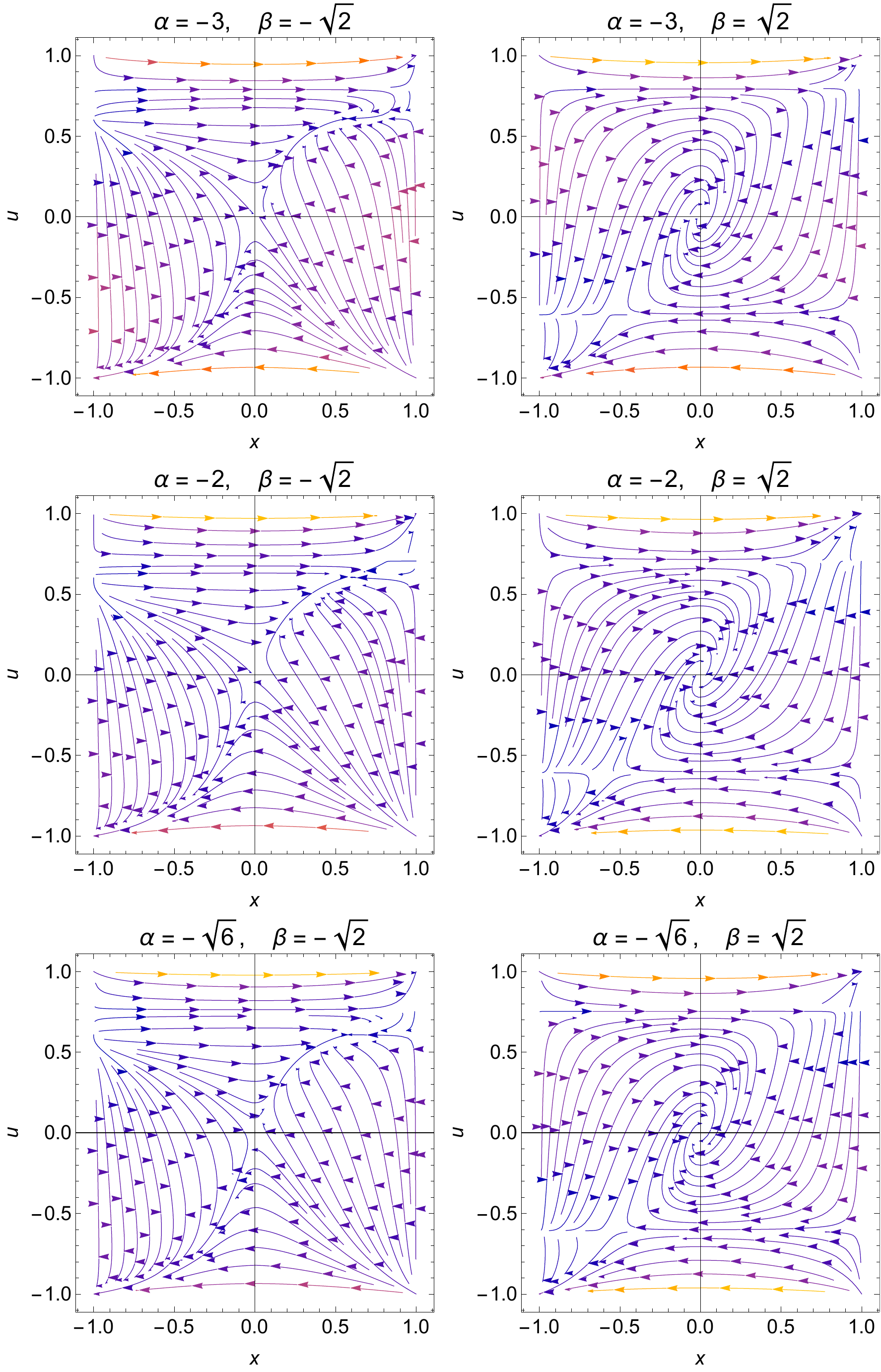}
    \caption{Flow of the system \eqref{background(eq:99)-double-exp} in the plane $\left(x, u\right)$ for different values of $\alpha$ and $\beta$.}
    \label{fig:double-exp-background2D}
\end{figure}

\begin{figure}
    \centering
    \includegraphics[scale=0.6]{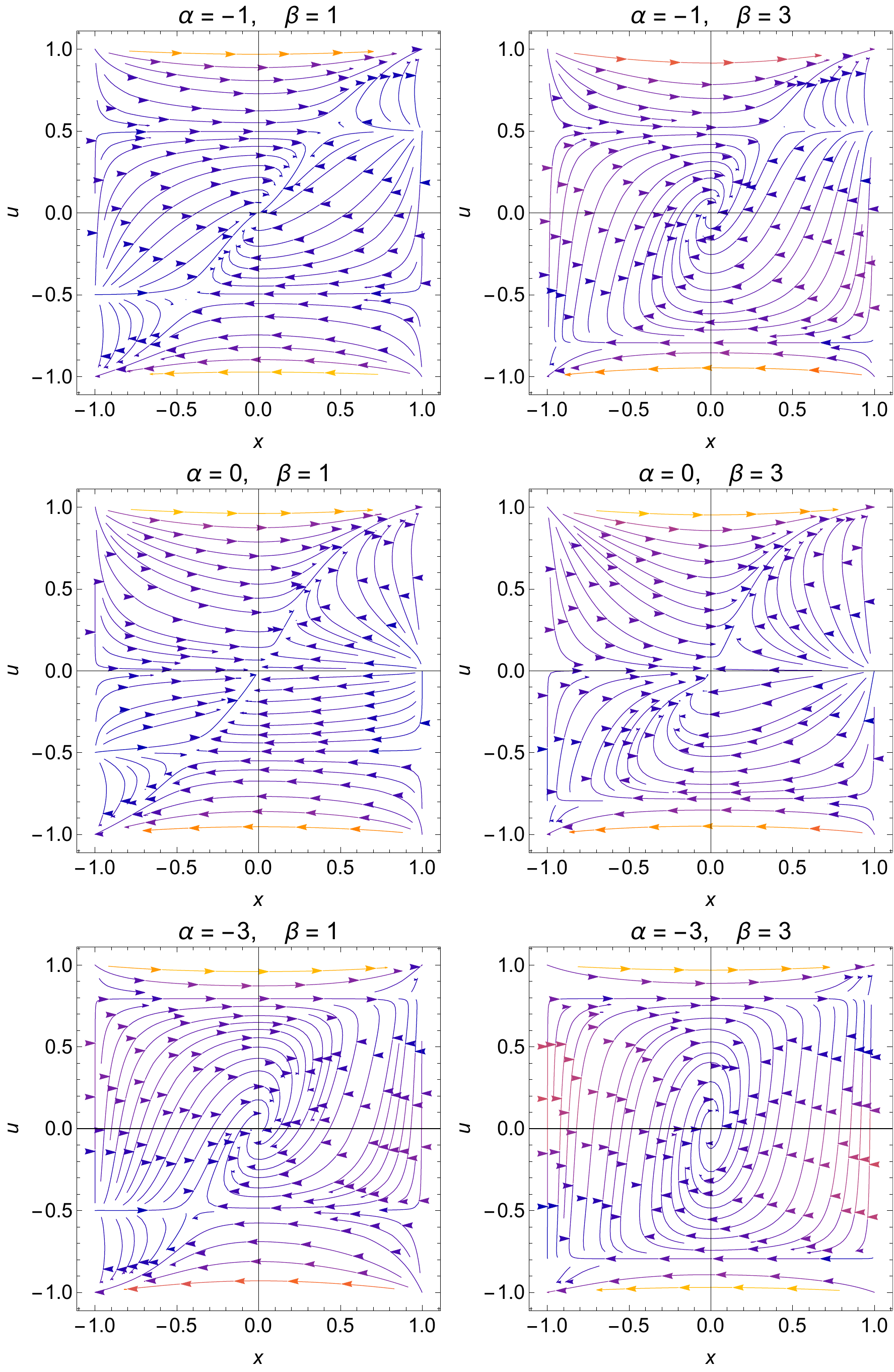}
    \caption{Flow of the system \eqref{background(eq:99)-double-exp} in the plane $\left(x, u\right)$ for different values of $\alpha$ and $\beta$.}
    \label{fig:double-exp-background22D}
\end{figure} 
Figs.  \ref{fig:double-exp-background2D} and \ref{fig:double-exp-background22D} represent the flow of the system \eqref{background(eq:99)-double-exp} in the plane $\left(x, u\right)$ for different values of $\alpha$ and $\beta$.
These numerical results illustrate the analytical results in Tab. \ref{Background_Zb}, therefore, the stability analysis is numerically confirmed.

\subsection{Extended phase space at background and perturbation levels: Bardeen Potential}

In equation \eqref{ptbn_bardeen3}, we first note that ${\Phi}_k$ is generally complex (as it came from Fourier transformation). 
So, we write ${\Phi}_k=F_1+iF_2$, where $F_1$ and $F_2$ are the real and imaginary parts of ${\Phi}_k$, respectively. Moreover, the resulting equation   has the structure 
\begin{equation}
  F{^{\prime \prime}}+P F^{\prime}+Q F =0, \label{pert-eq}
\end{equation}
where 
\begin{align}
    & P= \left[7-3x^{2}+\sqrt{6}\lambda\left(\frac{1-x^2}{x}\right)\right], \quad  Q= \left[6\left(1-x^2\right)+\sqrt{\frac{3}{2}}\lambda\left(\frac{1-x^2}{x}\right) + \frac{k^2}{a^2H^2}\right],
\end{align}
that is the same for $F_1$ and $F_2$. Generically, we denote $F_i=r_i\cos\theta_i$ and $F^{\prime}_i=r_i\sin\theta_i$ where $i=1,2$. So,
\begin{align}
   F^{\prime}=F  \tan\theta, \label{def-theta}
\end{align}
where $\frac{F^{\prime}}{F}=\mathcal{Y}$ has a period of $\pi$. Hence, the mapping $\mathcal{Y}=\tan\theta$ is two-to-one and, therefore, when $\theta$ makes one revolution ($0 \rightarrow  2\pi$)
$\mathcal{Y}$ has to be traversed twice $-\infty \rightarrow +\infty$ \cite{Alho:2020cdg}. Following this line, Equation \eqref{pert-eq} can be expressed then as 
\begin{equation}
{\mathcal{Y}}{^{\prime}}=-{\mathcal{Y}} ^2  -P {\mathcal{Y}} -Q \label{pert-eq_Y}, 
\end{equation}
or 
\begin{equation}
 \theta^{\prime} =-\sin^2\theta-P \sin\theta \cos\theta -Q\cos^2\theta.
\end{equation}
We also note that it is possible to get $F_i$ from $\mathcal{Y}_i$ through the expression
\begin{align}
    F_i(N)=F_i(0)\exp\left(\int_0^N \mathcal{Y}_i(\Tilde{N})d\Tilde{N}\right).
\end{align}
The sign of $\tan \theta$ denotes whether the $F_i(N)$ (for $i=1$ it is the real part) will grow or decay as $\theta$ ranges from $(-\pi,\pi]$.

Therefore, equation \eqref{EQ:(63)}  becomes 
\begin{align}
    Z^{\prime} = 2\left(3x^2 - 1\right)Z,
\end{align}
and equation \eqref{ptbn_bardeen3} lead for the \textbf{Bardeen potential $\Phi$} to 

\begin{align}\label{theta_bardeen}
   \theta^{\prime} &  = - \sin^{2}\theta - \left[7-3x^{2}+\sqrt{6}\lambda\left(\frac{1-x^2}{x}\right)\right] \sin\theta\cos\theta \nonumber \\
   & - \left[6\left(1-x^2\right)+\sqrt{\frac{3}{2}}\lambda\left(\frac{1-x^2}{x}\right) + Z\right]\cos^{2}\theta.
\end{align}

The final equations for the Bardeen potential are the background equations \eqref{(eq:99)} with the perturbation equation
\begin{align}
&\frac{d\theta}{d\bar{N}} = - \Bigg[\sin^2\theta + \left(7-3x^{2}+\sqrt{6}\lambda\left(\frac{1-x^2}{x}\right)\right) \sin\theta\cos\theta
    \nonumber\\
   & \qquad\qquad +\left(6\left(1-x^2\right)+\sqrt{\frac{3}{2}}\lambda\left(\frac{1-x^2}{x}\right)\right)\cos^2\theta\Bigg]\left(1-\bar{Z}\right)  - \bar{Z}\cos^2\theta, \label{(eq:99b)}
\end{align}
defined in the phase-space $B\times P$
modulo $n\pi, n\in\mathbb{Z}$, where  the background space is 
\begin{equation}
  B = \left\{\left(x, \lambda, \bar{Z}\right)\in [-1,1]\times \mathbb{R} \times [0,1]\right\}, \label{B-space}
\end{equation}
and the perturbation space is 
\begin{equation}
  P =  \left\{ \theta\in  [-\pi, \pi]\right\}. \label{P-space}
\end{equation}

The scalar field perturbations from Bardeen potentials, the comoving curvature perturbation, and the Sasaki-Mukhanov variable are growing up, decaying or frozen, according to the $\theta$-values. They are classified as super- or sub-horizon perturbations according to whether $\bar{Z}\rightarrow 0$ or  $\bar{Z}\rightarrow 1$. Therefore, we omitted the analysis of non-hyperbolic equilibrium points; the analysis of those points can be done numerically. 

\subsubsection{Sub-horizon boundary}

 Recall that the limit $k^2 \mathcal{H}^{-2}\gg 1$ corresponds to the short wavelength or sub-horizon boundary. It is related to the limit $\bar{Z}=1$.  
In this  limit \eqref{(eq:99)} and \eqref{(eq:99b)} becomes 
\begin{align}
& \frac{dx}{d\bar{N}} =0,  \;  \frac{d\lambda}{d\bar{N}} = 0, \;    \frac{d \theta}{d \bar{N}} = -\cos^2 \theta. \label{(eq:99c)}
\end{align}
Then, we have two asymptotic behaviours as $k^2 \mathcal{H}^{-2}\gg 1$, say there are two set of equilibrium points with constant $x, \lambda$ and $\theta= \pi/2 + n \pi, n=-1,0$. When $\cos^2 \theta>0$, $\theta$ is monotonically decreasing at constant $x, \lambda$. Then, the invariant set is spanned by a family of heteroclinic cycles with constant $x, \lambda$. They are denoted by       $A_{19}$  and $A_{20}$ in Tab. \ref{tab:III}. Because of their physical importance, we have distinguished some special points, say $A_{21}(\lambda^*)$, to $A_{30}$.

At $\theta= \pi/2 + n \pi, n=-1,0$ we have
\begin{equation}
    \frac{d \theta}{d \bar{N}}|_{\cos\theta=0}= -\left(1-\bar{Z}\right), \quad   \frac{d \theta}{d N}|_{\cos\theta=0}= -1. 
\end{equation}
which implies that the orbits near $\bar{Z}=1$ shadow the heteroclinic cycles and do not end at the equilibrium points in these cycles. 

Now passing to the  e-folding time, the stability of the set of equilibrium points on $\bar{Z}=1$ can be examined by analysing the  fast-slow system 
\begin{subequations}
\label{(eq:99sub-horizon)}
\begin{align}
 \frac{dx}{d{N}}& =-\left(3x - \sqrt{\frac{3}{2}} \lambda\right)\left(1-x^2\right), \\ 
 \frac{d\lambda}{d{N}} & = - \sqrt{6}xf, \\
 \frac{d\varepsilon}{d{N}} & = - 2\left(3x^2 - 1\right)\varepsilon \left(1-\varepsilon\right), \\
\varepsilon\frac{d\theta}{d{N}}& = -\cos^2\theta.
\end{align}
\end{subequations}
where $0<\left(1-\bar{Z}\right):=\varepsilon \leq 1$. We see, that, whenever $q:=(3x^2 - 1)>0$ the perturbation variable $\varepsilon$ monotonically tends to zero, so that the surface $\bar{Z}=1$ is approached as $q^*:=\left(3{x^*}^2 - 1\right)>0$ for a fixed value of $x$. On the other hand, for  $q^*<0$ for a fixed value of $x={x^*}$ the perturbation $\varepsilon$ is enhanced and  $\varepsilon\rightarrow 1$, whence, $\bar{Z}\rightarrow 0$. 

The angular variable produces an eigenvalue $-2\left(\cos\theta \sin\theta\right)/\varepsilon$ along the $\theta$-axis, that is zero at $\theta= \pi/2 + n \pi, n=-1,0$ as $\varepsilon\rightarrow 0$. Therefore, at the points with $\bar{Z}=1$, we have, in addition to the eigenvalues presented in section \ref{sect:A.1}, a zero eigenvalue corresponding to $\theta$. The stability conditions in the background space are the building blocks for the analysis of the extended phase space $B\times P$, modulo $n\pi, n\in\mathbb{Z}$, where  the background space is \eqref{B-space}
and the perturbation space is 
\eqref{P-space}.

\subsubsection{Stability analysis of the fixed points on the space $B\times P$}

\begin{table}[!ht]
    \centering
      \resizebox{\textwidth}{!}{%
    \begin{tabular}{|c|c|c|c|c|c|c|c|c|c|}
    \hline
    Label & $x$ & $\lambda$ & $\bar{Z}$ &$\theta$ & $k_1$ & $k_2$ & $k_3$ &$k_4$ & $a(t), H(t), \phi(t)$ \\\hline
    $A_{1}(\lambda^{*})$ &$ \frac{{{\lambda^*}}}{\sqrt{6}}$ &$ {{\lambda^*}}$ &$ 0$ & $-\cos ^{-1}(- \Delta_{1} )$ & $ \frac{1}{2} \left(\lambda^{*^2}-6\right)$ &$ \lambda^{*^2}-2 $& $ \Gamma_{1} +\left(8-\frac{3 \lambda^{*^2}}{2}\right) \sin \left(2 \sin ^{-1}( \Delta_{1} )\right) $& $ -{{\lambda^*}} f'({{\lambda^*}})$ & \eqref{A1},  \eqref{A1b}, \eqref{A1c} \\\hline
    $A_{2}(\lambda^{*})$ & $ \frac{{{\lambda^*}}}{\sqrt{6}}$ & ${{\lambda^*}}$ & $0$ & $\cos ^{-1}( \Delta_{1} )$ & $\frac{1}{2} \left(\lambda^{*^2}-6\right)$ & $\lambda^{*^2}-2$ & $ \Gamma_{1} +\left(8-\frac{3 \lambda^{*^2}}{2}\right) \sin \left(2 \cos ^{-1}( \Delta_{1} )\right)$ & $-{{\lambda^*}} f'({{\lambda^*}}) $& \eqref{A1},  \eqref{A1b}, \eqref{A1c}\\\hline
    $A_{3}(\lambda^{*})$ & $ \frac{{{\lambda^*}}}{\sqrt{6}}$ & ${{\lambda^*}}$ & $0$ &$ -\cos ^{-1}( \Delta_{1} )$ & $\frac{1}{2}\left(\lambda^{*^2}-6\right)$ &$ \lambda^{*^2}-2 $& $ \Gamma_{1} +\left(\frac{3 \lambda^{*^2}}{2}-8\right) \sin \left(2 \cos ^{-1}( \Delta_{1} )\right)$ & $-{{\lambda^*}} f'({{\lambda^*}}) $& \eqref{A1},  \eqref{A1b}, \eqref{A1c}\\\hline
    $A_{4}(\lambda^{*})$ &  $\frac{{{\lambda^*}}}{\sqrt{6}}$ & $ {{\lambda^*}}$ & $0$ & $\cos ^{-1}(- \Delta_{1} )$ & $\frac{1}{2}
   \left(\lambda^{*^2}-6\right)$ &$ \lambda^{*^2}-2$ & $ \Gamma_{1} +\left(\frac{3 \lambda^{*^2}}{2}-8\right) \sin \left(2 \sin ^{-1}( \Delta_{1} )\right) $& $ -{{\lambda^*}} f'({{\lambda^*}})$ & \eqref{A1},  \eqref{A1b}, \eqref{A1c} \\\hline
    $A_{5}(\lambda^{*})$ &  $\frac{{{\lambda^*}}}{\sqrt{6}}$ & ${{\lambda^*}}$ &$ 0$ & $-\cos ^{-1}(- \Delta_{2} )$ & $\frac{1}{2}
   \left(\lambda^{*^2}-6\right)$ & $\lambda^{*^2}-2$ &$  \Gamma_{2} +\left(8-\frac{3 \lambda^{*^2}}{2}\right) \sin \left(2 \sin ^{-1}( \Delta_{2} )\right)$ & $-{{\lambda^*}} f'({{\lambda^*}})$ & \eqref{A1},  \eqref{A1b}, \eqref{A1c}\\\hline
    $A_{6}(\lambda^{*})$ & $\frac{{{\lambda^*}}}{\sqrt{6}} $& ${{\lambda^*}}$ & $0$ &$ \cos ^{-1}( \Delta_{2} )$ & $\frac{1}{2}
   \left(\lambda^{*^2}-6\right)$ &$ \lambda^{*^2}-2$ & $ \Gamma_{2} +\left(8-\frac{3 \lambda^{*^2}}{2}\right) \sin \left(2 \cos ^{-1}( \Delta_{2} )\right)$ & $ -{{\lambda^*}} f'({{\lambda^*}}) $ & \eqref{A1},  \eqref{A1b}, \eqref{A1c} \\\hline
    $A_{7}(\lambda^{*})$ &$ \frac{{{\lambda^*}}}{\sqrt{6}}$ &$ {{\lambda^*}} $&$ 0$ & $-\cos ^{-1}( \Delta_{2} )$ & $\frac{1}{2}
   \left(\lambda^{*^2}-6\right)$ & $\lambda^{*^2}-2$ & $ \Gamma_{2} +\left(\frac{3 \lambda^{*^2}}{2}-8\right) \sin \left(2 \cos ^{-1}( \Delta_{2} )\right)$ &$ -{{\lambda^*}} f'({{\lambda^*}})$ & \eqref{A1},  \eqref{A1b}, \eqref{A1c}\\\hline
    $A_{8}(\lambda^{*})$ & $\frac{{{\lambda^*}}}{\sqrt{6}} $& ${{\lambda^*}}$ & $0 $&$ \cos ^{-1}(- \Delta_{2} )$ & $\frac{1}{2}
   \left(\lambda^{*^2}-6\right)$ & $\lambda^{*^2}-2$ &$  \Gamma_{2} +\left(\frac{3 \lambda^{*^2}}{2}-8\right) \sin \left(2 \sin ^{-1}( \Delta_{2} )\right)$ &$ -{{\lambda^*}} f'({{\lambda^*}})$ & \eqref{A1},  \eqref{A1b}, \eqref{A1c}\\\hline
    $A_{9}(\lambda^{*})$ & $-1$ &$ {{\lambda^*}}$ & $0$ & $0$ & $-4$ & $4$ & $\sqrt{6} {{\lambda^*}}+6$ & $\sqrt{6} f'({{\lambda^*}})$ & \eqref{A2}, \eqref{A2b}, \eqref{A2c}\\\hline
    $A_{10}(\lambda^{*})$ & $-1$ & ${{\lambda^*}}$ & $0 $&$ -\pi$  & $-4$ & $4$ & $\sqrt{6} {{\lambda^*}}+6$ & $\sqrt{6} f'({{\lambda^*}})$ & \eqref{A2}, \eqref{A2b}, \eqref{A2c}\\\hline
    $A_{11}(\lambda^{*})$ & $-1$ & ${{\lambda^*}}$ & $0$ & $\pi$  &$ -4$ &$ 4$ & $\sqrt{6} {{\lambda^*}}+6$ & $\sqrt{6} f'({{\lambda^*}})$ & \eqref{A2}, \eqref{A2b}, \eqref{A2c}\\\hline
    $A_{12}(\lambda^{*})$ & $-1$ & ${{\lambda^*}}$ & $0$ & $\sec ^{-1}\left(-\sqrt{17}\right)$ & $4$ & $4$ & $\sqrt{6} {{\lambda^*}}+6$ & $\sqrt{6}
   f'({{\lambda^*}})$ & \eqref{A2}, \eqref{A2b}, \eqref{A2c}\\\hline
    $A_{13}(\lambda^{*})$ & $ -1$ & ${{\lambda^*}}$ & 0 &$ -\sec ^{-1}\left(\sqrt{17}\right)$ & $4$ &$ 4$ & $\sqrt{6} {{\lambda^*}}+6$ & $\sqrt{6}
   f'({{\lambda^*}})$ & \eqref{A2}, \eqref{A2b}, \eqref{A2c}\\\hline
    $A_{14}(\lambda^{*})$ &  $1$ &$ {{\lambda^*}}$ & $0$ & $0$ &$ -4$ & $4$ & $6-\sqrt{6} {{\lambda^*}}$ & $-\sqrt{6} f'({{\lambda^*}})$& \eqref{A2}, \eqref{A2b}, \eqref{A2c}\\\hline
    $A_{15}(\lambda^{*})$ &  $1$ & ${{\lambda^*}}$ & $0$ &$ -\pi$  & $-4$ & $4$ & $6-\sqrt{6} {{\lambda^*}}$ & $-\sqrt{6} f'({{\lambda^*}})$ & \eqref{A2}, \eqref{A2b}, \eqref{A2c}\\\hline
    $A_{16}(\lambda^{*})$ &  $1$ & ${{\lambda^*}}$ &$ 0$ & $\pi$  &$ -4$ & $4 $&$ 6-\sqrt{6} {{\lambda^*}}$ & $-\sqrt{6} f'({{\lambda^*}}) $ & \eqref{A2}, \eqref{A2b}, \eqref{A2c}\\\hline
    $A_{17}(\lambda^{*})$ &  $1$ &$ {{\lambda^*}}$ & $0$ & $\sec ^{-1}\left(-\sqrt{17}\right)$ &$ 4$ & $4$ & $6-\sqrt{6} {{\lambda^*}} $&$ -\sqrt{6}
   f'({{\lambda^*}})$ & \eqref{A2}, \eqref{A2b}, \eqref{A2c}\\\hline
    $A_{18}(\lambda^{*})$ &  $1$ & ${{\lambda^*}}$ &$ 0 $& $-\sec ^{-1}\left(\sqrt{17}\right)$ & $4 $& $4 $& $6-\sqrt{6} {{\lambda^*}}$ &$ -\sqrt{6}
   f'({{\lambda^*}})$ & \eqref{A2}, \eqref{A2b}, \eqref{A2c}\\\hline
    $A_{19}$&  $x_c$ & $\lambda_c$  & $1$ & $-\frac{\pi }{2}$ & $0$ & $0$ & $0$ & $0$ & \eqref{CASE-A}, \eqref{CASE-Ab}, \eqref{CASE-Ac} \\\hline
    $A_{20}$ & $x_c$ & $\lambda_c$ & $1$ & $\frac{\pi }{2}$ & $0$ & $0$ & $0$ & $0$ & \eqref{CASE-A}, \eqref{CASE-Ab}, \eqref{CASE-Ac}\\\hline   
    $A_{21}(\lambda^{*})$ & $\frac{{{\lambda^*}}}{\sqrt{6}}$ & ${{\lambda^*}}$ & $1$ & $-\frac{\pi }{2}$ & $0$ &$0$ &$0$ &$0$ & \eqref{A1},  \eqref{A1b}, \eqref{A1c}\\\hline
    $A_{22}(\lambda^{*})$ & $\frac{{{\lambda^*}}}{\sqrt{6}}$ & ${{\lambda^*}}$ &$ 1$ & $\frac{\pi }{2}$ & $0$ &$0$ &$0$ & $0$ & \eqref{A1},  \eqref{A1b}, \eqref{A1c}\\\hline
    $A_{23}$ &$ -1 $& $\lambda_c $ & $1$ &$ -\frac{\pi }{2}$ & $0$ &$ 0 $& $ 0$ & $0$  & \eqref{A2}, \eqref{A2b}, \eqref{A2c} \\\hline
    $A_{24}$ & $1$ & $\lambda_c$  & $1$ & $-\frac{\pi }{2}$ & $0$ & $0$ & $0 $& $0$ & \eqref{A2}, \eqref{A2b}, \eqref{A2c} \\\hline
    $A_{25}$ &  $ -1$ & $\lambda_c$  & $1$ & $\frac{\pi }{2}$ & $0$ & $0$ & $0$ & $0$ & \eqref{A2}, \eqref{A2b}, \eqref{A2c} \\\hline
    $A_{26}$  & $1$ & $\lambda_c$  & $1$ & $\frac{\pi }{2}$ & $0$ & $0$ &$ 0 $& $0$ & \eqref{A2}, \eqref{A2b}, \eqref{A2c} \\\hline
    $A_{27}$   & $-\frac{1}{\sqrt{3}}$ & $\lambda_c$  & $1 $& $-\frac{\pi }{2}$ & $0$ &$0$ &$0$ &$0$ & \eqref{A3}, \eqref{A3b}, \eqref{A3c}\\\hline
    $A_{28}$ & $\frac{1}{\sqrt{3}}$ & $\lambda_c$  &  $1$ & $-\frac{\pi }{2}$ & $0$ &$0$ &$0$ &$0$ & \eqref{A3}, \eqref{A3b}, \eqref{A3c}\\\hline
    $A_{29}$ & $-\frac{1}{\sqrt{3}}$ & $\lambda_c $ & $1$ & $\frac{\pi }{2}$ & $0$ &$0$ &$0$ &$0$ & \eqref{A3}, \eqref{A3b}, \eqref{A3c}\\\hline
    $A_{30}$ & $\frac{1}{\sqrt{3}}$ & $\lambda_c $ & $1$ & $\frac{\pi }{2}$ & $0$ &$0$ &$0$ &$0$ & \eqref{A3}, \eqref{A3b}, \eqref{A3c}\\\hline
       \end{tabular}}
    \caption{Equilibrium points of the system \eqref{(eq:99)} and \eqref{(eq:99b)}. }
    \label{tab:III}
\end{table}
In Tab. \ref{tab:III} the equilibrium points of the system \eqref{(eq:99)} and \eqref{(eq:99b)}  are presented, where we denote by $\lambda^*$ any zeroes of $f(\lambda)$ and we define the quantities

\begin{align}
   \Delta_{1,2} & = \frac{2 \sqrt{2}}{\sqrt{9 {{\lambda^*}}^4\pm 3 \left(\sqrt{9 {{\lambda^*}}^4-132 {{\lambda^*}}^2+532} \mp 48\right) {{\lambda^*}}^2\mp 26 \sqrt{9 {{\lambda^*}}^4-132
   {{\lambda^*}}^2+532}+612}}, 
\end{align}
and 
\begin{align}
 \Gamma_{1,2}&= \frac{6 \left(3
   {{\lambda^*}}^2-26\right) \left(3 {{\lambda^*}}^2-23\right)}{-9 {{\lambda^*}}^4+108 {\lambda^{*}}^2\pm 3 {{\lambda^*}}^2\sqrt{9 {{\lambda^*}}^4-132 {{\lambda^*}}^2+532} \mp 26 \sqrt{9 {\lambda^{*}}^4-132 {{\lambda^*}}^2+532}-320}.
\end{align}

These equilibrium points and the stability conditions are summarised as follows:

 $A_{1,2}(\lambda^{*}):\left(\frac{{{\lambda^*}}}{\sqrt{6}}, {{\lambda^*}}, 0, \mp\cos ^{-1}(\mp\Delta_{1} )\right)$ exist for $-\sqrt{6}\leq \lambda^{*}\leq \sqrt{6}$, they are saddles  for $f'(\lambda^*)<0, -\sqrt{2}<\lambda^*<0$, or $f'(\lambda^*)>0, 0<\lambda^*<\sqrt{2}$, or $2 <\lambda^{*^2}<6$ or ${\lambda^{*}} f'({\lambda^{*}})<0$. They are non-hyperbolic otherwise. We have a cosmological solution for these equilibrium points with an asymptotic scale factor \eqref{A1}. Since $\frac{{\Phi}'}{\Phi}=\frac{\sqrt{1- \Delta_{1} ^2}}{ \Delta_{1} }$, and $ \Delta_{1}>0$, the amplitude of super-horizon Bardeen potential perturbation growth up an at an exponential rate. Using the procedures of section \ref{Section-6.3.1}, that is, under the transformation 
\begin{equation}
    \label{transform}
    \Phi_k = a^{-\left(6-\frac{\lambda^{*^2}}{2}\right)}v_k,
\end{equation} we obtain the equation  
\begin{equation}
\label{Bessel}
    \frac{d^2 v_k}{d\eta^2} +    v_k\left(k^2-\frac{2 ({\lambda^{*}} -3) ({\lambda^{*}} +3) \left({\lambda^{*}} ^2-6\right)}{\eta ^2 \left({\lambda^{*}} ^2-2\right)^2}\right)=0,
\end{equation}
with solution
\begin{equation}
    \label{sol}
 v_k(\eta)=  C_+ \sqrt{\eta } J_{\nu}(k \eta )+C_- \sqrt{\eta
   } Y_{\nu}(k \eta ),
\end{equation}
where 
\begin{equation}\label{nu}
 \nu=   \frac{\sqrt{9 {\lambda^{*}} ^4-124 {\lambda^{*}} ^2+436}}{2 \left({\lambda^{*}} ^2-2\right)},
\end{equation} and  $C_+$ and $C_-$ are complex constants depending on $k$.

 $A_{3,4}(\lambda^{*}):\left(\frac{{{\lambda^*}}}{\sqrt{6}}, {{\lambda^*}}, 0, \mp\cos ^{-1}(\pm \Delta_{1} )\right)$, with $-\sqrt{6}\leq \lambda^{*}\leq \sqrt{6}$, are sinks for $f'(\lambda^*)<0, -\sqrt{2}<\lambda^*<0$, or $f'(\lambda^*)>0, 0<\lambda^*<\sqrt{2}$.  For the range $2 <\lambda^{*^2}<6$ or ${\lambda^{*}} f'({\lambda^{*}})<0$  they are saddles. They are non-hyperbolic otherwise.      We have a cosmological solution for these equilibrium points with an asymptotic scale factor \eqref{A1}. Since $\frac{{\Phi}'}{\Phi}=-\frac{\sqrt{1- \Delta_{1} ^2}}{ \Delta_{1} }$, and $ \Delta_{1}>0$, the amplitude of super-horizon Bardeen potential perturbation exponentially decays. Introducing the transformation \eqref{transform}, we acquire the Bessel equation \eqref{Bessel} with solution \eqref{sol} where the parameter $\nu$ is defined by \eqref{nu}, and $C_+$ and $C_-$ are complex constants depending on $k$.

 $A_{5,6}(\lambda^{*}):\left(\frac{{{\lambda^*}}}{\sqrt{6}}, {{\lambda^*}}, 0, \mp\cos ^{-1}(\mp\Delta_{2} )\right)$,  with $-\sqrt{6}\leq \lambda^{*}\leq \sqrt{6}$, are saddles for $f'(\lambda^*)<0, -\sqrt{2}<\lambda^*<0$, or $f'(\lambda^*)>0, 0<\lambda^*<\sqrt{2}$, or   $2 <\lambda^{*^2}<6$ or ${\lambda^{*}} f'({\lambda^{*}})<0$. They are non-hyperbolic otherwise. We have a cosmological solution for these equilibrium points with an asymptotic scale factor \eqref{A1}. Since $\frac{{\Phi}'}{\Phi}=\frac{\sqrt{1- \Delta_{2} ^2}}{ \Delta_{2} }$, and $ \Delta_{2}>0$, the amplitude of super-horizon Bardeen potential perturbation growth up an at an exponential rate.   Introducing the transformation \eqref{transform}, we acquire the Bessel equation \eqref{Bessel} with solution \eqref{sol} where the parameter $\nu$ is defined by \eqref{nu}, and $C_+$ and $C_-$ are complex constants depending on $k$. 
   
 $A_{7,8}(\lambda^{*}):\left(\frac{{{\lambda^*}}}{\sqrt{6}}, {{\lambda^*}}, 0, \mp\cos ^{-1}( \pm\Delta_{2} )\right)$,  with $-\sqrt{6}\leq \lambda^{*}\leq \sqrt{6}$, are saddles for $f'(\lambda^*)<0, -\sqrt{2}<\lambda^*<0$, or $f'(\lambda^*)>0, 0<\lambda^*<\sqrt{2}$, or $2 <\lambda^{*^2}<6$ or ${\lambda^{*}} f'({\lambda^{*}})<0$.  They are non-hyperbolic otherwise. We have a cosmological solution for these equilibrium points with an asymptotic scale factor \eqref{A1}. Since $\frac{{\Phi}'}{\Phi}=-\frac{\sqrt{1- \Delta_{2} ^2}}{ \Delta_{2} }$, and $ \Delta_{2}>0$, the amplitude of super-horizon Bardeen potential perturbation exponentially decays. Introducing the transformation \eqref{transform}, we acquire the Bessel equation \eqref{Bessel} with solution \eqref{sol} where the parameter $\nu$ is defined by \eqref{nu}, and $C_+$ and $C_-$ are complex constants depending on $k$.

  $A_{9}(\lambda^*):\left(-1, {{\lambda^*}},0,0\right)$,  $A_{10}(\lambda^{*}):\left(-1, {{\lambda^*}}, 0, -\pi\right)$ and 
  $A_{11}(\lambda^{*}):\left(-1, {\lambda^*}, 0,\pi\right)$ are saddles. We have a cosmological solution for these equilibrium points with an asymptotic scale factor 
\eqref{A2}. Since $\frac{{\Phi}'}{\Phi}=0$, the amplitude of super-horizon Bardeen potential perturbation is frozen. 
Under the transformation 
\begin{equation}
    \label{transform1b}
    \Phi_k = a^{-3}v_k,
\end{equation} we obtain the equation  
\begin{equation}
\label{Bessel1b}
    \frac{d^2 v_k}{d\eta^2} +  k^2  v_k =0,
\end{equation}
with solution
\begin{equation}
    \label{sol1b}
 v_k(\eta)=  C_+ \cos(k \eta )+C_- \sin(k \eta )
\end{equation}
where  $C_+$ and $C_-$ are complex constants depending on $k$. 

  $A_{12,13}(\lambda^{*}):\left(-1, {{\lambda^*}}, 0, \pm\sec ^{-1}\left(\mp\sqrt{17}\right)\right)$  are sources for $ {{\lambda^*}}>- \sqrt{6}, 
   f'({{\lambda^*}})>0$. They are saddles for  $ {{\lambda^*}}<- \sqrt{6}$ or  
   $f'({{\lambda^*}})<0$. They are non-hyperbolic otherwise. We have a cosmological solution for this equilibrium point with an asymptotic scale factor
\eqref{A2}. Since $\frac{{\Phi}'}{\Phi}=-4$,  the amplitude of super-horizon Bardeen potential perturbation exponentially decays. Introducing the transformation \eqref{transform1b}, we acquire the equation \eqref{Bessel1b} with solution \eqref{sol1b} where  $C_+$ and $C_-$ are complex constants depending on $k$.  
    
  $A_{14}(\lambda^{*}):\left(1, {{\lambda^*}}, 0,0\right)$,  $A_{15}(\lambda^{*}):\left(1, {{\lambda^*}}, 0,  -\pi\right)$
 and $A_{16}(\lambda^{*}):\left(1, {{\lambda^*}},  0, \pi\right)$ are saddles. We have a cosmological solution for these equilibrium points with an asymptotic scale factor  
\eqref{A2}. Since $\frac{{\Phi}'}{\Phi}=0$, the amplitude of super-horizon Bardeen potential perturbation is frozen. Introducing the transformation \eqref{transform1b}, we acquire the equation \eqref{Bessel1b} with solution \eqref{sol1b} where  $C_+$ and $C_-$ are complex constants depending on $k$.     
 
  $A_{17,18}(\lambda^{*}):\left(1,  {{\lambda^*}}, 0, \pm\sec ^{-1}\left(\mp\sqrt{17}\right)\right)$  are sources for $ {{\lambda^*}}< \sqrt{6}, 
   f'({{\lambda^*}})<0$. They are saddles for  $ {{\lambda^*}}> \sqrt{6}$ or  
   $f'({{\lambda^*}})>0$.  They are non-hyperbolic otherwise. We have a cosmological solution for these equilibrium points with an asymptotic scale factor \eqref{A2}. Since $\frac{{\Phi}'}{\Phi}=-4$,  the amplitude of super-horizon Bardeen potential perturbation exponentially decays. Introducing the transformation \eqref{transform1b}, we acquire the equation \eqref{Bessel1b} with solution \eqref{sol1b} where  $C_+$ and $C_-$ are complex constants depending on $k$.

As we commented,  the invariant set $\bar{Z}=1$ is spanned by a family of heteroclinic cycles with constant $x, \lambda$. They are denoted by       $A_{19}$  and $A_{20}$ in Tab. \ref{tab:III}. Because of their physical importance, we have distinguished some special points, say $A_{21}(\lambda^*)$, to $A_{30}$. The eigenvalues of the linearisation of system \eqref{(eq:99)} are  $0, 0, 0, 0$ at those equilibrium points. Therefore they are non-hyperbolic.

  $A_{19}:\left(x_c, \lambda_c, 1, -\frac{\pi }{2}\right)$, with $-1\leq x_c\leq 1$. We have a cosmological solution for this equilibrium point with an asymptotic scale factor \eqref{CASE-A}. For  ${x_c}=0$ we have a de Sitter expansion with $a(t)= e^{H_0 \left(t-t_U\right)}$.  Since $\frac{{\Phi}'}{\Phi}\rightarrow-\infty$, the amplitude of sub-horizon Bardeen potential perturbation quickly decays.  

Assume $x_c\notin\{0, \pm \sqrt{3}/3\}$, then, under the transformation 
\begin{equation}
    \label{transform19}
    \Phi_k = a^{-\left(6-3 x_c^2\right)}v_k,
\end{equation} we obtain the equation  
\begin{equation}
\label{Bessel19}
    \frac{d^2 v_k}{d\eta^2} +    v_k\left(k^2-\frac{9 (x_c-1) (x_c+1) \left(2 x_c^2-3\right)}{\eta ^2 \left(1-3 x_c^2\right)^2}\right)=0,
\end{equation}
with solution
\begin{equation}
    \label{sol19}
 v_k(\eta)=  C_+ \sqrt{\eta } J_{\nu}(k \eta )+C_- \sqrt{\eta
   } Y_{\nu}(k \eta ),
\end{equation}
where 
\begin{equation}\label{nu19}
 \nu= \frac{\sqrt{81 x_c^4-186 x_c^2+109}}{2-6 x_c^2},
\end{equation} and  $C_+$ and $C_-$ are complex constants depending on $k$.

 $A_{20}:\left(x_c, \lambda_c,1, \frac{\pi }{2}\right)$.   For this equilibrium point, we have a cosmological solution with an asymptotic scale factor 
 \eqref{CASE-A}. For  ${x_c}=0$ we have a de Sitter expansion with $a(t)= e^{H_0 \left(t-t_U\right)}$.  Since $\frac{{\Phi}'}{\Phi}\rightarrow \infty$, the amplitude of sub-horizon Bardeen potential perturbation quickly diverges. Introducing the transformation \eqref{transform19}, we acquire the Bessel equation \eqref{Bessel19} with solution \eqref{sol19} where the parameter $\nu$ is defined by \eqref{nu19}. 

  $A_{21}(\lambda^{*}):\left(\frac{{{\lambda^*}}}{\sqrt{6}}, {{\lambda^*}}, 1, -\frac{\pi }{2}\right)$,  with $-\sqrt{6}\leq \lambda^{*}\leq \sqrt{6}$. We have a cosmological solution for this equilibrium point with an asymptotic scale factor \eqref{A1}. Since $\frac{{\Phi}'}{\Phi}\rightarrow-\infty$, the amplitude of sub-horizon Bardeen potential perturbation quickly decays. Introducing the transformation \eqref{transform}, we acquire the Bessel equation \eqref{Bessel} with solution \eqref{sol} where the parameter $\nu$ is defined by \eqref{nu}, and $C_+$ and $C_-$ are complex constants depending on $k$.  

  $A_{22}(\lambda^{*}):\left(\frac{{{\lambda^*}}}{\sqrt{6}}, {{\lambda^*}}, 1,\frac{\pi }{2}\right)$,  with $-\sqrt{6}\leq \lambda^{*}\leq \sqrt{6}$.   We have a cosmological solution for this equilibrium point with an asymptotic scale factor \eqref{A1}. Since $\frac{{\Phi}'}{\Phi}\rightarrow \infty$, the amplitude of sub-horizon Bardeen potential perturbation quickly diverges.   Introducing the transformation \eqref{transform}, we acquire the Bessel equation \eqref{Bessel} with solution \eqref{sol} where the parameter $\nu$ is defined by \eqref{nu}, and $C_+$ and $C_-$ are complex constants depending on $k$.  

 $A_{23}:\left( -1,\lambda_c,1, -\frac{\pi }{2}\right)$ and $A_{24}:\left(1,\lambda_c, 1,-\frac{\pi }{2}\right)$ always exist. We have a cosmological solution for these lines of equilibrium points with an asymptotic scale factor 
\eqref{A2}. Since $\frac{{\Phi}'}{\Phi}\rightarrow-\infty$, the amplitude of sub-horizon Bardeen potential perturbation quickly decays.   Introducing the transformation \eqref{transform1b}, we acquire the equation \eqref{Bessel1b} with solution \eqref{sol1b} where  $C_+$ and $C_-$ are complex constants depending on $k$.  

 $A_{25}:\left(-1,\lambda_c,1,\frac{\pi }{2}\right)$ and $A_{26}:\left(1,\lambda_c,1,\frac{\pi }{2}\right)$ always exist.  We have a cosmological solution for these lines of equilibrium points with an asymptotic scale factor \eqref{A2}. Since $\frac{{\Phi}'}{\Phi}\rightarrow \infty$, the amplitude of sub-horizon Bardeen potential perturbation quickly diverges.   Introducing the transformation \eqref{transform1b}, we acquire the equation \eqref{Bessel1b} with solution \eqref{sol1b} where  $C_+$ and $C_-$ are complex constants depending on $k$.  
 
 $A_{27}:\left(-\frac{1}{\sqrt{3}}, \lambda_c,1, -\frac{\pi }{2}\right)$ and $A_{28}:\left(\frac{1}{\sqrt{3}}, \lambda_c, 1, -\frac{\pi }{2}\right)$ always exist. We have a cosmological solution for these lines of equilibrium points with an asymptotic scale factor \eqref{A3}. Since $\frac{{\Phi}'}{\Phi}\rightarrow-\infty$, the amplitude of sub-horizon Bardeen potential perturbation quickly decays. Introducing the new variable \eqref{2ptbn_bardeen3-special}, equation \eqref{ptbn_bardeen3-special} becomes \eqref{3ptbn_bardeen3-special} with solution \eqref{4ptbn_bardeen3-special} where $\epsilon =-1, \lambda=\lambda_c$, and   $C_+$ and $C_-$ are complex constants depending on $k$.  

 $A_{29}:\left(-\frac{1}{\sqrt{3}},\lambda_c ,1,\frac{\pi }{2}\right)$ and $A_{30}:\left(\frac{1}{\sqrt{3}}, \lambda_c, 1,\frac{\pi }{2}\right)$ always exist. We have a cosmological solution for these lines of equilibrium points with an asymptotic scale factor \eqref{A3}. Since $\frac{{\Phi}'}{\Phi}=\tan\left(\frac{\pi}{2}\right)\rightarrow \infty$, the amplitude of sub-horizon Bardeen potential perturbation quickly diverges. 
Introducing the new variable \eqref{2ptbn_bardeen3-special}, equation \eqref{ptbn_bardeen3-special} becomes \eqref{3ptbn_bardeen3-special} with solution \eqref{4ptbn_bardeen3-special} where $\epsilon =1, \lambda=\lambda_c$, and   $C_+$ and $C_-$ are complex constants depending on $k$.

\subsection{Extended phase space at background and perturbation levels: comoving curvature perturbation} 
  
First, note that in equation \eqref{ptbn_comov}, $\mathcal{R}_k$ is generally complex (as it came from Fourier transformation). 
So, we write $\mathcal{R}_k=F_1+iF_2$, where $F_1$ and $F_2$ are the real and imaginary parts of the $\mathcal{R}_k$, respectively. Substituting  $\mathcal{R}_k=F_1+iF_2$ in  \eqref{ptbn_comov}, we would get the same equation for both real and imaginary parts. 
Denoting by $F$ in equation \eqref{ptbn_comov},  we get
\begin{align}
& F^{\prime\prime} + \sqrt{6}\lambda\left(\frac{1-x^2}{x}\right)F^{\prime}
+ \left(\frac{k^2}{a^2H^2}\right)F = 0. \label{eq:95}
\end{align}
This equation   has the structure of equation  \eqref{pert-eq},
where 
\begin{align}
    & P=\sqrt{6}\lambda\left(\frac{1-x^2}{x}\right), \quad Q=\left(\frac{k^2}{a^2H^2}\right).
\end{align}
As before, note that \eqref{eq:95} is a two-degree equation of $f$. We can expect to use phase space type analysis for this equation.
Using \eqref{def-theta} we obtain 
\begin{align}
   \theta^{\prime} =-\sin^2\theta - P \sin\theta \cos\theta -Q\cos^2\theta.
\end{align}
Replacing the equation \eqref{EQ:(63)} in $Q$ and considering equation \eqref{ptbn_comov}, we obtain for the \textbf{Comoving curvature perturbation $\mathcal{R}$} the expression
\begin{equation}\label{theta_comov}
    \theta^{\prime} = - \sin^2\theta - \sqrt{6}\lambda\left(\frac{1-x^2}{x}\right)\sin\theta\cos\theta - Z\cos^2 \theta.
\end{equation}

The final equations for comoving curvature perturbation are given by the background equations \eqref{(eq:99)} and the perturbation equation 
\begin{align}
& \frac{d\theta}{d\bar{N}} = - \left[\sin^2\theta + \sqrt{6}\lambda\left(\frac{1-x^2}{x}\right)\sin\theta\cos\theta\right]\left(1-\bar{Z}\right) - \bar{Z}\cos^2 \theta, \label{eq100}
\end{align}
defined in the phase-space $B\times P$, modulo $n\pi, n\in\mathbb{Z}$, where  the background space is \eqref{B-space}
and the perturbation space is 
\eqref{P-space}.

\subsubsection{Sub-horizon boundary}

In  the limit $\bar{Z}=1$, \eqref{(eq:99)} and \eqref{eq100} becomes \eqref{(eq:99c)}. As before,  we have two asymptotic behaviours as $k^2 \mathcal{H}^{-2}\gg 1$, say there are two set of equilibrium points with constant $x, \lambda$ and $\theta= \pi/2 + n \pi, n=-1,0$. When $\cos^2 \theta>0$, $\theta$ is monotonically decreasing at constant $x, \lambda$. Then, the invariant set is spanned by a family of heteroclinic cycles with constant $x, \lambda$. They are denoted by       $B_{14}$  and $B_{15}$ in Tab. \ref{tab:IV}. Because of their physical importance, we have distinguished some special points from these sets of equilibrium points ($B_{16}(\lambda^*)$  to $B_{25}$).

\subsubsection{Stability analysis of the fixed points on the space $B\times P$}

\begin{table}[!ht]
    \centering
      \resizebox{\textwidth}{!}{%
    \begin{tabular}{|c|c|c|c|c|c|c|c|c|c|}
    \hline
 Label & $x$ & $\lambda$ & $\bar{Z}$ &$\theta$ & $k_1$ & $k_2$ & $k_3$ &$k_4$ & $a(t), H(t), \phi(t)$\\\hline
$B_1({\lambda^{*}})$ & $ \frac{{\lambda^{*}}}{\sqrt{6}}$ & ${\lambda^{*}}$ &$ 0$& $-\cos ^{-1}\left(-\frac{1}{\sqrt{(\lambda^{*^2}-6)^2 +1}}\right)$ & $\frac{1}{2}
   \left(\lambda^{*^2}-6\right)$ & $6-\lambda^{*^2}$ & $\lambda^{*^2}-2$ & $-{\lambda^{*}} f'({\lambda^{*}}) $& \eqref{A1},  \eqref{A1b}, \eqref{A1c}\\\hline
$B_2({\lambda^{*}})$ & $\frac{{\lambda^{*}}}{\sqrt{6}}$ & ${\lambda^{*}}$ & $0$ & $\cos ^{-1}\left(\frac{1}{\sqrt{(\lambda^{*^2}-6)^2 +1}}\right)$ & $\frac{1}{2} 
 \left(\lambda^{*^2}-6\right)$ & $6-\lambda^{*^2}$ & $\lambda^{*^2}-2$  & $-{\lambda^{*}} f'({\lambda^{*}})$ & \eqref{A1},  \eqref{A1b}, \eqref{A1c}\\\hline
$B_3({\lambda^{*}})$ &   $ \frac{{\lambda^{*}}}{\sqrt{6}}$ &$ {\lambda^{*}}$ & $0$ & $-\cos ^{-1}\left(\frac{1}{\sqrt{(\lambda^{*^2}-6)^2 +1}}\right)$ &$ \frac{1}{2} \left(\lambda^{*^2}-6\right)$ & $(\lambda^{*^2}-6)\left[\frac{4}{(\lambda^{*^2}-6)^2 +1} -1\right]$ &$ \lambda^{*^2}-2$ &
  $ -{\lambda^{*}} f'({\lambda^{*}}) $& \eqref{A1},  \eqref{A1b}, \eqref{A1c}\\\hline
$B_4({\lambda^{*}})$ & $ \frac{{\lambda^{*}}}{\sqrt{6}}$ & ${\lambda^{*}}$ & $0$ & $\cos ^{-1}\left(-\frac{1}{\sqrt{(\lambda^{*^2}-6)^2 +1}}\right)$ & $\frac{1}{2}
   \left(\lambda^{*^2}-6\right)$ &$ (\lambda^{*^2}-6)\left[\frac{4}{(\lambda^{*^2}-6)^2 +1} -1\right]$  & $\lambda^{*^2}-2$& $-{\lambda^{*}} f'({\lambda^{*}})$ & \eqref{A1},  \eqref{A1b}, \eqref{A1c}\\\hline
$B_5({\lambda^{*}})$ &  $\frac{{\lambda^{*}}}{\sqrt{6}}$ & ${\lambda^{*}}$ & $0$ &$ 0$ &$ \frac{1}{2} \left(\lambda^{*^2}-6\right)$ & $\lambda^{*^2}-6$ & $\lambda^{*^2}-2$ &
   $-{\lambda^{*}} f'({\lambda^{*}})$ & \eqref{A1},  \eqref{A1b}, \eqref{A1c}\\\hline
$B_6({\lambda^{*}})$ &  $\frac{{\lambda^{*}}}{\sqrt{6}}$ & ${\lambda^{*}}$ & $0$ & $-\pi$  & $\frac{1}{2} \left(\lambda^{*^2}-6\right) $& $\lambda^{*^2}-6$ &$ \lambda^{*^2}-2$ &
   $-{\lambda^{*}} f'({\lambda^{*}})$ & \eqref{A1},  \eqref{A1b}, \eqref{A1c}\\\hline
$B_7({\lambda^{*}})$ & $\frac{{\lambda^{*}}}{\sqrt{6}}$ & ${\lambda^{*}}$ &$ 0 $& $\pi$  & $\frac{1}{2} \left(\lambda^{*^2}-6\right)$ & $\lambda^{*^2}-6$ & $\lambda^{*^2}-2$ &
   $-{\lambda^{*}} f'({\lambda^{*}})$ & \eqref{A1},  \eqref{A1b}, \eqref{A1c}\\\hline
$B_8({\lambda^{*}})$ &  $-1$ & ${\lambda^{*}}$ &$ 0 $& $0$ & $4$ &$ 0$ & $\sqrt{6} {\lambda^{*}}+6$ & $ \sqrt{6} f'({\lambda^{*}}) $  & \eqref{A2}, \eqref{A2b}, \eqref{A2c} \\\hline
$B_9({\lambda^{*}})$ &  $-1$ & ${\lambda^{*}}$ & $0$ & $-\pi$  & $4$ & $0$ & $\sqrt{6} {\lambda^{*}}+6 $& $\sqrt{6} f'({\lambda^{*}})$  & \eqref{A2}, \eqref{A2b}, \eqref{A2c} \\\hline
$B_{10}({\lambda^{*}})$ &  $-1$ & ${\lambda^{*}}$ &$ 0$ &$ \pi$  & $4$ &$ 0$ & $\sqrt{6} {\lambda^{*}}+6$ & $\sqrt{6} f'({\lambda^{*}})$  & \eqref{A2}, \eqref{A2b}, \eqref{A2c} \\\hline
$B_{11}({\lambda^{*}})$ & $ 1$ & ${\lambda^{*}}$ & $0$ &$ 0$ &$4 $& $0$ &$ 6-\sqrt{6} {\lambda^{*}} $& $-\sqrt{6} f'({\lambda^{*}}) $ & \eqref{A2}, \eqref{A2b}, \eqref{A2c} \\\hline
$B_{12}({\lambda^{*}})$ & $1$ & ${\lambda^{*}}$ & $0$ & $-\pi$  & $4$ &$ 0$ & $6-\sqrt{6} {\lambda^{*}}$ & $-\sqrt{6} f'({\lambda^{*}})$  & \eqref{A2}, \eqref{A2b}, \eqref{A2c} \\\hline
$B_{13}({\lambda^{*}})$ & $ 1 $& ${\lambda^{*}}$ & $0$ & $\pi$  & $4$ & $0$ & $6-\sqrt{6} {\lambda^{*}}$ & $-\sqrt{6} f'({\lambda^{*}})$ & \eqref{A2}, \eqref{A2b}, \eqref{A2c}  \\\hline
$B_{14}$ &$x_c$ & $\lambda_c$  & $1$ & $-\frac{\pi }{2}$ & $0$ & $0$ & $0$ & $0$ & \eqref{CASE-A}, \eqref{CASE-Ab}, \eqref{CASE-Ac}\\\hline
$B_{15}$ &$ x_c $& $\lambda_c$  & $1$ & $\frac{\pi }{2}$ & $0$ & $0$ & $0$ & $0$ & \eqref{CASE-A}, \eqref{CASE-Ab}, \eqref{CASE-Ac}\\\hline
$B_{16}({\lambda^{*}})$ & $ \frac{{\lambda^{*}}}{\sqrt{6}}$ & ${\lambda^{*}}$ &$ 1$ & $-\frac{\pi }{2}$ & $0$ &$0$& $0$ &$0$ & \eqref{A1},  \eqref{A1b}, \eqref{A1c}\\\hline
$B_{17}({\lambda^{*}})$ & $ \frac{{\lambda^{*}}}{\sqrt{6}}$ & $ {\lambda^{*}}$ & $1$ & $\frac{\pi}{2}$ & $0$ & $0$ & $0$ & $0$ & \eqref{A1},  \eqref{A1b}, \eqref{A1c}\\\hline
$B_{18}$ & $-1$ & $\lambda_c$  & $1$ &$ -\frac{\pi }{2}$ & $0$ &$ 0$ & $0$ &$ 0$ & \eqref{A2}, \eqref{A2b}, \eqref{A2c} \\\hline
$B_{19}$ & $1 $& $\lambda_c$  & $1$ & $-\frac{\pi }{2}$ & $0$ & $0$ & $0$ &$ 0$ & \eqref{A2}, \eqref{A2b}, \eqref{A2c} \\\hline
$B_{20}$ & $-1$ & $\lambda_c$  & $1$ & $\frac{\pi }{2}$ & $0$ &$ 0$ &$ 0$ &$ 0$ & \eqref{A2}, \eqref{A2b}, \eqref{A2c} \\\hline
$B_{21}$ & $1$ & $\lambda_c$  & $1$ & $\frac{\pi }{2}$ & $0$ & $0$ &$ 0$ &$ 0$ & \eqref{A2}, \eqref{A2b}, \eqref{A2c} \\\hline
$B_{22}$ & $ -\frac{1}{\sqrt{3}}$ & $\lambda_c $ &$1$ &$ -\frac{\pi }{2}$ & $0 $& $0$ & $0$ &$ 0$ & \eqref{A3}, \eqref{A3b}, \eqref{A3c}\\\hline
$B_{23}$ & $ \frac{1}{\sqrt{3}}$ & $\lambda_c$  & $1$ & $-\frac{\pi }{2}$ &$0$ & $0$ &$ 0 $& $0$ & \eqref{A3}, \eqref{A3b}, \eqref{A3c}\\\hline
$B_{24}$ & $-\frac{1}{\sqrt{3}}$ & $\lambda_c$  & $1$ & $\frac{\pi }{2}$ & $0 $& $0$ & $0 $& $0$ & \eqref{A3}, \eqref{A3b}, \eqref{A3c}\\\hline
$B_{25}$ & $ \frac{1}{\sqrt{3}}$ & $\lambda_c $ & $1 $& $\frac{\pi }{2}$ & $0$ &$0$ & $0$ & $0$ & \eqref{A3}, \eqref{A3b}, \eqref{A3c}\\\hline
    \end{tabular}}
    \caption{Equilibrium points of system \eqref{(eq:99)} and \eqref{eq100}.}
    \label{tab:IV}
 \end{table}

In Tab.  \ref{tab:IV}, the equilibrium points of system \eqref{(eq:99)} and \eqref{eq100} are presented.

These equilibrium points and the stability conditions are summarised as follows:

 $B_{1,2}({\lambda^{*}}):\left(\frac{{\lambda^{*}}}{\sqrt{6}}, {\lambda^{*}}, 0, \mp\cos ^{-1}\left(\mp\frac{1}{\sqrt{(\lambda^*-6)^2 +1}}\right)\right)$ exist for $-\sqrt{6}\leq \lambda^{*}\leq \sqrt{6}$. They are saddles. For a perturbation $k$-mode, this corresponds  to the super-horizon limit of a cosmology with an asymptotic scale factor \eqref{A1}. 
   Since $\frac{\mathcal{R}'}{\mathcal{R}}=|\lambda^*-6|$, the amplitude of super-horizon comoving curvature perturbation is exponentially increasing.
   
    Using the procedures of section \ref{Section-6.3.2}, that is, under the transformation 
\begin{equation}
    \label{transform2}
   \mathcal{R}_k = a^{-\left(\frac{5}{2}-\frac{1}{4}{\lambda^{*^2}}\right)}v_k,
\end{equation} we obtain the equation  
\begin{equation}
\label{Bessel2}
    \frac{d^2 v_k}{d\eta^2} +    v_k\left(k^2-\frac{\left({\lambda^{*}}^2-10\right) \left(3 {\lambda^{*}}^2-10\right)}{4 \eta ^2 \left({\lambda^{*}}^2-2\right)^2}\right)=0,
\end{equation}
with solution
\begin{equation}
    \label{sol2}
 v_k(\eta)=  C_+ \sqrt{\eta } J_{\nu}(k \eta )+C_- \sqrt{\eta
   } Y_{\nu}(k \eta ),
\end{equation}
where 
\begin{equation}\label{nu2}
 \nu=  \frac{\sqrt{{\lambda^{*}} ^4-11 {\lambda^{*}} ^2+26}}{{\lambda^{*}} ^2-2},
\end{equation}
and $C_+$ and $C_-$ are complex constants depending on $k$. 

 $B_{3,4}({\lambda^{*}}):\left(\frac{{\lambda^{*}}}{\sqrt{6}},   {\lambda^{*}}, 0, \mp\cos^{-1}\left(\pm \frac{1}{\sqrt{(\lambda^*-6)^2 +1}}\right)\right)$ exist for $-\sqrt{6}\leq \lambda^{*}\leq \sqrt{6}$. They are saddles for $f'(\lambda^*)<0, -\sqrt{2}<\lambda^*<0$, or $f'(\lambda^*)>0, 0<\lambda^*<\sqrt{2}$ or $2 <\lambda^{*^2}<6$ or ${\lambda^{*}} f'({\lambda^{*}})<0$. They are non-hyperbolic otherwise. For a perturbation $k$-mode, this corresponds  to the super-horizon limit of a cosmology with an asymptotic scale factor \eqref{A1}.    
   Since $\frac{\mathcal{R}'}{\mathcal{R}}=-|\lambda^*-6|$, the amplitude of super-horizon comoving curvature perturbation is exponentially decreasing. Introducing the transformation \eqref{transform2}, we acquire the Bessel equation \eqref{Bessel2} with solution \eqref{sol2} where the parameter $\nu$ is defined by \eqref{nu2}, and $C_+$ and $C_-$ are complex constants depending on $k$. 

 $B_5({\lambda^{*}}):\left(\frac{{\lambda^{*}}}{\sqrt{6}}, {\lambda^{*}}, 0,  0\right)$, $B_6({\lambda^{*}}):\left(\frac{{\lambda^{*}}}{\sqrt{6}},{\lambda^{*}},0, -\pi\right)$ and $B_7({\lambda^{*}}):\left(\frac{{\lambda^{*}}}{\sqrt{6}}, {\lambda^{*}}, 0, \pi\right)$, exist for $-\sqrt{6}\leq \lambda^{*}\leq \sqrt{6}$. They are sinks for $f'(\lambda^*)<0, -\sqrt{2}<\lambda^*<0$, or $f'(\lambda^*)>0, 0<\lambda^*<\sqrt{2}$. They are saddles for $2 <\lambda^{*^2}<6$ or ${\lambda^{*}} f'({\lambda^{*}})<0$.   They are non-hyperbolic otherwise.  
   For a perturbation $k$-mode, this corresponds to the super-horizon limit of cosmology with an asymptotic scale factor \eqref{A1}. Since $\frac{\mathcal{R}'}{\mathcal{R}}=0$, the amplitude of super-horizon comoving curvature perturbation is frozen. Introducing the transformation \eqref{transform2}, we acquire the Bessel equation \eqref{Bessel2} with solution \eqref{sol2} where the parameter $\nu$ is defined by \eqref{nu2}, and $C_+$ and $C_-$ are complex constants depending on $k$.
  
 $B_8({\lambda^{*}}): \left(-1, {\lambda^{*}},   0 ,  0\right)$, $B_9({\lambda^{*}}):\left(-1, {\lambda^{*}}, 0, -\pi\right)$ 
and $B_{10}({\lambda^{*}}):\left(-1, {\lambda^{*}},   0,   \pi\right)$  are non-hyperbolic with a three-dimensional unstable manifold for  $\lambda^{*}>-\sqrt{6}$ and $f'({\lambda^{*}})>0$. For a perturbation $k$-mode, this corresponds to the super-horizon limit with an asymptotic scale factor  \eqref{A2}. Since $\frac{\mathcal{R}'}{\mathcal{R}}=0$, the amplitude of super-horizon comoving curvature perturbation is frozen. Under the transformation 
\begin{equation}
    \label{transform18}
     \mathcal{R}_k = a^{-1}v_k,
\end{equation} we obtain the equation  
\begin{equation}
\label{Bessel18}
    \frac{d^2 v_k}{d\eta^2} +    v_k\left(k^2+\frac{1}{2 \eta ^2}\right)=0,
\end{equation}
with solution
\begin{equation}
    \label{sol18}
 v_k(\eta)=  \sqrt{\eta } \left(C_+ J_{\frac{i}{2}}(k \eta )+C_- Y_{\frac{i}{2}}(k \eta )\right), 
\end{equation}
where  $C_+$ and $C_-$ are complex constants depending on $k$.
 
 $B_{11}({\lambda^{*}}):\left(1, {\lambda^{*}}, 0,   0\right)$,  $B_{12}({\lambda^{*}}):\left(1, {\lambda^{*}}, 0, -\pi\right)$ 
 and $B_{13}({\lambda^{*}}):\left(1 ,  {\lambda^{*}}, 0, \pi\right)$ are non-hyperbolic with a three-dimensional unstable manifold for  $\lambda^{*}<\sqrt{6}$ and $f'({\lambda^{*}})<0$. For a perturbation $k$-mode, this corresponds to the super-horizon limit of a cosmology of the form \eqref{A2}. Since $\frac{\mathcal{R}'}{\mathcal{R}}=0$, the amplitude of super-horizon comoving curvature perturbation is frozen. Introducing the transformation \eqref{transform18}, we acquire the Bessel equation \eqref{Bessel18} with solution \eqref{sol18}, and $C_+$ and $C_-$ are complex constants depending on $k$.

As we commented,  the invariant set $\bar{Z}=1$ is spanned by a family of heteroclinic cycles with constant $x, \lambda$. They are denoted by       $B_{14}$  and $B_{15}$ in Tab. \ref{tab:IV}. Because of their physical importance, we have distinguished some special points from these sets of equilibrium points ($B_{16}(\lambda^*)$  to $B_{25}$). The eigenvalues of the linearisation of system \eqref{eq100} are  $0, 0, 0, 0$ at those equilibrium points. Therefore they are non-hyperbolic.
 
 $B_{14}:(x_c,\lambda_c,1,-\frac{\pi}{2})$, with $-1\leq x_c\leq 1$. For a perturbation $k$-mode, this corresponds to the sub-horizon limit with an asymptotic scale factor  \eqref{CASE-A}. For  ${x_c}=0$ we have a de Sitter expansion with $a(t)= e^{H_0 \left(t-t_U\right)}$.  Since $\frac{\mathcal{R}'}{\mathcal{R}}\rightarrow-\infty$, the amplitude of sub-horizon comoving curvature perturbation decays quickly. 
Assume $x_c\notin\{0, \pm \sqrt{3}/3\}$, then, under the transformation 
\begin{equation}
    \label{transform14}
     \mathcal{R}_k = a^{-\frac{1}{2} \left(5-3 x_c^2\right)}v_k,
\end{equation} we obtain the equation  
\begin{equation}
\label{Bessel14}
    \frac{d^2 v_k}{d\eta^2} +    v_k\left(k^2-\frac{\left(3 x_c^2-5\right) \left(9 x_c^2-5\right)}{4 \eta ^2 \left(1-3 x_c^2\right)^2}\right)=0,
\end{equation}
with solution
\begin{equation}
    \label{sol14}
 v_k(\eta)=  C_+ \sqrt{\eta } J_{\nu}(k \eta )+C_- \sqrt{\eta
   } Y_{\nu}(k \eta ),
\end{equation}
where 
\begin{equation}\label{nu14}
 \nu=\frac{\sqrt{9 x_c^4-\frac{33 x_c^2}{2}+\frac{13}{2}}}{1-3 x_c^2},
\end{equation} and  $C_+$ and $C_-$ are complex constants depending on $k$.

   $B_{15}:(x_c,\lambda_c,1,\frac{\pi}{2})$, with $-1\leq x_c\leq 1$. For a perturbation $k$-mode, this corresponds to the sub-horizon limit with an asymptotic scale factor  \eqref{CASE-A}. For  ${x_c}=0$ we have a de Sitter expansion with $a(t)= e^{H_0 \left(t-t_U\right)}$. Since $\frac{\mathcal{R}'}{\mathcal{R}}\rightarrow\infty$, the amplitude of sub-horizon comoving curvature perturbation diverges quickly. Introducing the transformation \eqref{transform14}, we acquire the Bessel equation \eqref{Bessel14} with solution \eqref{sol14} where the parameter $\nu$ is defined by \eqref{nu14}. 
 
 $B_{16}({\lambda^{*}}):\left(\frac{{\lambda^{*}}}{\sqrt{6}}, {\lambda^{*}},   1, -\frac{\pi }{2}\right)$, with $-\sqrt{6}\leq \lambda^{*}\leq \sqrt{6}$. For a perturbation $k$-mode, this corresponds to the sub-horizon limit with an asymptotic scale factor \eqref{A1}. Since $\frac{\mathcal{R}'}{\mathcal{R}}\rightarrow-\infty$, the amplitude of sub-horizon comoving curvature perturbation decays quickly. Introducing the transformation \eqref{transform2}, we acquire the Bessel equation \eqref{Bessel2} with solution \eqref{sol2} where the parameter $\nu$ is defined by \eqref{nu2}, and $C_+$ and $C_-$ are complex constants depending on $k$.
 
$B_{17}({\lambda^{*}}):\left(\frac{{\lambda^{*}}}{\sqrt{6}},  {\lambda^{*}}, 1, \frac{\pi}{2}\right)$, with $-\sqrt{6}\leq \lambda^{*}\leq \sqrt{6}$. For a perturbation $k$-mode, this corresponds to the sub-horizon limit with an asymptotic scale factor \eqref{A1}. Since $\frac{\mathcal{R}'}{\mathcal{R}}\rightarrow\infty$, the amplitude of sub-horizon comoving curvature perturbation diverges quickly. Introducing the transformation \eqref{transform2}, we acquire the Bessel equation \eqref{Bessel2} with solution \eqref{sol2} where the parameter $\nu$ is defined by \eqref{nu2}, and $C_+$ and $C_-$ are complex constants depending on $k$.
 
  $B_{18}:(-1,\lambda_c,1,-\frac{\pi}{2})$ and  $B_{19}:(1,\lambda_c,1,-\frac{\pi}{2})$. For a perturbation $k$-mode, this corresponds to the sub-horizon limit with an asymptotic scale factor \eqref{A2}. Since $\frac{\mathcal{R}'}{\mathcal{R}}\rightarrow-\infty$, the amplitude of sub-horizon comoving curvature perturbation decays quickly. Introducing the transformation \eqref{transform18}, we acquire the Bessel equation \eqref{Bessel18} with solution \eqref{sol18}, and $C_+$ and $C_-$ are complex constants depending on $k$.
 
 $B_{20}:(-1,\lambda_c,1,\frac{\pi}{2})$ and  $B_{21}:(1,\lambda_c,1,\frac{\pi}{2})$. For a perturbation $k$-mode, this corresponds to the sub-horizon limit with an asymptotic scale factor \eqref{A2}. Since $\frac{\mathcal{R}'}{\mathcal{R}}\rightarrow\infty$, the amplitude of sub-horizon comoving curvature perturbation diverges quickly. Introducing the transformation \eqref{transform18}, we acquire the Bessel equation \eqref{Bessel18} with solution \eqref{sol18}, and $C_+$ and $C_-$ are complex constants depending on $k$.
 
 $B_{22}:(-\frac{1}{\sqrt{3}},\lambda_c,1,-\frac{\pi}{2})$ and  $B_{23}:(\frac{1}{\sqrt{3}},\lambda_c,1,-\frac{\pi}{2})$. For a perturbation $k$-mode, this corresponds to the sub-horizon limit with an asymptotic scale factor \eqref{A3}. Since $\frac{\mathcal{R}'}{\mathcal{R}}\rightarrow-\infty$, the amplitude of sub-horizon comoving curvature perturbation decays quickly. 
Introducing the new variable \eqref{2ptbn_comov3-special}, equation \eqref{ptbn_comov3-special} becomes \eqref{3ptbn_comov33-special} with solution \eqref{4ptbn_comov3-special} where  $C_+$ and $C_-$ are complex constants depending on $k$.  
 
 $B_{24}:(-\frac{1}{\sqrt{3}},\lambda_c,1,\frac{\pi}{2})$ and  $B_{25}:(\frac{1}{\sqrt{3}},\lambda_c,1,\frac{\pi}{2})$. For a perturbation $k$-mode, this corresponds to the sub-horizon limit with an asymptotic scale factor \eqref{A3}. Since $\frac{\mathcal{R}'}{\mathcal{R}}\rightarrow\infty$, the amplitude of sub-horizon comoving curvature perturbation diverges quickly. Introducing the new variable \eqref{2ptbn_comov3-special}, equation \eqref{ptbn_comov3-special} becomes \eqref{3ptbn_comov33-special} with solution \eqref{4ptbn_comov3-special} where  $C_+$ and $C_-$ are complex constants depending on $k$.

\subsection{Extended phase space at background and perturbation levels: Sasaki-Mukhanov variable}

First, note that $\varphi_{ck}$ in \eqref{eq:Uggla} is generally complex (as it came from Fourier transformation). 
So, we write $\varphi_{ck}=F_1+iF_2$, where $F_1$ and $F_2$ are the real and imaginary parts of the ${\Phi}_k$, respectively. Following the same procedures as before, the resulting equation  has the structure of the equation \eqref{pert-eq}, where 
\begin{equation}
 P=3 \left( 1-x^2\right),  \quad  Q=18\left(1-x^2\right) \left[  \frac{f}{6}+\left(x-\frac{\lambda }{\sqrt{6}}\right)^2\right] +\frac{k^2}{a^2H^2},
\end{equation}
that is the same for $F_1$ and $F_2$.

As before, note that \eqref{eq:Uggla} is a two-degree equation of $f$, so we can expect to use phase space type analysis for this equation. Using \eqref{def-theta} 
we obtain 
\begin{align}
   \theta^{\prime} =-\sin^2\theta - P \sin\theta \cos\theta -Q\cos^2\theta,
\end{align}
where the replacement of the \eqref{EQ:(63)} in $Q$ leads to
\begin{align}
     \theta^{\prime} & =-\sin^2\theta - 3\left(1-x^2\right)\sin\theta\cos\theta \nonumber \\
     & - 18\left(1-x^2\right) \left[  \frac{f}{6}+\left(x-\frac{\lambda }{\sqrt{6}}\right)^2\right]\cos^2\theta - Z
\cos^2\theta,
\end{align}
with $f\equiv \lambda^2(\Gamma-1)$. For the exponential potential, $f\equiv 0$, and with the re-definitions  $\left(\frac{\lambda }{\sqrt{6}}, x\right)\mapsto \left(\lambda, \Sigma_\varphi\right)$ we recover Eq. (30) of \cite{Alho:2020cdg}.

For the scalar field perturbation in the uniform curvature gauge, the evolution of background quantities and perturbations leads to a dynamical system given by 
the background equations \eqref{(eq:99)} and  the perturbation equation 

\begin{align}
& \frac{d\theta}{d\bar{N}} =- \Bigg[ \sin^2\theta + 3\left(1-x^2\right)\sin\theta\cos\theta \nonumber \\
& + 18\left(1-x^2\right) \left(  \frac{f}{6}+\left(x-\frac{\lambda }{\sqrt{6}}\right)^2 \right)\cos^2\theta\Bigg]\left(1-\bar{Z}\right)   -\bar{Z} \cos^2\theta,\label{(eq:103)}
\end{align}

defined in the phase-space $B\times P$, modulo $n\pi, n\in\mathbb{Z}$, where  the background space is \eqref{B-space}
and the perturbation space is 
\eqref{P-space}.

\subsubsection{Sub-horizon boundary}

Recall that the limit $k^2 \mathcal{H}^{-2}\gg 1$ corresponds to the short wavelength or sub-horizon boundary. It is related to the limit $\bar{Z}=1$.  
In this  limit \eqref{(eq:99)} and \eqref{(eq:103)} becomes \eqref{(eq:99c)}. As before,  we have two asymptotic behaviours as $k^2 \mathcal{H}^{-2}\gg 1$, say there are two set of equilibrium points with constant $x, \lambda$ and $\theta= \pi/2 + n \pi, n=-1,0$. When $\cos^2 \theta>0$, $\theta$ is monotonically decreasing at constant $x, \lambda$. Then, the invariant set is spanned by a family of heteroclinic cycles with constant $x, \lambda$. They are denoted by       $C_{14}$  and $C_{15}$ in Tab. \ref{tab:V}. Because of their physical importance, we have distinguished some special points from these sets of equilibrium points ($C_{16}(\lambda^*)$  to $C_{21}$).  

\subsubsection{Stability analysis of the fixed points on the space $B\times P$}

\begin{table}[!ht]
    \centering
      \resizebox{\textwidth}{!}{%
    \begin{tabular}{|c|c|c|c|c|c|c|c|c|c|}
    \hline
 Label & $x$ & $\lambda$ & $\bar{Z}$ &$\theta$ & $k_1$ & $k_2$ & $k_3$ &$k_4$ & $a(t), H(t), \phi(t)$\\\hline
$C_{1}({\lambda^{*}})$&$ \frac{\lambda ^*}{\sqrt{6}}$ & $\lambda ^* $&$ 0$ &$ -\cos ^{-1}\left(-\frac{2}{\sqrt{\lambda ^{*^4}-12 \lambda ^{*^2}+40}}\right)$ &$ \frac{1}{2}
   \left(\lambda ^{*^2}-6\right)$ & $\lambda ^{*^2}-2 $& $3-\frac{\lambda ^{*^2}}{2}$ & $-\lambda ^* f'\left(\lambda ^*\right) $& \eqref{A1},  \eqref{A1b}, \eqref{A1c}\\\hline
$C_{2}({\lambda^{*}})$ &$ \frac{\lambda ^*}{\sqrt{6}}$ &$ \lambda ^*$ & $0$ &$ \cos ^{-1}\left(\frac{2}{\sqrt{\lambda ^{*^4}-12 \lambda ^{*^2}+40}}\right)$ & $\frac{1}{2} \left(\lambda ^{*^2}-6\right)$ & $\lambda ^{*^2}-2$ & $3-\frac{\lambda ^{*^2}}{2} $& $-\lambda ^* f'\left(\lambda ^*\right)$ & \eqref{A1},  \eqref{A1b}, \eqref{A1c}\\\hline
$C_{3}({\lambda^{*}})$ & $\frac{\lambda ^*}{\sqrt{6}}$ & $\lambda ^*$ & $0 $&$ -\cos ^{-1}\left(\frac{2}{\sqrt{\lambda ^{*^4}-12 \lambda ^{*^2}+40}}\right)$ &$ \frac{1}{2} \left(\lambda ^{*^2}-6\right)$ & $\lambda ^{*^2}-2$ & $-\frac{1}{2} \lambda ^{*^2}+\frac{8 \left(\lambda ^{*^2}-6\right)}{\lambda ^{*^4}-12 \lambda ^{*^2}+40}+3 $& $-\lambda ^* f'\left(\lambda ^*\right)$& \eqref{A1},  \eqref{A1b}, \eqref{A1c} \\\hline
$C_{4}({\lambda^{*}})$& $\frac{\lambda ^*}{\sqrt{6}}$ &$\lambda ^* $& $0$ & $\cos ^{-1}\left(-\frac{2}{\sqrt{\lambda ^{*^4}-12 \lambda ^{*^2}+40}}\right)$ &$ \frac{1}{2} \left(\lambda ^{*^2}-6\right)$ & $\lambda ^{*^2}-2$ & $-\frac{1}{2} \lambda ^{*^2}+\frac{8 \left(\lambda ^{*^2}-6\right)}{\lambda ^{*^4}-12 \lambda ^{*^2}+40}+3$ &$ -\lambda ^* f'\left(\lambda ^*\right)$ & \eqref{A1},  \eqref{A1b}, \eqref{A1c}\\\hline
$C_{5}({\lambda^{*}})$ &$ \frac{\lambda ^*}{\sqrt{6}}$ & $\lambda ^*$& $0$ & $0$ & $\frac{1}{2} \left(\lambda ^{*^2}-6\right)$ & $\frac{1}{2} \left(\lambda ^{*^2}-6\right)$ &$ \lambda ^{*^2}-2$ & $-\lambda ^* f'\left(\lambda ^*\right) $& \eqref{A1},  \eqref{A1b}, \eqref{A1c}\\\hline
$C_{6}({\lambda^{*}})$ & $\frac{\lambda ^*}{\sqrt{6}}$ & $\lambda ^*$ & $0 $& $-\pi$  & $\frac{1}{2} \left(\lambda ^{*^2}-6\right)$ &$ \frac{1}{2} \left(\lambda ^{*^2}-6\right)$ & $\lambda ^{*^2}-2$ &$ -\lambda ^* f'\left(\lambda ^*\right)$& \eqref{A1},  \eqref{A1b}, \eqref{A1c} \\\hline 
$C_{7}({\lambda^{*}})$ &$ \frac{\lambda ^*}{\sqrt{6}}$ & $\lambda ^*$ &$ 0$ & $\pi$  &$ \frac{1}{2} \left(\lambda ^{*^2}-6\right) $& $\frac{1}{2} \left(\lambda ^{*^2}-6\right)$ & $\lambda ^{*^2}-2$ &$ -\lambda ^* f'\left(\lambda ^*\right)$ & \eqref{A1},  \eqref{A1b}, \eqref{A1c}\\\hline
$C_{8}({\lambda^{*}})$& $-1 $& $\lambda ^* $& $0 $&$ 0$ &$ 4$ &$ 0$ & $\sqrt{6} \lambda ^*+6$ &$ \sqrt{6} f'\left(\lambda ^*\right)$ & \eqref{A2}, \eqref{A2b}, \eqref{A2c} \\\hline
$C_{9}({\lambda^{*}})$ & $-1$ & $\lambda ^*$& $0$ & $-\pi $ & $4$ &$0$ &$ \sqrt{6} \lambda ^*+6$ & $\sqrt{6} f'\left(\lambda ^*\right)$& \eqref{A2}, \eqref{A2b}, \eqref{A2c} \\\hline
$C_{10}({\lambda^{*}}) $& $-1$ &$ \lambda ^*$ &$ 0$ & $\pi$  &$ 4$ & $0$ & $\sqrt{6} \lambda ^*+6 $& $\sqrt{6} f'\left(\lambda ^*\right)$& \eqref{A2}, \eqref{A2b}, \eqref{A2c}  \\\hline
$C_{11}({\lambda^{*}})$ &$ 1$ & $\lambda ^*$ & $0$ & $0$ & $4$ &$ 0$ &$ 6-\sqrt{6} \lambda ^*$ & $-\sqrt{6} f'\left(\lambda ^*\right) $& \eqref{A2}, \eqref{A2b}, \eqref{A2c} \\\hline
$C_{12}({\lambda^{*}})$ & $1$ & $\lambda ^* $& $0$ & $-\pi$  &$4$ & $0 $&$ 6-\sqrt{6} \lambda ^*$ & $-\sqrt{6} f'\left(\lambda ^*\right)$ & \eqref{A2}, \eqref{A2b}, \eqref{A2c} \\\hline
$C_{13}({\lambda^{*}}) $& $1$ & $\lambda ^*$ & $0$ &$ \pi$  &$ 4$ &$ 0 $&$ 6-\sqrt{6} \lambda ^*$ &$ -\sqrt{6} f'\left(\lambda ^*\right)$ & \eqref{A2}, \eqref{A2b}, \eqref{A2c} \\\hline
$C_{14}$ & $x_c$ &$ \lambda_c$ & $1$ & $-\frac{\pi }{2}$ & $0$ & $0$ & $0$ & $0$ & \eqref{CASE-A}, \eqref{CASE-Ab}, \eqref{CASE-Ac} \\\hline
$C_{15}$ & $x_c$ & $\lambda_c$ & $1$ &$ \frac{\pi }{2}$ & $0$ &$ 0$ & $0$ &$ 0$ & \eqref{CASE-A}, \eqref{CASE-Ab}, \eqref{CASE-Ac} \\\hline
$C_{16} $& $0$ & $0$ & $1$ & $-\frac{\pi }{2}$ &$ 0$ & $0$ & $0$ & $0$ & $ e^{H_0\left(t-t_U\right)}, H_0, \phi_0$\\\hline
$C_{17}$& $0$ &$ 0$ & $1 $& $\frac{\pi }{2}$ & $0$ & $0$ &$ 0$ &$ 0$ & $ e^{H_0\left(t-t_U\right)}, H_0, \phi_0$ \\\hline
$C_{18}$ & $-\frac{1}{\sqrt{3}}$ & $-\sqrt{2}$ &$ 1$ & $-\frac{\pi }{2}$ & $0$ &$ 0$& $0$ & $0$ & \eqref{A3}, \eqref{A3b}, \eqref{A3c}\\\hline
$C_{19}$ &$ -\frac{1}{\sqrt{3}} $& $-\sqrt{2}$ &$ 1$ & $\frac{\pi }{2}$ & $0$ & $0$ &$ 0$ & $0$ & \eqref{A3}, \eqref{A3b}, \eqref{A3c}\\\hline
$C_{20}$ & $\frac{1}{\sqrt{3}} $& $\sqrt{2}$ & $1$ & $-\frac{\pi }{2}$ & $0$ & $0$ &$ 0$ &$ 0$ & \eqref{A3}, \eqref{A3b}, \eqref{A3c}\\\hline
$C_{21}$ &$ \frac{1}{\sqrt{3}}$ & $\sqrt{2}$ &$ 1$ & $\frac{\pi }{2}$ & $0$ & $0$ & $0 $& $0$ & \eqref{A3}, \eqref{A3b}, \eqref{A3c}\\\hline
    \end{tabular}}
    \caption{Equilibrium points of system \eqref{(eq:99)} and \eqref{(eq:103)}.}
    \label{tab:V}
\end{table}
In Tab. \ref{tab:V}, the equilibrium points of system \eqref{(eq:99)} and \eqref{(eq:103)} are presented.

These equilibrium points and the stability conditions are summarised as follows:

 $C_{1,2}({\lambda^{*}}):\left(\frac{\lambda ^*}{\sqrt{6}}, \lambda ^*, 0, \mp\cos ^{-1}\left(\mp\frac{2}{\sqrt{\lambda ^{*^4}-12 \lambda ^{*^2}+40}}\right)\right)$ exist for $-\sqrt{6}\leq \lambda^{*}\leq \sqrt{6}$, and are saddles. The scale factor has an asymptotic form \eqref{A1}. Since $\frac{{\phi_{ck}}'}{\phi_{ck}}= \frac{1}{2} |\lambda ^{*^2}-6| $, the amplitude of super-horizon Sasaki-Mukhanov variable is exponentially increasing.
Using the procedures of section \ref{Section-6.3.3}, that is, under the transformation 
\begin{equation}
    \label{transform3}
  \varphi_{ck} =  a^{-1}v_k,
\end{equation} we obtain the equation  
\begin{equation}
\label{Bessel3}
    \frac{d^2 v_k}{d\eta^2} +    v_k\left(k^2 + \eta ^{-2} \left| 1-\frac{{\lambda ^{*2}}}{2}\right|^{-1}\right)=0,
\end{equation}
with solution
\begin{equation}
    \label{sol3}
 v_k(\eta)=  C_+ \sqrt{\eta } J_{\nu}(k \eta )+C_- \sqrt{\eta
   } Y_{\nu}(k \eta ),
\end{equation}
where 
\begin{equation}\label{nu3}
 \nu= \frac{1}{2} \sqrt{1-4 \left| 1-\frac{\lambda ^{*2}}{2}\right|^{-1}}.
\end{equation}

 $C_{3,4}({\lambda^{*}}):\left(\frac{\lambda ^*}{\sqrt{6}}, \lambda ^*, 0, \mp\cos ^{-1}\left(\pm \frac{2}{\sqrt{\lambda ^{*^4}-12 \lambda ^{*^2}+40}}\right)\right)$ exist for $-\sqrt{6}\leq \lambda^{*}\leq \sqrt{6}$. They are saddles for $f'(\lambda^*)<0, -\sqrt{2}<\lambda^*<0$, or $f'(\lambda^*)>0, 0<\lambda^*<\sqrt{2}$ or $2 <\lambda^{*^2}<6$ or ${\lambda^{*}} f'({\lambda^{*}})<0$. They are non-hyperbolic otherwise. The scale factor has an asymptotic form \eqref{A1}. Since $\frac{{\phi_{ck}}'}{\phi_{ck}}=-\frac{1}{2} |\lambda ^{*^2}-6| $, the amplitude of super-horizon Sasaki-Mukhanov variable is exponentially decreasing. Introducing the transformation \eqref{transform3}, we acquire the Bessel equation \eqref{Bessel3} with solution \eqref{sol3} where the parameter $\nu$ is defined by \eqref{nu3}, and $C_+$ and $C_-$ are complex constants depending on $k$. 

 $C_{5}({\lambda^{*}}): \left(\frac{\lambda ^*}{\sqrt{6}}, \lambda ^*, 0, 0\right)$, $C_{6}({\lambda^{*}}):\left(\frac{\lambda ^*}{\sqrt{6}}, \lambda ^*, 0,-\pi\right)$ and $C_{7}({\lambda^{*}}):\left(\frac{\lambda ^*}{\sqrt{6}}, \lambda ^*,  0, \pi\right)$, exist for $-\sqrt{6}\leq \lambda^{*}\leq \sqrt{6}$. They are sinks for $f'(\lambda^*)<0, -\sqrt{2}<\lambda^*<0$, or $f'(\lambda^*)>0, 0<\lambda^*<\sqrt{2}$. They are saddles for $2 <\lambda^{*^2}<6$ or ${\lambda^{*}} f'({\lambda^{*}})<0$. They are non-hyperbolic otherwise. The scale factor has an asymptotic form \eqref{A1}. Since $\frac{{\phi_{ck}}'}{\phi_{ck}}=0$, the amplitude of super-horizon Sasaki-Mukhanov variable is frozen. Introducing the transformation \eqref{transform3}, we acquire the Bessel equation \eqref{Bessel3} with solution \eqref{sol3} where the parameter $\nu$ is defined by \eqref{nu3}, and $C_+$ and $C_-$ are complex constants depending on $k$. 

 $C_{8}({\lambda^{*}}):\left(-1, \lambda ^*, 0, 0\right)$, $C_{9}({\lambda^{*}}):\left(-1, \lambda ^*, 0, -\pi\right)$ and $C_{10}({\lambda^{*}}):\left(-1, \lambda ^*, 0, \pi\right)$ are non-hyperbolic with a three-dimensional unstable manifold for  $\lambda^{*}>-\sqrt{6}$ and $f'({\lambda^{*}})>0$. The scale factor has an asymptotic form   \eqref{A2}. Since $\frac{{\phi_{ck}}'}{\phi_{ck}}=0$, the amplitude of super-horizon comoving curvature perturbation is frozen. 

 Using the procedures of section \ref{Section-6.3.3}, that is, under the transformation \eqref{transform3}, we obtain the equation  
\begin{equation}
\label{Bessel3b}
    \frac{d^2 v_k}{d\eta^2} +    v_k\left(k^2 + \frac{1}{2 \eta ^2}\right)=0,
\end{equation}
with solution
\begin{equation}
    \label{sol3b}
 v_k(\eta)=  C_+ \sqrt{\eta } J_{\frac{i}{2}}(k \eta )+C_- \sqrt{\eta } Y_{\frac{i}{2}}(k \eta),
\end{equation}
and $C_+$ and $C_-$ are complex constants depending on $k$.
  
 $C_{11}({\lambda^{*}}):\left(1, \lambda ^*, 0, 0\right)$, $C_{12}({\lambda^{*}}):\left(1, \lambda ^* ,0, -\pi\right)$ and $C_{13}({\lambda^{*}}):\left(1, \lambda ^*, 0, \pi\right)$ are non-hyperbolic with a three-dimensional unstable manifold for  $\lambda^{*}<\sqrt{6}$ and $f'({\lambda^{*}})<0$.  The scale factor has an asymptotic form  \eqref{A2}. Since $\frac{{\phi_{ck}}'}{\phi_{ck}}=0$, the amplitude of super-horizon Sasaki-Mukhanov variable is frozen.  Introducing the transformation \eqref{transform3}, we acquire the Bessel equation \eqref{Bessel3b} with solution \eqref{sol3b}, and $C_+$ and $C_-$ are complex constants depending on $k$.

As we commented,  the invariant set $\bar{Z}=1$ is spanned by a family of heteroclinic cycles with constant $x, \lambda$. They are denoted by       $C_{14}$  and $C_{15}$ in Tab. \ref{tab:V}. Because of their physical importance, we have distinguished some special points from these sets of equilibrium points ($C_{16}(\lambda^*)$  to $C_{21}$). The eigenvalues of the linearisation of system \eqref{(eq:103)} are  $0 0, 0, 0$ at those equilibrium points. Therefore they are non-hyperbolic.

$C_{14}({\lambda^{*}}):\left(x_c, \lambda ^*, 1, -\frac{\pi }{2}\right)$, with $-1\leq x_c\leq 1$.   The scale factor has the asymptotic form  \eqref{CASE-A}. For  ${x_c}=0$ we have a de Sitter expansion with $a(t)= e^{H_0 \left(t-t_U\right)}$. Since $\frac{{\phi_{ck}}'}{\phi_{ck}}\rightarrow-\infty$, the amplitude of sub-horizon Sasaki-Mukhanov variable quickly decays. 
 Introducing the transformation \eqref{transform3}, we obtain the Bessel equation 
\begin{align}\label{Bessel3c}
 \frac{d^2 v_k}{d\eta^2}+    v_k  \left(k^2+\eta ^{-2} |3 x_c^2-1|^{-1}\right)=0,
 \end{align} where $x_c \neq0$, with the solution 
\begin{align}\label{sol3c} 
  v_k(\eta )= C_+ \sqrt{\eta } J_{\nu}(k \eta )+C_- \sqrt{\eta } Y_{\nu}(k \eta ),
\end{align}
where 
 \begin{equation}\label{nu3c}
   \nu=\frac{1}{2} \sqrt{1- {4}{|3 x_c^2-1|^{-1}}}.
 \end{equation}
 $C_+$ and $C_-$ are complex constants depending on $k$. 

 $C_{15}({\lambda^{*}}):\left(x_c, \lambda ^*, 1, \frac{\pi }{2}\right)$, with $-1\leq x_c\leq 1$. The scale factor has the asymptotic form  \eqref{CASE-A}. For  ${x_c}=0$ we have a de Sitter expansion with $a(t)= e^{H_0 \left(t-t_U\right)}$.  Since $\frac{{\phi_{ck}}'}{\phi_{ck}}\rightarrow\infty$, the amplitude of sub-horizon Sasaki-Mukhanov variable quickly diverges.   Introducing the transformation \eqref{transform3}, we acquire the Bessel equation \eqref{Bessel3c} with solution \eqref{sol3c} where the parameter $\nu$ is defined by \eqref{nu3c}. 

 $C_{16}:\left(0, 0, 1, -\frac{\pi }{2}\right)$. The scale factor has the asymptotic form $a(t)= e^{H_0 \left(t-t_U\right)}$, which corresponds to de Sitter expansion. Since $\frac{{\phi_{ck}}'}{\phi_{ck}}\rightarrow-\infty$, the amplitude of sub-horizon Sasaki-Mukhanov variable quickly decays. 

 $C_{17}:\left(0, 0, 1, \frac{\pi }{2}\right)$. The scale factor has the asymptotic form $a(t)= e^{H_0 \left(t-t_U\right)}$, which corresponds to de Sitter expansion. Since $\frac{{\phi_{ck}}'}{\phi_{ck}}\rightarrow\infty$, the amplitude of sub-horizon Sasaki-Mukhanov variable quickly diverges. 

 $C_{18}:\left(-\frac{1}{\sqrt{3}}, -\sqrt{2}, 1, -\frac{\pi }{2}\right)$ with $f(-\sqrt{2})=0$. The scale factor has an asymptotic form  \eqref{A3}. Since $\frac{{\phi_{ck}}'}{\phi_{ck}}\rightarrow-\infty$, the amplitude of sub-horizon Sasaki-Mukhanov variable quickly decays. 
Introducing the new variable \eqref{2eq:Uggla-special}, equation \eqref{eq:Uggla-special} becomes \eqref{3eq:Ugglaspecial} with solution \eqref{4eq:Uggla-special} where  $C_+$ and $C_-$ are complex constants depending on $k$.  

 $C_{19}:\left( -\frac{1}{\sqrt{3}},-\sqrt{2}, 1, \frac{\pi }{2}\right)$  with $f(-\sqrt{2})=0$. The scale factor has an asymptotic form   \eqref{A3}. Since $\frac{{\phi_{ck}}'}{\phi_{ck}}\rightarrow\infty$, the amplitude of sub-horizon Sasaki-Mukhanov variable quickly diverges.  
Introducing the new variable \eqref{2eq:Uggla-special}, equation \eqref{eq:Uggla-special} becomes \eqref{3eq:Ugglaspecial} with solution \eqref{4eq:Uggla-special} where  $C_+$ and $C_-$ are complex constants depending on $k$.  

 $C_{20}:\left(\frac{1}{\sqrt{3}},\sqrt{2}, 1, -\frac{\pi }{2}\right)$  with $f(\sqrt{2})=0$. The scale factor has an asymptotic form  \eqref{A3}. Since $\frac{{\phi_{ck}}'}{\phi_{ck}}\rightarrow-\infty$, the amplitude of sub-horizon Sasaki-Mukhanov variable quickly decays. 
Introducing the new variable \eqref{2eq:Uggla-special}, equation \eqref{eq:Uggla-special} becomes \eqref{3eq:Ugglaspecial} with solution \eqref{4eq:Uggla-special} where  $C_+$ and $C_-$ are complex constants depending on $k$.   

 $C_{21}:\left(\frac{1}{\sqrt{3}}, \sqrt{2}, 1, \frac{\pi }{2}\right)$  with $f(\sqrt{2})=0$. The scale factor has an asymptotic form   \eqref{A3}. Since $\frac{{\phi_{ck}}'}{\phi_{ck}}\rightarrow\infty$, the amplitude of sub-horizon Sasaki-Mukhanov variable quickly diverges. 
Introducing the new variable \eqref{2eq:Uggla-special}, equation \eqref{eq:Uggla-special} becomes \eqref{3eq:Ugglaspecial} with solution \eqref{4eq:Uggla-special} where  $C_+$ and $C_-$ are complex constants depending on $k$.

\section{Dynamical system analysis of matter perturbations on top of equilibrium points}
\label{sect:new}
To complement our analysis, we investigate cosmological perturbations with two matter components, e.g. a perfect fluid and a scalar field. A widespread practice in literature concentrates on a particular cosmological epoch when only one matter component is dominant. In that sense, even though not generic, our subsequent analysis is still relevant when the Universe is a scalar field dominated, e.g. during the early inflationary epoch or the late-time acceleration.

In a general non-interacting scenario, which includes dust matter and dynamical dark energy, the scalar perturbations in the Newtonian gauge are determined by the equations \cite{Ma:1995ey}: 

\begin{subequations}
\begin{eqnarray}
&&\dot{\delta}_m+\frac{\theta_m}{a}=0,\;  \label{eq:line1} \\
&&\dot{\delta}_\phi+(1+{{w_\phi})\frac{\theta_\phi}{a}+3H(c_{\mathrm{eff}}^{2}-w_\phi)%
\delta_\phi=0,\;}  \label{eq:line2} \\
&&\dot{\theta}_m+H\theta_m-\frac{k^{2}\Phi }{a}=0,\;  \label{eq:line3} \\
&&\dot{\theta}_\phi+H\theta_\phi-\frac{k^{2}c_{\mathrm{eff}}^{2}\delta_\phi}{(1+%
{{w_\phi})a}}-\frac{k^{2}\Phi }{a}=0.\;  \label{eq:line4}
\end{eqnarray}%
where $k$ is the wavenumber of Fourier modes, and $\Phi$ is 
the scalar metric perturbation assuming zero anisotropic stress, and the dot means derivative with respect to time. Additionally, $\delta_{i} \equiv \delta \rho_i/\rho_i$, $i\in\{m, \phi\}$ are the densities perturbations and $\theta_i$,  $i\in\{m, \phi\}$ are the velocity perturbations \cite{Ma:1995ey}. Furthermore, $c_{\mathrm{eff}}^{2}$ is the effective sound speed of the dark energy perturbations (the corresponding quantity for matter is zero in the dust case), which determines the amount of
dark-energy clustering. Note that the above equations can be simplified by considering the Poisson equation,
which in sub-horizon scales becomes \cite{Ma:1995ey}:

\end{subequations}
\begin{equation}
-\frac{k^{2}}{a^{2}}\Phi =\frac{3}{2}H^{2}\left[\Omega_m\delta_{m}+\left(1+3c_{\mathrm{%
eff}}^{2}\right)\Omega_\phi\delta_\phi\right].  \label{eq:poisson}
\end{equation}
In the above equations, we have introduced the density parameters 
\begin{equation}
\Omega_i=\frac{\rho_i}{3 H^2}.
\end{equation}

The amount of DE clustering depends on the magnitude of its effective sound
speed $c_{\mathrm{eff}}^{2}$ and for $c_{\mathrm{eff}}^{2}=0$ DE clusters in
a similar manner to dark matter. However, due to the presence of the DE
pressure, one may expect that the amplitude of the DE perturbations is
relatively low with respect to that of dark matter. Notice, that bellow we
set $c_{\mathrm{eff}}^{2}=0$.

In the current work, we treat DE as a perfect fluid  which implies that the
effective sound speed coincides with the adiabatic sound speed 
\begin{equation}
c_{\mathrm{a}}^{2}={{w_\phi}-\frac{a d{w_\phi}/da}{3(1+{w_\phi})}.\;}
\label{eq:c_a}
\end{equation} 

Now eliminating $\theta $ from the system of equations \eqref{eq:line1}, \eqref{eq:line2}, \eqref{eq:line3} and \eqref{eq:line4}, and using $\frac{d}{dt}=aH \frac{d}{da}$, $\frac{d}{dt}=H\frac{d}{d\ln a},\frac{d}{da}=\frac{1}{a}\frac{d}{d\ln a}$, $\frac{d^{2}}{da^{2}}=\frac{d}{da}\left(\frac{1}{a}\frac{d}{d\ln a}\right)=-\frac{1}{a^{2}}\frac{d}{d\ln a}+\mathrm{\;}\frac{1}{a^{2}}\frac{d^{2}}{d\ln a^{2}}=
\frac{1}{a^{2}}\left(\frac{d^{2}}{d\ln a^{2}}-\frac{d}{d\ln a}\right)$, we obtain after some calculations the following second order
differential equations which describe the evolution of matter and DE
perturbations respectively: 
\begin{subequations}
\begin{eqnarray}
&&\delta_m^{\prime \prime }+(A_{m}-1)\delta_m^{\prime }+B_{m}\delta_m=\frac{3%
}{2}(\Omega_m\delta_{\mathrm{m }}+\Omega_\phi\delta_\phi),  \label{perteq1} \\
&& \delta_\phi^{\prime \prime } +(A_\phi-1)\delta_\phi^{\prime}+B_\phi\delta_\phi=%
\frac{3}{2}(1+{{w_\phi} )(\Omega_m\delta _m+\Omega_\phi\delta_\phi)},
\label{perteq2}
\end{eqnarray}
where the coefficients are 
\end{subequations}
\begin{eqnarray}
A_m &=&\frac{3}{2}(1-\Omega_\phi{w_\phi}), \;  
B_m =0, \notag \\
A_\phi &=& -3{w_\phi}-\frac{{{w_\phi^{\prime }}}}{1+{{w_\phi}}}+\frac{3}{2}%
(1-\Omega_\phi{{w_\phi})}, \; 
B_\phi = -{{w_\phi^{\prime }}+\frac{{w_\phi^{\prime }}{w_\phi}}{1+{w_\phi}}-%
\frac{1}{2}{w_\phi}(1-3\Omega_\phi{w_\phi})},
\end{eqnarray}
where the prime means derivative with respect to $\ln a$.

\subsection{Application to  $\Lambda$CDM}

Let us first examine the simple $\Lambda$CDM example. In this case, and
assuming dust matter, the background equations are 
\begin{align}
& H^{2}=\frac{1}{3}(\rho_m+\Lambda), \\
& {H}'=-\frac{\rho_m}{2H}, \\
& {\rho}_m'+3\rho_m=0,  
\end{align}%
and the equation for the matter perturbations is
\begin{eqnarray}
&&\delta_m^{\prime \prime }+ \left(2-\frac{3}{2} \Omega_m \right)\delta_m^{\prime }=\frac{3%
}{2} \Omega_m\delta_{\mathrm{m }},  \label{Aperteq1} 
\end{eqnarray}
where the prime means derivative with respect to $N=\ln a$. 
Using the above equations, and defining the the ratio $U_m=\frac{\delta_m^{\prime }(N)}{\delta_m}$, we obtain the autonomous system:
\begin{subequations}
\label{LCDM-perts}
\begin{align}
&\Omega_m'= 3 (\Omega_{m}-1)\Omega_{m}, \quad
U_m'= \frac{3}{2}(U_{m}+1) \Omega_{m}-U_m (U_m+2). 
\end{align}
\end{subequations}
The system is integrable, depending on $N$ and  $\Omega_{m0}$ and $U_{m0}$ are the values of $\Omega_m$ and $U_m$ today ($N=0, a=1$).

 As we will discuss in more detail later, one can think of $U_{m}$ as the phase of the matter
perturbation. If $U_{m}>0$ during the evolution, it follows by
definition that the perturbations $\delta_{m}$ are growing with time
(since $\delta_{m}>0$ and $\delta'_{m}>0$ or $\delta_{m}<0$ and $%
\delta_{m}'<0$), while if $U_{m}<0$ at that time, the
perturbation is decaying. 
Now, we can analyze the stability of the equilibrium points in the plane $(\Omega_m, U_m)$. In this extended phase space, we can see both the background equations' stability and the perturbations' stability.  

There are identified the equilibrium points of the system \eqref{LCDM-perts} are:
\begin{enumerate}
\item $P_1: (\Omega_m, U_m)=(0, -2)$. The phase of the perturbations $U_m$ is negative; therefore, the perturbation $\delta_m$ is decaying with time. The eigenvalues of the linear matrix evaluated at the equilibrium point are $\{-3,2\}$; that is, the equilibrium point is a saddle. That corresponds to the Universe dominated by the cosmological constant, with matter perturbations scaling as $\delta_m\propto e^{-2 N}=a^{-2}$ when the equilibrium point is approached along the stable direction.

\item $P_2: (\Omega_m, U_m)=\left(1, -\frac{3}{2}\right)$. The phase of the perturbations $U_m$ is negative; therefore, the perturbation $\delta_m$ is decaying with time. 
 The eigenvalues of the linear matrix evaluated at the equilibrium point are $\left\{3,\frac{5}{2}\right\}$. That is, the equilibrium point is a source. That corresponds to the matter-dominated Universe, with matter perturbations scaling as $\delta_m\propto e^{-\frac{3}{2} N}=a^{-\frac{3}{2}}$ as $N\rightarrow -\infty$ ($a \rightarrow 0$).  

\item $P_3: (\Omega_m, U_m)=(0, 0)$.  The phase of the perturbations $U_m$  is zero. Therefore, the perturbation $\delta_m$ remains constant. The eigenvalues of the linear matrix evaluated at the equilibrium point are $\{-3,-2\}$. Then, this equilibrium point is a sink. That corresponds to the Universe dominated by the cosmological constant with $\delta_m=\text{const.}$ as $N\rightarrow +\infty$ ($a \rightarrow +\infty$).   It is stable in the extended phase space.

\item $P_4:  (\Omega_m, U_m)=(1, 1)$. The phase of the perturbations $U_m$ is positive; therefore, the perturbation $\delta_m$ is growing with time. The eigenvalues of the linear matrix evaluated at the equilibrium point are $\left\{3,-\frac{5}{2}\right\}$. Then, it is a saddle. That corresponds to a matter-dominated universe, with matter perturbations scaling as $\delta_m\propto e^{-\frac{5}{2} N}=a^{-\frac{5}{2}}$ when the equilibrium point is approached along the stable direction.

\end{enumerate}

\begin{figure*}[t!]
    \centering
    \begin{subfigure}{(a)}
        \centering
        \includegraphics[scale=0.7]{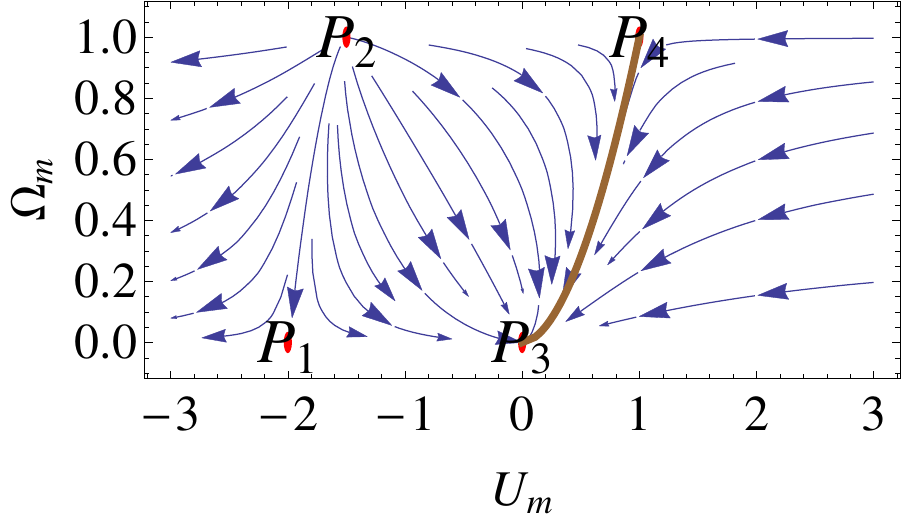}
				           \end{subfigure}%
    ~ 
    \begin{subfigure}{(b)}
        \centering
        \includegraphics[scale=0.7]{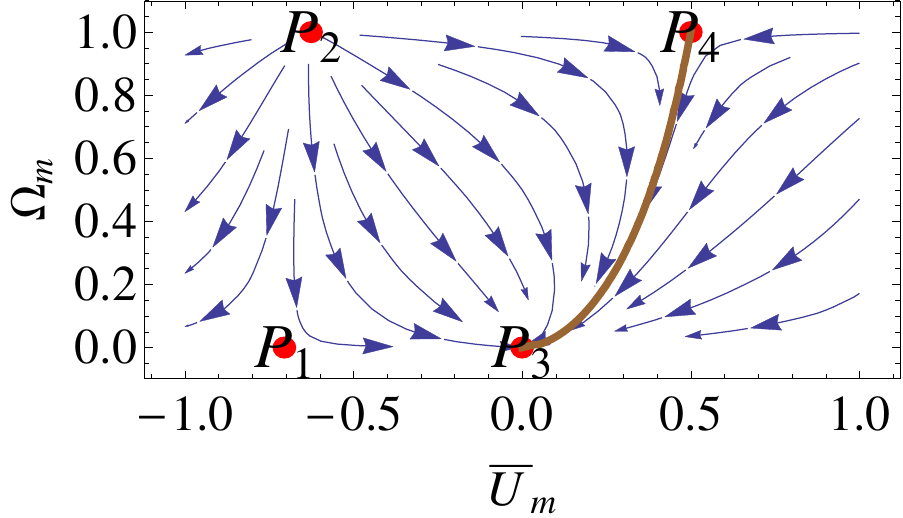}
            \end{subfigure}
    \caption{	\label{fig:LCDM_perts} Phase plane for the system \eqref{LCDM-perts} showing both the stability of the background equations as well as the stability of the perturbations for the $\Lambda$CDM model. (a) in the original fractional energy density $\Omega_m$ vs the phase of matter perturbations $U_m$. (b) Using the compact variables $\Omega_m$ vs $\frac{2}{\pi}\arctan(U_m)$.}
\end{figure*}

Figure \ref{fig:LCDM_perts} presents a flow for the system \eqref{LCDM-perts} showing both the stability of the background equations and the stability of the perturbations for the $\Lambda$CDM model. (a) in the original fractional energy density $\Omega_m$ vs the phase of matter perturbations $U_m$. (b) Using the compact variables  $\left(\Omega_m, \frac{2}{\pi}\arctan(U_m)\right)$ 
we see that there are no equilibrium points at infinity.

This numerical elaboration suggests that the unstable manifold of $P_4$ connects $P_4$ and $P_3$, which is stable. To find an estimate of this solution, we proceed as follows. We propose the polytropic law $U_m=\Omega_m^\Gamma$.  Using the equations  \eqref{LCDM-perts} we obtain the equation
\begin{equation}
3 \Gamma (\Omega_m-1) \Omega_m^\Gamma+\left(\Omega_m^\Gamma+2\right) \Omega_m^\Gamma-\frac{3}{2}\Omega_m \left(\Omega_m^\Gamma+1\right)=0.
\end{equation}
It is required that the solution passes through $P_4$, therefore, expanding the above equation in Taylor series around $\Omega_m=1$ we obtain the approximation
\begin{equation}
-\left(\frac{11 \Gamma}{2}-3\right) (1-\Omega_m)+O\left((1-\Omega_m)^2\right)=0.
\end{equation}
This implies $\Gamma=\frac{6}{11}$. That is, $U_m\simeq \Omega_m^{\frac{6}{11}}$ for the matter dominated universe. 
This line is represented in the figure \ref{fig:LCDM_perts} by a thick (brown) line and incidentally coincides with the stable manifold of the matter-dominated solution $P_4$.

\begin{figure*}[t!]
    \centering
		\includegraphics[width=1.00\textwidth]{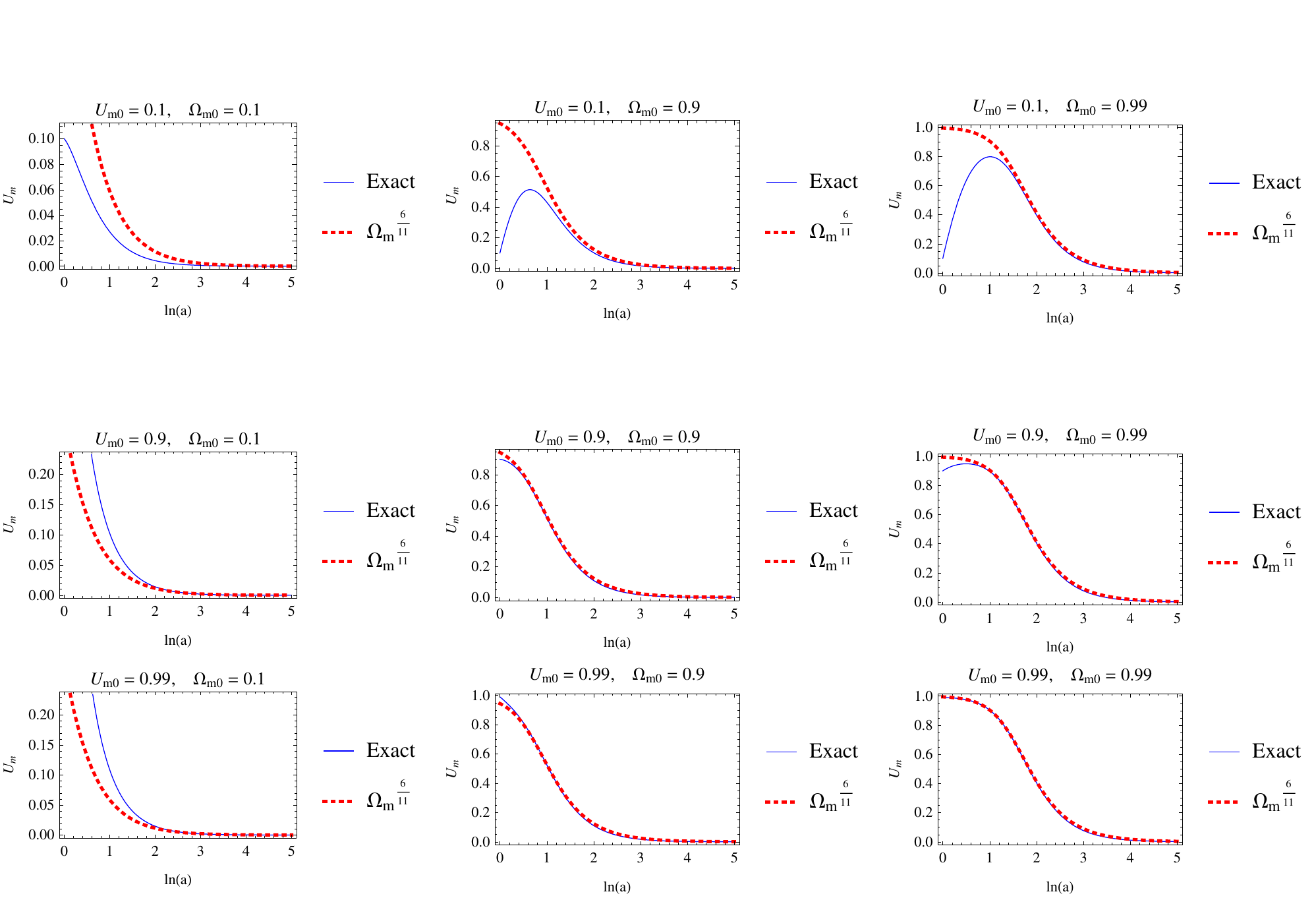}
	\caption{	\label{fig:Exact-vs-App} $U_m$ vs $\ln a$. The plot shows the exact solution  (solid blue line) and the approximated solution $U_m=\Omega_m^{\frac{6}{11}}$ (the red dotted line) for different values of the initial conditions $U_{m0}, \Omega_{m0}$. The closer the initial conditions to the equilibrium point $P_4$, the more accurate the approximation will be.}
\end{figure*}

In the figure \ref{fig:Exact-vs-App} $U_m$ vs $\ln a$ are presented the exact solution (solid blue line) and the approximated solution $U_m=\Omega_m^{\frac{6}{11}}$ (the red dotted line) for different values of the initial conditions $U_{m0}, \Omega_{m0}$. The closer the initial conditions to the equilibrium point $P_4$, the more accurate the approximation will be.

\subsection{Application to Quintessence}

As a  second example, let us examine the simple quintessence example. In this case, and
assuming dust matter, the background equations are 

\begin{align}
&H^{2} =\frac{1}{3}(\rho_m+\rho_\phi), \\
&  \dot{H}=-\frac{1}{2} (\rho_m+\rho_\phi+p_\phi), \\
&\dot{\rho}_m+3H\rho_m =0,  \\
&\dot{\rho}_\phi+3H(1+w_\phi)\rho_\phi =0, 
\end{align}
where 
\begin{eqnarray}
\rho_\phi=\frac{\dot{\phi}^2}{2}+V(\phi), \quad 
p_\phi=\frac{\dot{\phi}^2}{2}-V(\phi), \quad 
w_\phi=\frac{p_\phi}{\rho_\phi},
\end{eqnarray}
while the perturbation equations are (\ref{perteq1}),(\ref{perteq2}). Let us
take for simplicity the usual case 
\begin{eqnarray}
V(\phi)=V_0 e^{-\lambda \phi}.
\end{eqnarray}

We introduce the following auxiliary variables $(x,y)$ given by \eqref{(9)} \cite{Copeland:1997et}
such that 
\begin{equation}
\Omega_\phi= x^2+y^2, \quad \Omega_m=1-\Omega_\phi, \quad w_\phi=\frac{x^2-y^2}{%
x^2+y^2}.
\end{equation}
As before, the background equations recast as  \eqref{(18)}-\eqref{(19)}.

The equations of matter perturbations become 
\begin{subequations}
\begin{align}
\delta_{m}^{\prime \prime }&= \frac{1}{2} \delta_{m}^{\prime }\left(1-3
x^2+3 y^2\right)+\frac{3}{2} \delta_\phi \left(x^2+y^2\right)-\frac{3}{2}
\delta_{m} \left(x^2+y^2-1\right),  \label{eqn.27} \\
\delta_\phi^{\prime \prime }&=\frac{1}{2} \delta_\phi^{\prime }\left(x^2
\left(\frac{24}{x^2+y^2}+3\right)+\frac{2 \sqrt{6} \lambda y^2}{x}-3
y^2-19\right)  \notag \\
& +\frac{1}{2} \delta_\phi \left(\frac{2 \left(7-6 x^2\right) x^2}{x^2+y^2}%
+15 x^2+\frac{y^2 \left(2 \sqrt{6} \lambda -3 x\right)}{x}-13\right)  \nonumber \\
& + 3 \delta_{m} x^2 \left(\frac{1}{x^2+y^2}-1\right).  \label{eqn.28}
\end{align}
\end{subequations}
The stability of the equilibrium point in the plane $(x,y)$ was examined in 
\cite{Copeland:1997et}.

\begin{enumerate}
\item At the equilibrium point $A: (x,y)=(0,0)$, the equation for the matter
perturbation is written as 
\begin{equation}  \label{perts1}
{\delta_m}^{\prime }+3 {\delta_m}=2 {\delta_m}^{\prime \prime },
\end{equation}
where prime means derivative with respect to  $N=\ln a$. \newline
Hence, the perturbations evolve as 
\begin{equation}  \label{solperts1}
\delta_{m}(N )= \frac{1}{5} {\delta_m}_{0} e^{-N } \left(2 e^{5 N
/2}+3\right)+\frac{2}{5} {\delta_m^{\prime }}_{0} e^{-N } \left(e^{5 N
/2}-1\right),
\end{equation}
for the initial conditions ${\delta_m}_{0}={\delta_m}|_{N=0}, {%
\delta_m^{\prime }}_{0}={\delta_m}^{\prime }|_{N=0}$.

For $x^2+y^2=0$, we have no scalar field such that the perturbations $%
\delta_\phi $ are not defined.

Defining the ratio $U_m=\frac{\delta_m^{\prime }(N)}{\delta_m}$%
, we have the equation 
\begin{equation}
U_m^{\prime }=\frac{1}{2} \left(-2 U_m^2+U%
_m+3\right).
\end{equation}

The equilibrium points of the above system are 

\begin{enumerate}
\item $A_1: U_m=-1$ and 
\item $A_2: U_m=\frac{3}{2}$. 
 \end{enumerate}

Since 
\begin{equation}
\lambda(U_m):= \frac{d U_m^{\prime }}{d U_m}=%
\frac{1}{2} (1-4 U_m), \quad \lambda(-1)=\frac{5}{2},\quad
\lambda(3/2)=-\frac{5}{2}.
\end{equation}
That is, $U_m=-1$ is unstable, such that the decaying mode $%
\delta_m\propto e^{-N}$ is important as $N\rightarrow -\infty$.
Besides $U_m=\frac{3}{2}$ is stable, such that the growing mode $%
\delta_m\propto e^{\frac{3}{2}N}$ is the dominant at late times, that is,
as $N\rightarrow +\infty$. Using this qualitative analysis, we can
anticipate the result that is expected from \eqref{solperts1} that matter
perturbations decays as $\delta_m\propto e^{-N}$ in the past, but they
grows as $\delta_m\propto e^{\frac{3}{2}N}$ latter on the evolution,
without having solved the original equation \eqref{perts1}. The aim of 
this section is to formulate the standard quintessence model as an extended
dynamical system (in scale-invariant variables), with the lower dimension as
possible, that comprised both the background equations as well the equations of the perturbations. 
Inspired by the analysis
done in Chapter 14 of Wainwright \& Ellis book \cite{Ellis} we extent 
this analysis for scalar field cosmologies could also be extended to modified gravity theories. 
The method consists in defining
$U_m=\frac{\delta_m^{\prime }(N)}{\delta_m}, \quad U_d=%
\frac{\delta_\phi^{\prime }(N)}{\delta_\phi}$ 
(and some other variables with physical interpretation).
In the standard method to write second-order equations like %
\eqref{eqn.27} or \eqref{eqn.28} is to define $(X_{(n)},Y_{(n)})=\left(%
\delta_{(n)}, \delta^{\prime }_{(n)}\right)$ as variables. Hence, we should
regard this $U_{(n)}$ as $\tan\theta_{(n)}$ where this $%
\theta_{(n)}$ is the usual polar angle in the $\left(\delta_{(n)},
\delta^{\prime }_{(n)}\right)$-plane with $0\leq \theta_{(n)}<2\pi$.
Therefore, one can think of $U_{(n)}$ as the phase of the
perturbation. If $U_{(n)}>0$ during the evolution, it follows by
definition that the perturbations $\delta_{(n)}$ are growing with time
(since $\delta_{(n)}>0$ and $\delta'_{(n)}>0$ or $\delta_{(n)}<0$ and $%
\delta_{(n)}'<0$), while if $U_{(n)}<0$ at that time, the
perturbation is decaying. If an orbit of the flow is asymptotic to an
equilibrium point, then the perturbation approaches a stationary state,
decaying to zero if $U_{(n)}<0$ or growing if $U_{(n)}>0$. If the orbits are asymptotic to a periodic solution, then the perturbation
propagates as sound waves.

\item At the equilibrium points $B, C: (x,y)=(\pm 1, 0)$,  the evolution of perturbations
is: 
\begin{align}  \label{perts2}
{\delta_m}^{\prime \prime }=\frac{3 {\delta_\phi}}{2}-{\delta_m}^{\prime },
\quad {\delta_\phi}^{\prime \prime }=4 {\delta_\phi}^{\prime }+2 {\delta_\phi}.
\end{align}
Hence, the perturbations evolve as
\begin{subequations}
\begin{align}
& \delta_m=\frac{1}{8} {\delta_\phi}_{0} e^{\left(2-\sqrt{6}\right) N }
\left(e^{\left(\sqrt{6}-3\right) N } \left(24 e^{N }+\left(9 \sqrt{6}%
-22\right) e^{\left(3+\sqrt{6}\right) N }+20\right)-9 \sqrt{6}-22\right) 
\notag \\
& +\frac{1}{8} {\delta_\phi^{\prime }}_{0} e^{\left(2-\sqrt{6}\right) N }
\left(e^{\left(\sqrt{6}-3\right) N } \left(-6 e^{N }+\left(5-2 \sqrt{6}%
\right) e^{\left(3+\sqrt{6}\right) N }-4\right)+2 \sqrt{6}+5\right)\nonumber \\
&  +{%
\delta_m}_{0}  +{\delta_m^{\prime }}_{0} (\sinh (N )-\cosh (N )+1) , \\
&\delta_\phi= \frac{{\delta_\phi^{\prime }}_{0} e^{2 N } \sinh \left(\sqrt{6}
N \right)}{\sqrt{6}}-\frac{1}{3} {\delta_\phi}_{0}e^{2 N } \left(\sqrt{6}
\sinh \left(\sqrt{6} N \right)-3 \cosh \left(\sqrt{6} N \right)\right),
\end{align}
\end{subequations}
for the initial conditions ${\delta_m}_{0}={\delta_m}|_{N=0}, {\delta_\phi}%
_{0}={\delta_\phi}|_{N=0}, {\delta_m^{\prime }}_{0}={\delta_m}^{\prime
}|_{N=0}, {\delta_\phi^{\prime }}_{0}={\delta_\phi}^{\prime }|_{N=0}$.

With the above perturbation equations \eqref{perts2}, we construct a system
of differential equations for the quantities 
\begin{equation}
\quad V_m=\frac{%
\delta_m^{\prime }(N)}{\delta_\phi}, \quad U_d=\frac{%
\delta_\phi^{\prime }(N)}{\delta_\phi},
\end{equation}
as given by 
\begin{subequations}
\begin{align}
& V_m^{\prime }=\frac{3}{2}-(U_d+1) V_m, \\
& U_d^{\prime }= 2-(U_d-4) U_d.
\end{align}
\end{subequations}

We now try and research its stability and integrability in the reduced phase plane  $(V_m, U_d)$. 
We obtain the equilibrium points: 
\begin{enumerate}
\item $B_1: (V_m,U_d)=\left(\frac{1}{2} \left(3+\sqrt{6}\right),2-\sqrt{6}\right)$ 
The eigenvalues are $\left\{2 \sqrt{6},\sqrt{6}-3\right\}$. Therefore, the equilibrium point is saddle (unstable). 
The phase of the matter perturbation at the equilibrium point is  $U_m^*=U_d^*=2-\sqrt{6}$ are both negative; therefore, the perturbations $\delta_m$ and $\delta_\phi$ decay with time. 
\item $B_2: (V_m,U_d)=\left(\frac{1}{2} \left(3-\sqrt{6}\right),2+\sqrt{6}\right)$. 
The eigenvalues are $\left\{-3-\sqrt{6},-2 \sqrt{6}\right\}$. Therefore, the equilibrium point is stable. The phase of the matter perturbation at the equilibrium point is  $U_m^*=U_d^*=2+\sqrt{6}$ are both positive; therefore, the perturbations $\delta_m$ and $\delta_\phi$ are growing with time. 
\end{enumerate}

Introducing the compact variables
\begin{equation}
\bar{V}_m=\frac{2}{\pi}\arctan\left(V_m
\right)
,\quad \bar{U}_d=\frac{2}{\pi}\arctan\left(U_d\right),
\end{equation}
we obtain the dynamical system
\begin{subequations}
\label{SCF_perts_3_4}
\begin{align}
&\bar{V}_m'=\frac{\sin (\pi  \bar{V}_m) \left(3 \cot \left(\frac{\pi  \bar{V}_m}{2}\right)-2 \left(\tan \left(\frac{\pi  \bar{U}_d}{2}\right)+1\right)\right)}{2 \pi },\\
&\bar{U}_d'=\frac{4 \sin (\pi  \bar{U}_d)+3 \cos (\pi  \bar{U}_d)+1}{\pi}.
\end{align}
\end{subequations}

\begin{figure}[]
\centering
        \includegraphics[scale=0.9]{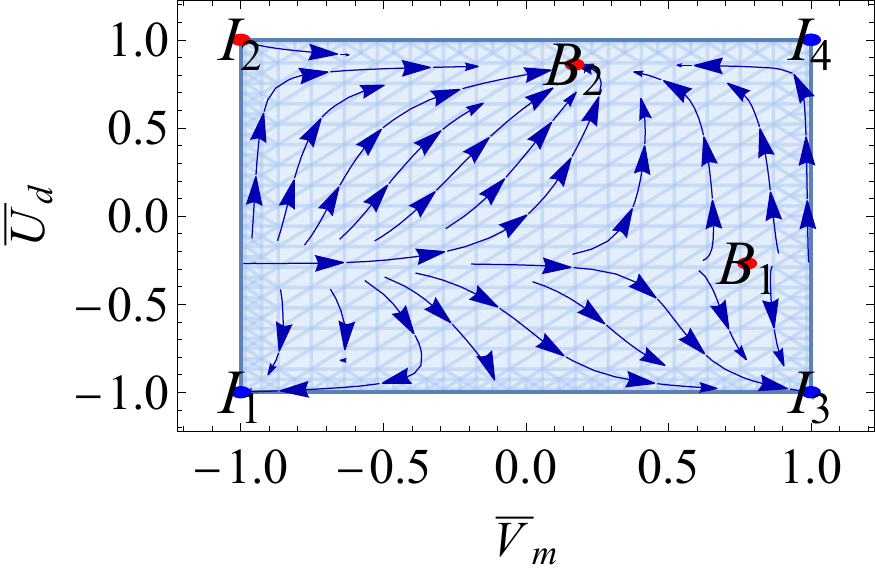}
      
    \caption{	\label{fig:SCF_perts_3_4} Phase plane  for the system \eqref{SCF_perts_3_4} showing both the stability of the background equations as well as the stability of the perturbations for the Kinetic dominated points for the quintessence model using the compact variables $\bar{V}_m=\frac{2}{\pi}\arctan\left(V_m
\right)$ vs. $\bar{U}_d=\frac{2}{\pi}\arctan\left(U_d\right)$.}
\end{figure}

Figure \ref{fig:SCF_perts_3_4} it is presented a flow for the system \eqref{SCF_perts_3_4} showing both the stability of the background equations as well as the stability of the perturbations for the  Kinetic dominated points $(x,y)=(\pm 1, 0)$ for the quintessence model using the compact variables $\bar{V}_m=\frac{2}{\pi}\arctan\left(V_m
\right)$ vs. $\bar{U}_d=\frac{2}{\pi}\arctan\left(U_d\right)$. There are some configurations $I_{1,2}:(\bar{V}_m, \bar{U}_d)=(-1, \pm 1)$ and $I_{3,4}:(\bar{V}_m, \bar{U}_d)=(1, \pm 1)$ which are not equilibrium points due to $\bar{U}_d'=-\frac{2}{\pi}\neq 0$ at the fixed points but they are essential for the dynamics at infinity. The figure suggests that the attractor at the infinity region are $I_1$ and $I_3$, and the other points at infinity are saddle points. 
Indeed, if we choose $\bar{V}_m=\pm 1$, then $\bar{U}_d'= -\frac{2}{\pi} \implies \bar{U}_d=-\frac{2}{\pi} N+c$, but by definition $-1< \bar{U}< 1$, therefore any solution starting on the line $\bar{V}_m=\pm 1$ hits the boundary $\bar{U}_d=+1$ in a finite time-lapse to the past and the boundary $\bar{U}_d=-1$ in a finite time-lapse to the future.

\item For the equilibrium points $D: (x,y)=\left(\frac{\lambda }{\sqrt{6}}, \sqrt{1-%
\frac{\lambda^2}{6}}\right)$, we have 
\begin{small}
\begin{align}  \label{perts3}
\delta_{m}^{\prime \prime }= \frac{1}{2} \left(3 {\delta_\phi}-\left(\lambda
^2-4\right) {\delta_m}^{\prime }\right), \quad {\delta_\phi}^{\prime \prime
}= \frac{1}{2} \left(3 \lambda ^2-10\right) {\delta_\phi}^{\prime }-\frac{1}{6}
\left(\lambda ^4-10 \lambda ^2+12\right) {\delta_\phi}.
\end{align}
\begin{align}
&\delta_m={\delta_m}_{0}+\frac{2 {\delta_m^{\prime }}_{0} \left(1-e^{-\frac{1%
}{2} (\lambda^2 -4) N }\right)}{\lambda^2 -4}  \notag \\
& +9{\delta_\phi}_{0} \Big(\frac{e^{N_{-}} (k (\lambda^2 ((401-43 \lambda^2 )
\lambda^2 -1348)+1644)+(\lambda^2 (17 \lambda^2 -113)+198) (\lambda^2 (19 \lambda^2
-100)+204))}{2 ((\lambda^2 -10) \lambda^2 +12) (\lambda^2 (7 \lambda^2 -55)+96)
(\lambda^2 (19 \lambda^2 -100)+204)}  \notag \\
& +\frac{e^{N_{+}} (k (\lambda^2 (\lambda^2 (43 \lambda^2
-401)+1348)-1644)+(\lambda^2 (17 \lambda^2 -113)+198) (\lambda^2 (19 \lambda^2
-100)+204))}{2 ((\lambda^2 -10) \lambda^2 +12) (\lambda^2 (7 \lambda^2 -55)+96)
(\lambda^2 (19 \lambda^2 -100)+204)}  \notag \\
&+\frac{2 (2 \lambda^2 -7) e^{-\frac{1}{2} (\lambda^2 -4) N }}{(\lambda^2 -4)
(\lambda^2 (7 \lambda^2 -55)+96)}+\frac{10-3 \lambda^2 }{(\lambda^2 -4) ((\lambda^2
-10) \lambda^2 +12)}\Big)  \notag \\
& +\frac{9{\delta_\phi^{\prime }}_{0}}{(\lambda^2 -4) ((\lambda^2 -10) \lambda^2 +12)}
\Big(2+ \frac{e^{-\frac{\lambda^2 N }{2}} }{(\lambda^2 (7 \lambda^2 -55)+96)
(\lambda^2 (19 \lambda^2 -100)+204)} \times  \notag \\
& \Big\{-(\lambda^2 -4) (k ((103-16 \lambda^2 ) \lambda^2 -186)+3 (2 \lambda^2 -7)
(\lambda^2 (19 \lambda^2 -100)+204)) e^{\frac{\lambda^2 N }{2}+N_{-}}  \notag
\\
& -(\lambda^2 -4) (k (\lambda^2 (16 \lambda^2 -103)+186)+3 (2 \lambda^2 -7) (\lambda^2
(19 \lambda^2 -100)+204)) e^{\frac{\lambda^2 N }{2}+N_{+}}  \notag \\
& -2 ((\lambda^2 -10) \lambda^2 +12) (\lambda^2 (19 \lambda^2 -100)+204) e^{2 N }%
\Big\}\Big), \\
&\delta_\phi=\frac{{\delta_\phi}_{0}\left(k (3 \lambda^2 -10)
\left(e^{N_{+}}-e^{N_{-}}\right)+(\lambda^2 (19 \lambda^2 -100)+204)
\left(e^{N_+}+e^{N_{-}}\right)\right)}{2 (\lambda^2 (19 \lambda^2
-100)+204)}  \nonumber \\
& -\frac{2 {\delta_\phi^{\prime }}_{0} \left(e^{N_{+}} -e^{N _{-}}\right)%
}{\sqrt{\frac{1}{3} \lambda^2 (19 \lambda^2 -100)+68}},
\end{align}
\end{small}
for the initial conditions ${\delta_m}_{0}={\delta_m}|_{N=0}, {\delta_\phi}%
_{0}={\delta_\phi}|_{N=0}, {\delta_m^{\prime }}_{0}={\delta_m}^{\prime
}|_{N=0}, {\delta_\phi^{\prime }}_{0}={\delta_\phi}^{\prime }|_{N=0}$, where 
$N_+=-\frac{1}{12} \left(-9 \lambda^2 +30+k\right) N$, $N_-=-\frac{1}{%
12} \left(-9 \lambda^2 +30-k\right) N$, \newline $k=\sqrt{57 \lambda^4-300 \lambda^2
+612}$.

With the above perturbation equations \eqref{perts3}, we construct a system
of differential equations for the quantities 
\begin{equation}
V_m=\frac{%
\delta_m^{\prime }(N)}{\delta_\phi}, \quad U_d=\frac{%
\delta_\phi^{\prime }(N)}{\delta_\phi},
\end{equation}
as given by 
\begin{subequations}
\begin{align}
&Q^{\prime }=-Q (U_d-U_m), \\
&U_m^{\prime }= \frac{1}{2} \left(\frac{3}{Q}-U_m
\left(\lambda ^2+2 U_m-4\right)\right), \\
&U_d^{\prime }= -\frac{\lambda ^4}{6}+\frac{5 \lambda ^2}{3}+\frac{%
3 \lambda ^2 U_d}{2}-U_d(U_d+5)-2
\end{align}
\end{subequations}
For further simplification, we define $%
V_m={U_m}{Q}$, so that we acquire the reduced
dynamical system 
\begin{subequations}
\label{systemC}
\begin{align}
& V_m^{\prime }=\frac{1}{2} \left(3-V_{m} \left(\lambda
^2+2 U_d-4\right)\right), \\
&U_d^{\prime }= -\frac{\lambda ^4}{6}+\frac{5 \lambda ^2}{3}+\frac{%
3 \lambda ^2 U_d}{2}-U_d(U_d+5)-2
\end{align}
\end{subequations}

For arbitrary $\lambda$, we obtain the equilibrium points: 
\begin{enumerate}
\item $D_1: (V_m,U_d)=\left(-\frac{18}{-15 \lambda ^2+\sqrt{57 \lambda ^4-300 \lambda ^2+612}+54}, \frac{1}{12} \left(9 \lambda ^2-\sqrt{57 \lambda ^4-300 \lambda ^2+612}-30\right)\right)$. For $-\sqrt{\frac{1}{14} \left(55+\sqrt{337}\right)}<\lambda <\sqrt{\frac{1}{14} \left(55+\sqrt{337}\right)}\approx 2.28907$, both phases of the perturbations are negative, implying that the perturbations $\delta_m$ and $\delta_\phi$ are decaying with time.  The eigenvalues  of the linear matrix are \newline $\left\{\frac{\sqrt{19 \lambda ^4-100 \lambda ^2+204}}{2 \sqrt{3}},\frac{1}{12} \left(-15 \lambda ^2+\sqrt{3} \sqrt{19 \lambda ^4-100 \lambda ^2+204}+54\right)\right\}$. For the above interval, the point is a source. For $\lambda ^2 >{\frac{1}{14} \left(55+\sqrt{337}\right)}$  it is a saddle. 

\item $D_2: (V_m,U_d)=\left(\frac{18}{15 \lambda ^2+\sqrt{57 \lambda ^4-300 \lambda ^2+612}-54}, \frac{1}{12} \left(9 \lambda ^2+\sqrt{57 \lambda ^4-300 \lambda ^2+612}-30\right)\right)$.  For  $-\sqrt{5-\sqrt{13}}<\lambda <\sqrt{5-\sqrt{13}}\approx 1.18087$, both phases of the perturbations are negative, implying that the perturbations $\delta_m$ and $\delta_\phi$ are decaying with time. The eigenvalues  of the linear matrix are \newline
$\left\{-\frac{\sqrt{19 \lambda ^4-100 \lambda ^2+204}}{2 \sqrt{3}},\frac{1}{12} \left(-15 \lambda ^2-\sqrt{3} \sqrt{19 \lambda ^4-100 \lambda ^2+204}+54\right)\right\}$. 

When $-\sqrt{\frac{55}{14}-\frac{\sqrt{337}}{14}}<\lambda <\sqrt{\frac{55}{14}-\frac{\sqrt{337}}{14}}$ the point is a saddle. It is stable (sink) when $\lambda <-\sqrt{\frac{55}{14}-\frac{\sqrt{337}}{14}}\lor \lambda >\sqrt{\frac{55}{14}-\frac{\sqrt{337}}{14}}$. In this region, both phases of the perturbations are positive, implying that the perturbations $\delta_m$ and $\delta_\phi$ are growing with time. Therefore, one of the scalar field-dominated solutions can have stable phases of the perturbations, but the perturbations $\delta_m$ and $\delta_\phi$ themselves are, in this case growing with time.  
\end{enumerate}

For $\lambda=0$, the points above correspond to the de Sitter point (exact $\Lambda$CDM model) that has $(x,y)=(0,1)$.
That is,  the de Sitter solution is represented in the extended phase space $(V_m, U_d)$ by two equilibrium points:
\begin{enumerate}
\item $D_1: (V_m,U_d)=\left( -\frac{18}{54+6 \sqrt{17}},  \frac{1}{12} \left(-30-6 \sqrt{17}\right)\right)$.  The eigenvalues of the linear matrix are $\left\{\frac{1}{2} \left(9+\sqrt{17}\right),\sqrt{17}\right\}$. Hence, the point is unstable (source). 
\item $D_2: (V_m,U_d)=\left(\frac{18}{6 \sqrt{17}-54}, \frac{1}{12} \left(6\sqrt{17}-30\right)\right)$. The eigenvalues of the linear matrix are  $\left\{-\sqrt{17},\frac{1}{2} \left(9-\sqrt{17}\right)\right\}$ and the point is a saddle point. 
\end{enumerate}
For these equilibrium points, the  phases of the perturbations $U_m$ and $U_d$ 
are negative therefore, the perturbations $\delta_m$ and $\delta_\phi$ are decaying with time. That is expected for the exact $\Lambda$CDM model with $\delta_m\rightarrow 0, \delta_\phi=0$. 
Although in the phase plane, $(x,y)$ the de Sitter solution is stable, when the directions along the ``phase of the perturbations'' are considered in the dynamics, the de Sitter solution (and their two representations) one becomes unstable. The other becomes saddle in the phase plane  $(V_m, U_d)$. Summarizing, on top of the fixed point $D$, the phases of the perturbations are not stable, although the perturbations $\delta_m$ and $\delta_\phi$ are decaying with time.  

Introducing the compact variables
\begin{equation}
\quad \bar{V}_m=\frac{2}{\pi}\arctan\left(V_m
\right)
,\quad \bar{U}_d=\frac{2}{\pi}\arctan\left(U_d\right),
\end{equation}
we obtain the dynamical system
\begin{subequations}
\label{SCF_perts}
\begin{align}
&\bar{V}_m'=-\frac{\cos ^2\left(\frac{\pi  \bar{V}_m}{2}\right) \left(\tan \left(\frac{\pi  \bar{V}_m}{2}\right) \left(2 \tan \left(\frac{\pi  \bar{U}_d}{2}\right)+\lambda^2-4\right)-3\right)}{\pi },\\
&\bar{U}_d'=-\frac{\left(30-9 \lambda ^2\right) \sin \left(\pi  \bar{U}_d\right)+\left(\lambda ^4-10 \lambda ^2+6\right) \cos \left(\pi  \bar{U}_d\right)+\lambda
   ^4-10 \lambda ^2+18}{6 \pi }.
\end{align}
\end{subequations}

Figure \ref{fig:SFC} it is presented a flow  for the system \eqref{SCF_perts} showing  both the stability of the background equations as well as the stability of the perturbations for the scalar field dominated point  $(x,y)=\left(\frac{\lambda }{\sqrt{6}}, \sqrt{1-%
\frac{\lambda^2}{6}}\right)$ for the quintessence model with exponential potential with $\lambda=0, 1, 2, 3$ using the compact variables $\bar{V}_m=\frac{2}{\pi}\arctan\left(V_m
\right)$ vs. $\bar{U}_d=\frac{2}{\pi}\arctan\left(U_d\right)$. There are some configurations $I_{1,2}:(\bar{V}_m, \bar{U}_d)=(-1, \pm 1)$ and $I_{3,4}:(\bar{V}_m, \bar{U}_d)=(1, \pm 1)$ which are not equilibrium points due to $\bar{U}_d'=-\frac{2}{\pi}\neq 0$ at the fixed points but they are important for the dynamics at infinity. The figure suggests that the attractor at the infinity region are $I_1$ and $I_3$, and the other points at infinity are saddle points. 
Indeed, if we choose $\bar{V}_m=\pm 1$, then $\bar{U}_d'= -\frac{2}{\pi} \implies \bar{U}_d=-\frac{2}{\pi} N+c$, but by definition $-1< \bar{U}< 1$, therefore any solution starting on the line $\bar{V}_m=\pm 1$ hits the boundary $\bar{U}_d=+1$ in a finite time-lapse to the past and the boundary $\bar{U}_d=-1$ in a finite time-lapse to the future.  

\begin{figure*}
	\centering
		\includegraphics[scale=0.65]{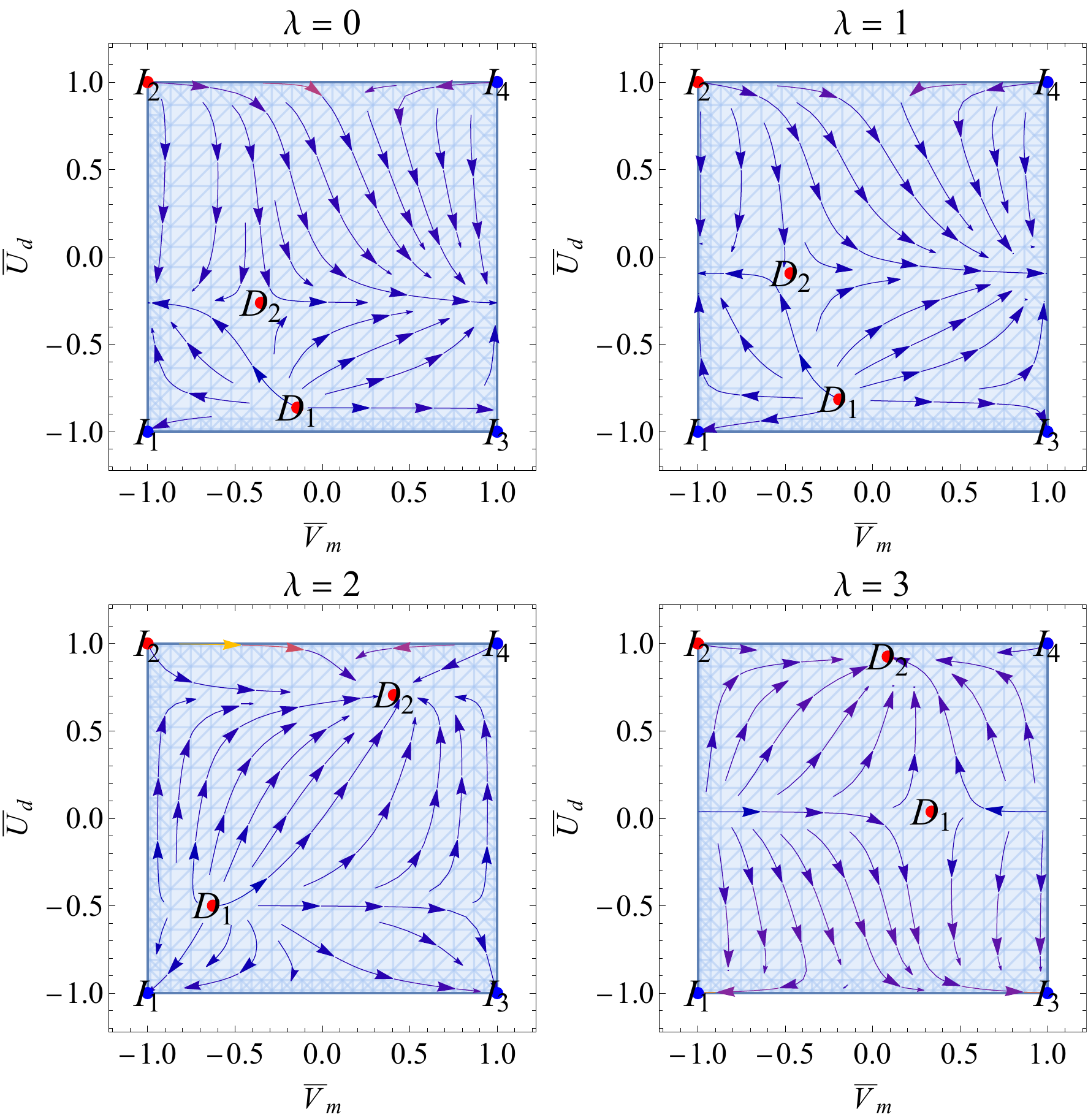}
	\caption{\label{fig:SFC} Phase plane  for the system \eqref{SCF_perts} showing  both the stability of the background equations as well as the stability of the perturbations for the scalar field dominated point  $(x,y)=\left(\frac{\lambda }{\sqrt{6}}, \sqrt{1-%
\frac{\lambda^2}{6}}\right)$ for the quintessence model with exponential potential with $\lambda=0, 1, 2, 3$ using the compact variables $\bar{V}_m=\frac{2}{\pi}\arctan\left(V_m
\right)$ vs. $\bar{U}_d=\frac{2}{\pi}\arctan\left(U_d\right)$.}
\end{figure*}

\item For the equilibrium points $E:= (x,y)=\left(\frac{\sqrt{\frac{3}{2}}}{\lambda }, 
\sqrt{\frac{3}{2 \lambda^2}}\right)$, we have 
\begin{align}  \label{perts4}
{\delta_m}^{\prime \prime }= \frac{9 {\delta_\phi}+\lambda ^2 {\delta_m}%
^{\prime }+3 \left(\lambda ^2-3\right) {\delta_m}}{2 \lambda ^2},\quad {%
\delta_\phi}^{\prime \prime }= \frac{-\lambda ^2 {\delta_\phi}^{\prime }+9 {%
\delta_\phi}+3 \left(\lambda ^2-3\right) {\delta_m}}{2 \lambda ^2}
\end{align}

With the above perturbation equations \eqref{perts4}, we construct a system
of differential equations for the quantities 
\begin{equation}
Q=\frac{\delta_m}{\delta_\phi}, \quad V_m=\frac{%
\delta_m^{\prime }(N)}{\delta_\phi}, \quad U_d=\frac{%
\delta_\phi^{\prime }(N)}{\delta_\phi},
\end{equation}
as given by 
\begin{align}
& V_m'=\frac{\lambda ^2 (3 Q-2 U_\phi V_{m}+ V_{m})-9 Q+9}{2 \lambda ^2}, \\
& U_d'= \frac{3 \left(\lambda ^2-3\right) Q-\lambda ^2 U_\phi (2 U_\phi+1)+9}{2 \lambda ^2},\\
& Q'=V_{m}-Q  U_\phi.
\end{align}

In this case, the real-valued equilibrium points are 
\begin{enumerate}
\item $E_1: (V_m, U_d, Q)= \left( 0,  0,  -\frac{3}{\lambda ^2-3}\right)$
and 
\item $E_2: (V_m, U_d, Q)=$ \newline
$\Bigg( \frac{4 \Delta ^{2/3} \left(8 \lambda ^2-27\right) \lambda ^2+\sqrt[3]{3} \Delta 
   \left(7 \lambda ^2-18\right)+2 \sqrt[6]{3} \sqrt[3]{\Delta } \left(29 \sqrt{3} \lambda ^2+\sqrt{-25 \lambda ^4-729 \lambda ^2+2187}-90 \sqrt{3}\right) \lambda ^4+49 \sqrt[3]{3} \lambda
   ^8}{36 \Delta ^{2/3} \lambda ^2 \left(\lambda ^2-3\right)}$,\\
   $ \frac{\Delta ^{2/3}+7 \sqrt[3]{3} \lambda ^4}{2\ 3^{2/3} \sqrt[3]{\Delta } \lambda ^2},$\\
   $\frac{3 \Delta ^{2/3}
   \left(14 \lambda ^2+\sqrt[3]{81 \lambda ^6+6 \left(\sqrt{-75 \lambda ^4-2187 \lambda ^2+6561}-81\right) \lambda ^4}-54\right) \lambda ^2+3^{2/3} \Delta ^{4/3}+21\ 3^{2/3} \sqrt[3]{\Delta
   } \lambda ^6+147 \sqrt[3]{3} \lambda ^8}{54 \Delta ^{2/3} \lambda ^2 \left(\lambda ^2-3\right)}\Bigg)$, \\
where $\Delta =27 \lambda ^6+2 \left(\sqrt{-75 \lambda ^4-2187 \lambda ^2+6561}-81\right) \lambda ^4$.
\end{enumerate}
The first point $E_1$ has constant densities perturbations with $\delta_m=-\frac{3}{\lambda ^2-3}\delta_\phi$. They remain constant with evolution.
The eigenvalues of $E_1$ are $\left\{\frac{ k_{1}}{2 \lambda ^2 \left(\lambda ^2-3\right)},\frac{k_{2}}{2 \lambda ^2 \left(\lambda ^2-3\right)},\frac{ k_{3}}{2 \lambda ^2
   \left(\lambda ^2-3\right)}\right\}$, where $k_{1,2,3}$ are the three roots of the polynomial $P(k)=-6 \lambda ^{12}+90 \lambda ^{10}-486 \lambda ^8+1134 \lambda ^6-972 \lambda ^4+k^3+k \left(-7 \lambda ^8+42 \lambda ^6-63 \lambda ^4\right)$. 
   
   \begin{figure}[]
\centering
        \includegraphics[scale=0.8]{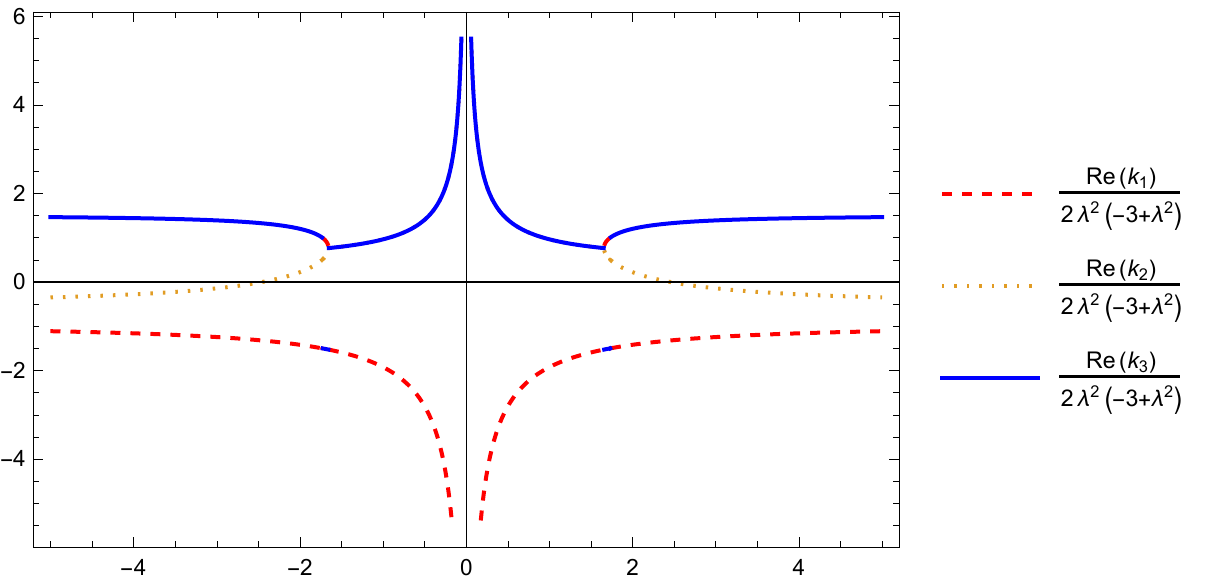}
      
    \caption{	\label{fig:EigenE1} Real parts of the eigenvalues of  $E_1$ }
\end{figure}
   Figure \eqref{fig:EigenE1} is presented the real parts of the eigenvalues of $E_1$. At least two eigenvalues have different signs for all the values of $\lambda$. Therefore, $E_1$ is a saddle. 

For $E_2$, we  have \newline $\delta_m=\frac{3 \Delta ^{2/3}
   \left(14 \lambda ^2+\sqrt[3]{81 \lambda ^6+6 \left(\sqrt{-75 \lambda ^4-2187 \lambda ^2+6561}-81\right) \lambda ^4}-54\right) \lambda ^2+3^{2/3} \Delta ^{4/3}+21\ 3^{2/3} \sqrt[3]{\Delta
   } \lambda ^6+147 \sqrt[3]{3} \lambda ^8}{54 \Delta ^{2/3} \lambda ^2 \left(\lambda ^2-3\right)} \delta_\phi$. Moreover,  $U_d>0$  implies that the perturbation  $\delta_\phi$ (therefore, $\delta_m$) is growing with time. 
   
      \begin{figure}[]
\centering
        \includegraphics[scale=0.8]{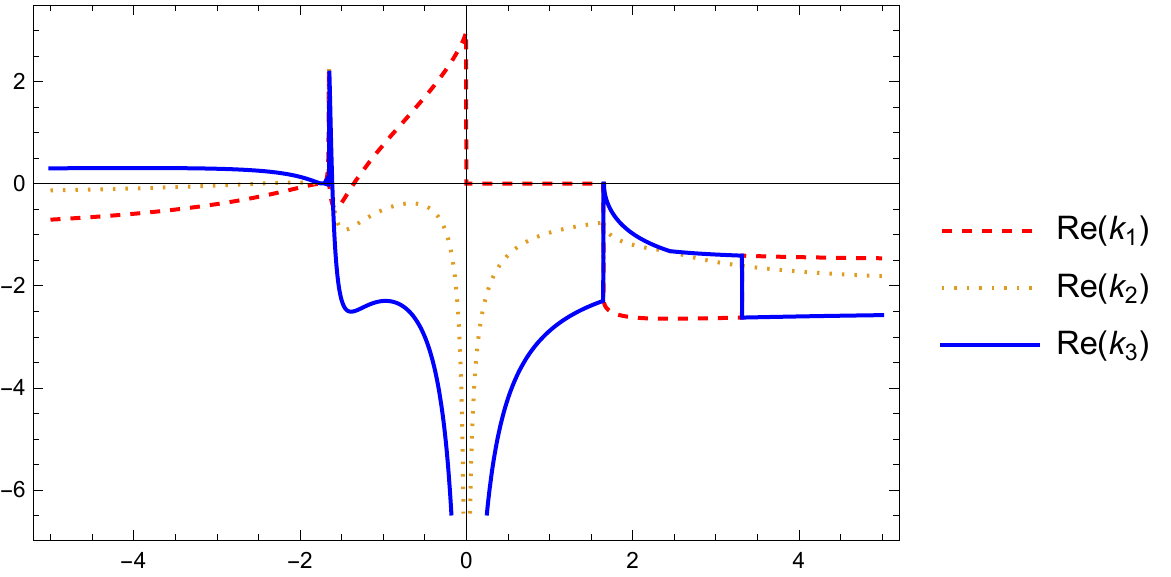}
          \caption{\label{fig:EigenE2} Real parts of the eigenvalues of  $E_2$ }
\end{figure}

Figure \ref{fig:EigenE2} presents the real parts of the eigenvalues of $E_2$. For $\lambda \gtrsim  1.65014$ or $- 1.65014 \lesssim \lambda \lesssim -1.35169$, $E_2$ is a sink; otherwise, it is a saddle. 

\begin{figure}[]
	\centering
		\includegraphics[scale=0.65]{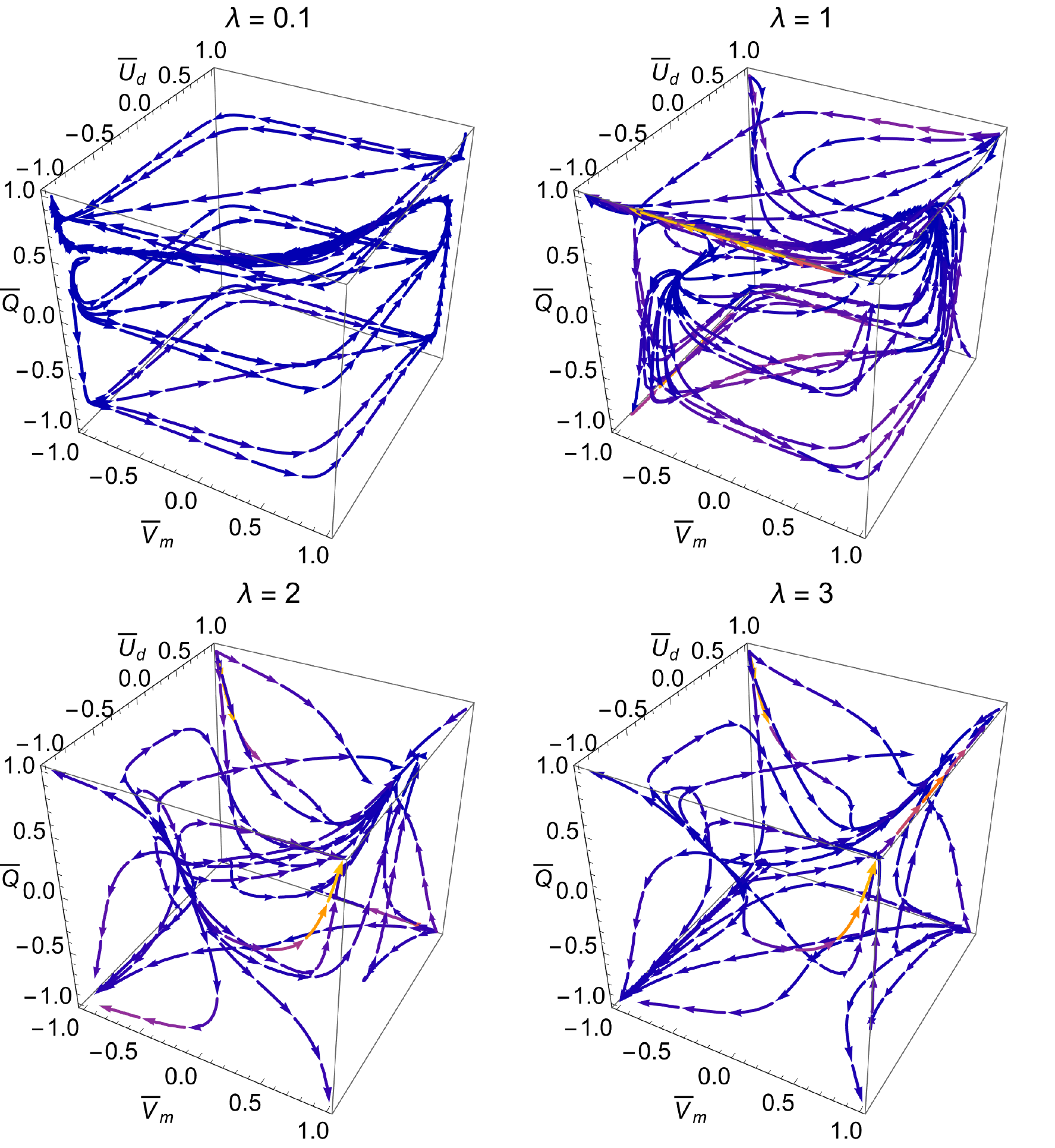}
	\caption{\label{fig:SFC3D} Phase space for the system \eqref{SCF_perts3D} showing  both the stability of the background equations as well as the stability of the perturbations for the scalar field dominated point  $(x,y)=\left(\frac{\lambda }{\sqrt{6}}, \sqrt{1-%
\frac{\lambda^2}{6}}\right)$ for the quintessence model with exponential potential with $\lambda=0.1, 1, 2, 3$ using the compact variables $\bar{V}_m=\frac{2}{\pi}\arctan\left(V_m\right),\quad \bar{U}_d=\frac{2}{\pi}\arctan\left(U_d\right), \quad \bar{Q}=\frac{2}{\pi}\arctan\left(\frac{ \delta_m%
}{\delta_\phi}\right)$.}
\end{figure}

Introducing the compact variables
\begin{equation}
\bar{V}_m=\frac{2}{\pi}\arctan\left(V_m
\right)
,\quad \bar{U}_d=\frac{2}{\pi}\arctan\left(U_d\right), \quad \bar{Q}=\frac{2}{\pi}\arctan\left(\frac{ \delta_m%
}{\delta_\phi}\right)
\end{equation}
we obtain the dynamical system
\begin{subequations}
\label{SCF_perts3D} 
\begin{align}
&\bar{V}_m'=\frac{\cos ^2\left(\frac{\pi  \bar{V}_m}{2}\right) \left(\lambda ^2 \left(1-2 \tan \left(\frac{\pi  \bar{U}_d}{2}\right)\right) \tan \left(\frac{\pi  \bar{V}_m}{2}\right)+3
   \left(\lambda ^2-3\right) \tan \left(\frac{\pi  \bar{Q}}{2}\right)+9\right)}{\pi  \lambda ^2},\\
&\bar{U}_d'=\frac{\cos ^2\left(\frac{\pi  \bar{U}_d}{2}\right) \left(\lambda ^2 \left(-\tan
   \left(\frac{\pi  \bar{U}_d}{2}\right)\right) \left(2 \tan \left(\frac{\pi  \bar{U}_d}{2}\right)+1\right)+3 \left(\lambda ^2-3\right) \tan \left(\frac{\pi 
   \bar{Q}}{2}\right)+9\right)}{\pi  \lambda ^2},\\
&\bar{Q}'=\frac{2 \cos ^2\left(\frac{\pi  \bar{Q}}{2}\right) \left(\tan \left(\frac{\pi  \bar{V}_m}{2}\right)-\tan \left(\frac{\pi 
   \bar{Q}}{2}\right) \tan \left(\frac{\pi  \bar{U}_d}{2}\right)\right)}{\pi }.
\end{align}
\end{subequations}

\end{enumerate}

Fig. \ref{fig:SFC3D} shows a phase space for the system \eqref{SCF_perts3D} showing  both the stability of the background equations as well as the stability of the perturbations for the scalar field dominated point  $(x,y)=\left(\frac{\lambda }{\sqrt{6}}, \sqrt{1-%
\frac{\lambda^2}{6}}\right)$ for the quintessence model with exponential potential with $\lambda=0.1, 1, 2, 3$ using the compact variables $\bar{V}_m=\frac{2}{\pi}\arctan\left(V_m\right),\quad \bar{U}_d=\frac{2}{\pi}\arctan\left(U_d\right), \quad \bar{Q}=\frac{2}{\pi}\arctan\left(\frac{ \delta_m}{\delta_\phi}\right)$.
Both plots shows that there are late-time attractors at $(\bar{V}_m, \bar{U}_d, \bar{Q})= (-1, -1, + 1)$ (top panels) and at $(\bar{V}_m, \bar{U}_d, \bar{Q})= (-1, -1, + 1)$, or at $(\bar{V}_m, \bar{U}_d, \bar{Q})= (-1, -1, - 1)$, or both (bottom panels). Some orbits is the bottom panels approaches $(\bar{V}_m, \bar{U}_d, \bar{Q})= (+1, -1, -1)$. 

\section{Concluding Remarks}
\label{sect:6}
This article extensively analysed dynamical systems in cosmological models involving a scalar field with an arbitrary potential. Our research covers a broad class of potentials and examines the system's behaviour at the background and perturbation levels. We utilised the $f$-deviser method to study a non-interacting scalar field cosmology with arbitrary potential. This approach facilitates a comprehensive analysis of the system's dynamics, extending the existing literature's findings. We analysed the background quantities using Hubble-normalized variables. We provided two examples: the monomial potential and the double exponential potential, which includes the hyperbolic cosine, exponential potential, and Cosmological Constant. These two classes of potentials, monomial and double exponential, comprise the asymptotic behaviour of several classes of scalar field potentials. Therefore, they provide the skeleton for the typical behaviour of arbitrary potentials.

As mentioned earlier, technical papers such as \cite{Alho:2020cdg} have derived a new regular dynamical system on a three-dimensional compact state space. This system describes linear scalar perturbations of spatially flat RW geometries for relativistic models with a minimally coupled scalar field with exponential potential. That allows them to create a global solution space where known solutions reside on specific invariant sets. Their dynamical systems approach has obtained new findings about the comoving and uniform density curvature perturbations. Additionally, they have extended this approach to more general scalar field potentials, resulting in state spaces where the models' exponential potential state space appears as invariant boundary sets, demonstrating their significance as building blocks in a hierarchy of increasingly complex cosmological models. 

Our research paper employs a similar approach as Ref. \cite{Alho:2020cdg} by assuming the matterless scenario for simplicity. Using dynamical systems methods, we analysed the dynamics of linear scalar cosmological perturbations for a generic scalar field model. We focused on three scalar perturbations: the evolution of the Bardeen potentials, the comoving curvature perturbation, and the Sasaki-Mukhanov variable (the scalar field perturbation in uniform curvature gauge). To achieve this, we created three autonomous nonlinear first-order ordinary differential equations with a product structure for the state space $S=B\times P$, a product space of the background state space $B$ that describes the dynamics of a Robertson-Walker background, and $P$ which contains Fourier decomposed gauge invariant variables that describe linear cosmological perturbations. Our investigation employed methodologies to explore scalar field theories at the background level for exact spacetimes. Specifically,  an exhaustive dynamical system analysis for each scalar perturbation was presented, and we have integrated the different subsystems numerically. These are powerful tools for investigating homogeneous scalar field cosmologies with arbitrary potential. 

To finish this section, we compare the procedures presented in \cite{Alho:2020cdg} for generic potential with the method of $f$-devisers used here. 
In Ref. \cite{Alho:2020cdg}, are defined for an arbitrary potential the quantities  
\begin{equation}
    \lambda_\phi = - \frac{\sqrt{6}}{6} \frac{d \ln V(\phi)}{d \phi }, \quad \Upsilon_\phi= \frac{V''(\phi)}{6 V(\phi)}. 
\end{equation}
Comparing with  \eqref{sdef} and \eqref{fdef}, we have 
\begin{equation}
  \lambda= \sqrt{6}  \lambda_\phi, \quad f = 6(\Upsilon_\phi-\lambda_\phi^2).  \label{(eq190)}
\end{equation}
Therefore, the conditions upon the scalars   $\lambda_\phi, \Upsilon_\phi$ to satisfy 
\begin{equation}
    \lim_{\phi \rightarrow \pm \infty} \lambda_\phi(\phi)= \lambda_{\pm}, \quad  \lim_{\phi \rightarrow \pm \infty}  \Upsilon_\phi (\phi)= \lambda_{\pm}^2
\end{equation}
are translated in our scenario to  
\begin{equation}
    \lim_{\phi \rightarrow \pm \infty} \lambda(\phi)= \sqrt{6} \lambda_{\pm}, \quad    \lim_{\phi \rightarrow \pm \infty} f (\phi)= 0. 
\end{equation}
Both cases reduce to a Potential that is asymptotically an exponential one as $\phi \rightarrow \pm \infty$. These asymptotically exponential potentials can be studied using the formalism of \cite{Alho:2020cdg} or the procedures previously introduced in \cite{Foster:1998sk}. Following the nomenclature and formalism
introduced in \cite{Foster:1998sk}, let $V:\mathbb{R}\rightarrow \mathbb{R}$ be a $C^2$ non-negative
function. Let there exist some $\phi_0>0$ for which $V(\phi)> 0$
 for all $\phi>\phi_0$ and some number $N$ such that the function
$W_V:[\phi_0,\infty)\rightarrow \mathbb{R},  W_V(\phi)=\frac{V^{\prime}(\phi)}{ V(\phi)} - N $
 satisfies $\lim_{\phi\rightarrow\infty}W_V(\phi)=0$.
Then we say that $V$ is Well Behaved at Infinity (WBI) of
exponential order $N$.

Assume that there are $\phi_{0}> 0$, and a coordinate transformation $\varphi=h(\phi)$,
with inverse $h^{(-1)}(\varphi)$, which maps the interval $[\phi_{0},\infty)$ onto
$(0, \delta]$, where $\delta=h(\phi_{0})$, satisfying $\lim
_{\phi\rightarrow+\infty}h(\phi)=0$, and has the following additional properties:
\begin{enumerate}
\item $h$ is $C^{k+1}$ and strictly decreasing,
\item
\begin{equation}
\overline{h^{\prime}}(\varphi)=\left\{
\begin{array}
[c]{cc}%
h^{\prime}(h^{(-1)}(\varphi)), & \varphi>0,\\
\lim_{\varphi\rightarrow\infty} h^{\prime}(\varphi), & \varphi=0
\end{array}
\right.  \label{eq23}%
\end{equation}
is $C^{k}$ on the closed interval $[0, \delta]$ and
\item $\frac{d \overline{h^{\prime}}}{d \varphi}(0)$ and higher derivatives
$\frac{d^{m}\overline{h^{\prime}}}{d \varphi^{m}}(0)$ satisfy
\begin{equation}
\frac{d \overline{h^{\prime}}}{d \varphi}(0)=\frac{d^{m}\overline{h^{\prime}}}{d \varphi^{m}%
}(0)=0.
\end{equation}
\end{enumerate}

\begin{table}
\begin{center}
\bigskip
\begin{tabular}{|l|l|l|l|l|}
\hline
 $V(\phi)$&$W_V(\phi)$ & $\varphi=h(\phi)$&$\overline{W_V}(\varphi)$ &$\overline{h^{\prime}}(\varphi)$ \\ \hline
$\left|\frac{\lambda}{n}\right|\phi^{n}$&$n\phi^{-1}$
 &$\phi^{-\frac{1}{2}}$
 &$n \varphi^2$&
$-\frac{1}{2}\varphi^3$\\[3pt]
$e^{\lambda\phi}$&0&$\phi^{-1}$&0&$-\varphi^2$\\[3pt]
$2e^{\lambda\sqrt{\phi}}$ &$\lambda\phi^{-\frac{1}{2}}$
&$\phi^{-\frac{1}{4}}$ &$\lambda \varphi^2$&$-\frac{1}{4}\varphi^5$\\[3pt]
$\left(A+(\phi-B)^2\right)e^{-\mu\phi}$ &
$\frac{2\left(\phi-B\right)}{A+(B-\phi)^2}$&$\phi^{-\frac{1}{2}}$
 &$-\frac{2\varphi^2\left(B\varphi^2-1\right)}{A \varphi^4+\left(B\varphi^2-1\right)^2}$&
$-\frac{1}{2}\varphi^3$\\[3pt]
$\left(1-e^{-\lambda^2\phi}\right)^2$ & $ -\frac{2 \lambda
^2}{1-e^{\lambda ^2 \phi }}$
&$\phi^{-1}$& $-\frac{2 \lambda ^2}{1-e^{\frac{\lambda ^2}{\varphi }}}$&$-\varphi^2$\\[3pt]
 $\ln{\phi}$&$(\phi\ln\phi)^{-1}$&$(\ln\phi)^{-1}$&$\varphi e^{-\frac{1}{\varphi}}$&$-\varphi e^{-\frac{2}{\varphi}}$\\[3pt]
$\phi^2\ln{\phi}$&$2\phi^{-1} +(\phi\ln\phi)^{-1}$&
$(\ln\phi)^{-1}$&$(2+\varphi)e^{-\frac{1}{\varphi}}$&$-\varphi e^{-\frac{2}{\varphi}}$\\[3pt]
$\begin{cases}
   V_0(\phi^4+M^4)   &  \textrm{if} \quad \phi<0 \\
   \frac{V_0 M^8}{\phi^4+M^4}  & \textrm{if} \quad \phi \geq 0
\end{cases}$ & $ -\frac{4|\phi|^3}{(\phi^4+M^4)}$ & $|\phi|^{-1}$ & $  -\frac{4|\varphi|}{(1+M^4\varphi^4)}  $ & $- \sgn (\phi) \varphi^2 $\\[3pt]
\hline
\end{tabular}
\caption{\label{WBItransf} Simple examples of WBI behavior at large $\phi$. $n$ and $\lambda$ are arbitrary constants. Adapted
from \cite{Foster:1998sk}.}
\end{center}
\end{table}
Table \ref{WBItransf} displays simple examples of WBI behaviour at large $\phi.$

We see that, as $|\phi|\rightarrow \infty$, a potential function $V(\phi)$ satisfying \eqref{(eq190)} is a function Well-Behaved at Infinity (WBI)  of exponential order $N_\pm= -\sqrt{6} \lambda_{\pm}$. 
Hence, we have to transform the system \eqref{(18)}, \eqref{(19)} and \eqref{(20)} to a system well suited for the analysis at infinity  \cite{Foster:1998sk,Leon:2008de,Fadragas:2014mra}.  Then, using the procedures of \cite{Foster:1998sk,Leon:2008de,Fadragas:2014mra}
for $\phi>0$, and taking the scalar field transformation  $\varphi= h(\phi)$ (satisfying conditions 1, 2 and 3 before), and by replacing 
\begin{equation}
    \lambda \mapsto -\left(\overline{W_V} +N\right), \label{lambda_varphi}
\end{equation} we obtain
\begin{align}
\frac{d x}{{d \bar{N}} }& =- (1-x^2)\left(1-\bar{Z}\right) \left[3 x + \sqrt{\frac{3}{2}} \left(\overline{W_V} +N\right)\right], \label{(57)}\\
\frac{d\varphi}{{d \bar{N}} }& = \sqrt{6} x \overline{h^{\prime}}(\varphi) \left(1-\bar{Z}\right), \label{(59)}\\
 \frac{d\bar{Z}}{d \bar{N}} & = 2\left(3x^2 - 1\right)\bar{Z}\left(1-\bar{Z}\right)^2
\end{align}
with the restriction $x^2+y^2=1$ and considering one of the following perturbation equations for $\theta$. 
From \eqref{(eq:99b)}, 
\begin{align}
&\frac{d\theta}{d\bar{N}} = - \Bigg[\sin^2\theta + \left(7-3x^{2}- \sqrt{6}\left(\overline{W_V}(\varphi) +N\right)\left(\frac{1-x^2}{x}\right)\right) \sin\theta\cos\theta
    \nonumber\\
   & \qquad\qquad +\left(6\left(1-x^2\right)-\sqrt{\frac{3}{2}}\left(\overline{W_V} +N\right)\left(\frac{1-x^2}{x}\right)\right)\cos^2\theta\Bigg]\left(1-\bar{Z}\right)  - \bar{Z}\cos^2\theta.
\end{align}

From \eqref{eq100}, the comoving curvature perturbation evolve as  
\begin{align}
& \frac{d\theta}{d\bar{N}} = - \left[\sin^2\theta - \sqrt{6}\left(\overline{W_V} +N\right)\left(\frac{1-x^2}{x}\right)\sin\theta\cos\theta\right]\left(1-\bar{Z}\right) - \bar{Z}\cos^2 \theta. 
\end{align}
From \eqref{(eq:103)}, by using the transformation \eqref{lambda_varphi}, and 
\begin{equation}
f(\lambda) \mapsto \bar{f}(\varphi)=\left\{\begin{array}{cc}
f\left(-\left(\overline{W_V}(\varphi) +N\right)\right), & \varphi>0,\\
0, & \varphi=0\end{array}\right., 
\end{equation}  and we obtain 
\begin{align}
& \frac{d\theta}{d\bar{N}} =- \Bigg[ \sin^2\theta + 3\left(1-x^2\right)\sin\theta\cos\theta  \nonumber \\
& + 18\left(1-x^2\right) \left(  \frac{\bar{f}(\varphi)}{6}+\left(x+\frac{\left(\overline{W_V}(\varphi) +N\right)}{\sqrt{6}}\right)^2 \right)\cos^2\theta\Bigg]\left(1-\bar{Z}\right)   -\bar{Z} \cos^2\theta,  
\end{align}
defined in the phase-space $B\times P$, modulo $n\pi, n\in\mathbb{Z}$, where  the background space is 
\begin{equation}
B =  \left\{ (x, \varphi, \bar{Z})\in [-1,1] \times [0, h(\phi_{0})] \times [0,1]\right\}, 
\end{equation}
and the perturbation space is 
\begin{equation}
  P =  \left\{ \theta\in  [-\pi, \pi]\right\}. 
\end{equation}
The three possible dynamical systems presented here have the same asymptotic behaviour as those investigated in Sect. \ref{sect:4}. To investigate the potential $V_* \left(e ^ {\kappa \beta (1 - \tanh{(\phi/\beta)})}-1\right)$, $\beta>0$ of \cite{Alho:2020cdg}, we relax the condition $\frac{d \overline{h^{\prime}}}{d \varphi}(0)=0$, and define $h(\phi)=1-\tanh \left(\frac{\phi }{\beta }\right)$  to obtain   $\overline{W_V}(\varphi)=\frac{e^{\beta  \kappa  \varphi} (\beta  \kappa  (\varphi-2) \varphi+2)-2}{\beta  \left(e^{\beta  \kappa  \varphi}-1\right)}$, $\overline{h^{\prime}}(\varphi)=\frac{(\varphi-2) \varphi}{\beta }$ and $N=-2/\beta$ as $\phi\rightarrow \infty$. 

Finally, we have investigated cosmological perturbations in the presence of two matter components, e.g. a perfect fluid and a scalar field with exponential potential. As a drawback of this approach, we must concentrate on a particular cosmological epoch when only one matter component is dominant. In that sense, even though not generic, our subsequent analysis is still relevant when the Universe is a scalar field dominated, e.g. during the early inflationary epoch or the late-time acceleration.  

Our future aim is to evaluate the viability of cosmological models by utilising observational data from various sources such as Supernovae Ia, Cosmic Chronometers, baryon acoustic oscillation, and cosmic microwave background. We will comprehensively analyse beyond the traditional linear stability approach, exploring various gravitational and cosmological models. This analysis will include studying multiple-scale, slow-fast dynamics, averaging theory, and non-smooth dynamical systems. The results of our research will be presented in a forthcoming paper.

\section*{Conflict of Interest Statement} 

The authors declare to have no conflict of interest.

\section*{Author Contributions}

Genly Leon was the main contributor (40\%), followed by Saikat Chakraborty (20\%). The rest authors equally contributed: Sayantan Ghosh (10\%), Raja Solanki (10\%), P.K. Sahoo (10\%) and Esteban Gonz\'alez (10\%). The author's order reflects their contributions. 

\section*{Acknowledgments} \label{sec10}
GL was funded by Vicerrectoría de
Investigación y Desarrollo Tecnológico (Vridt) at Universidad Católica del Norte through
Concurso De Pasantías De Investigación Año 2022, Resolución Vridt No. 040/2022 and
through Resolución Vridt No. 054/2022. He also thanks the support of Núcleo de Investigación Geometría Diferencial y Aplicaciones, Resolución Vridt No. 096/2022. SC acknowledges funding support from the NSRF via the Program Management Unit for Human Resources and Institutional Development, Research and Innovation [grant number B01F650006]. SG acknowledges the Council of Scientific and Industrial Research (CSIR), Government of India, New Delhi, for a junior research fellowship (File no.09/1026(13105)/2022-EMR-I). RS acknowledges UGC, New Delhi, India, for providing a Senior Research Fellowship (UGC-Ref. No.: 191620096030). PKS acknowledges IUCAA, Pune, India, for providing support through the visiting Associateship program. EG acknowledges the support of Direcci\'on de Investigaci\'on y Postgrado at Universidad de Aconcagua.

\bibliographystyle{unsrtnat}
\bibliography{refs.bib}

\end{document}